\documentclass[%
reprint,
 amsmath,amssymb,
 aps %,linenumbers %tightenlines,
]{revtex4-2}

\usepackage{graphicx} % Include figure files
\usepackage{subfigure}
\usepackage{array}
\usepackage{makecell}
\usepackage{comment}
\usepackage{dcolumn}% Align table columns on decimal point
\usepackage{bm}% bold math
\usepackage{stmaryrd}
\usepackage{algpseudocode}
\usepackage{algorithm}
\usepackage{multirow}
\usepackage{xcolor}

\begin{document}

\title{
Modeling framework unifying contact and social networks
}

\author{Didier Le Bail}
\affiliation{Aix Marseille Univ, Universit\'e de Toulon, CNRS, CPT, Marseille, France}
\author{Mathieu G\'enois}%
\affiliation{Aix Marseille Univ, Universit\'e de Toulon, CNRS, CPT, Marseille, France}
\author{Alain Barrat}
\affiliation{Aix Marseille Univ, Universit\'e de Toulon, CNRS, CPT, Marseille, France}
\email{alain.barrat@cpt.univ-mrs.fr}

\date{\today}% It is always \today, today,

\begin{abstract}
Temporal networks of face-to-face interactions between individuals are useful proxies of the dynamics of social systems on fast time scales. Several empirical statistical properties of these networks have been shown to be robust across a large variety of contexts. 
In order to better grasp the role of various mechanisms of social interactions in the emergence of these properties, models in which schematic implementations of such mechanisms can be carried out have proven useful. 
Here, we put forward a new framework to model temporal networks of human interactions, based on the idea of a co-evolution and feedback between (i) an observed network of instantaneous interactions and (ii) an underlying unobserved social bond network: social bonds partially drive interaction opportunities, and in turn are reinforced by interactions and weakened or even removed by the lack of interactions.
Through this co-evolution, we also integrate in the model well-known mechanisms such as triadic closure, but also the impact of {shared social context} and {non-intentional (casual) interactions}, with several tunable parameters. 
We then propose a method to 
compare the statistical properties of each version of the model with empirical face-to-face interaction data sets, to determine which sets of mechanisms lead to realistic social temporal networks within this modeling framework. 
\end{abstract}

\maketitle

%\tableofcontents

\section{\label{sec:level1}Introduction}

Social systems evolve at many different spatio-temporal scales, from individual decision-making or interactions to the history of civilisations. The study of social networks, where individuals are represented by the nodes of the networks and links (ties) are summaries of their social interactions, has proven to be a valuable framework to understand the structure and evolution of these interactions 
\cite{Granovetter:1973,Hinde:1976,wasserman1994social}. To this aim, empirical data on social interactions have largely been collected through surveys \cite{Mossong:2008,Conlan:2011} or direct observation \cite{malik2018bias,schaible2021sensing}.

Recent technological developments have made data available at high temporal and spatial resolution \cite{cattuto2010dynamics,salathe2010high,stehle2011high,barrat2013empirical,stopczynski2014measuring,toth2015role,sapiezynski2019interaction}, providing new proxies of social relationships and making it possible to describe social networks of face-to-face interactions at the spatial scale of a single place such as a conference, a school or a workplace, at time scales ranging from one minute to several days, even if such proxies do not include information about possible discussions or even physical contact, nor about which partner initiated the interaction \cite{schaible2021sensing}.

The resulting data are typically represented as temporal networks \cite{holme2015modern,holme2012temporal}, where we associate a node to each social agent and we draw an edge between $i$ and $j$ at time $t$ if $i$ and $j$ were interacting at time $t$: this 
has allowed to study the statistical properties of a number of relevant observables such as the duration of interactions, or the time elapsed between consecutive interactions. The
resulting distributions are typically broad with robust functional shapes across contexts \cite{cattuto2010dynamics,barrat2013empirical,sapiezynski2019interaction}.
Aggregating the interactions along the temporal dimension can also make structures at larger time scales visible: aggregated interaction networks typically exhibit a small world topology, a high clustering coefficient and broad distributions of edge weights (the edge weight being defined as the aggregated duration of interactions along that edge), with similar shapes in different social contexts \cite{barrat2013empirical}.
 
The robustness of these properties
has motivated the search for models of temporal networks that could reproduce the observed statistical distributions at diverse time scales
\cite{stehle2010dynamical,Karsai2012,vestergaard2014how,karsai2014time,Laurent_2015}, with a dual aim: on the one hand, understanding which social mechanisms lead to the emergence of these properties, and, on the other hand, producing synthetic realistic data sets that can be of use to study dynamical processes on temporal networks.   

The main social mechanisms implemented in such models include (i) reinforcement processes, where the probability for two nodes to interact with each other increases after each interaction, leading to broad distributions of contact durations and edge weights in the aggregated network
\cite{stehle2010dynamical,vestergaard2014how,karsai2014time}; (ii) triadic closure, which states that a node is more likely to interact with a neighbour of a neighbour, and has been shown to account for the high clustering coefficient of the aggregated network;
(iii) ``memory loss process'', which can be random or target unused social ties \cite{gelardi2021temporal,Laurent_2015}, and
contributes to the emergence of community structure in the aggregated network of social systems \cite{jo2011emergence,kumpula2007emergence,Laurent_2015}.

%%% % 

In this paper, we extend the modeling of temporal networks of face-to-face interactions in two main directions. On the one hand, we go beyond the commonly considered observables mentioned above, as they do not cover the entire complexity of the empirical networks' structures. 
We do not intend to answer the question of which list of observables would fully characterize a social system represented as a temporal network, as this question is not fully answered even for static network representations \cite{orsini2015quantifying}. However we extend the set of commonly used observables: we consider the distributions of
the node activity duration and interduration, and of the duration of newly established edges,
as well as structural patterns such as the size of connected components in the instantaneous graph of interactions, and spatio-temporal patterns like Egocentric Temporal Networks (ETN) \cite{Longa2022}, which have recently been shown to be useful building blocks to decompose a temporal network \cite{longa2022neighbourhood}.

On the other hand, we propose a modeling framework based on a core hypothesis: the existence of an underlying (not observable) directed temporal network called the social bond graph $B$, which co-evolves with the observed temporal network of interactions denoted $G$. 
The weight of an edge in $B$, $B_{ij}(t)$, represents how much $i$ is inclined to interact with $j$ at time $t$ ($B$ is thus directed as the inclination of $i$ towards $j$ can differ from the inclination of $j$ towards $i$), while the undirected temporal edge
$G_{ij}(t)$ is simply $1$ if $i$ and $j$ interact at $t$ and $0$ otherwise.
The evolutions of $B$ and $G$ follow two feedback mechanisms.
First, $B(t)$ guides the interactions that will take place at $t$, i.e., influences the edges of $G(t)$. Second, interactions have an impact on social bonds
through a reinforcement mechanism \cite{gelardi2021temporal}: if an interaction occurs between $i$ and $j$, then $B_{ij}$ increases. Moreover, we take into account that the time and energy spent to maintain the tie with an individual is taken from a finite interaction capacity and is thus time not spent with others \cite{dunbar2009time,miritello2013limited}. Therefore, if $i$ and $j$ do not interact but $i$ interacts with another agent $k$ at $t$, $B_{ij}$ decreases \cite{gelardi2021temporal}.

We integrate this framework within a well-known framework for temporal network modelling, the Activity Driven (AD) model  \cite{perra2012activity}: in this model, nodes representing social agents are endowed with an intrinsic ``activity'' quantifying their propensity to form edges at each time step. The initial model \cite{perra2012activity} has been refined to introduce memory of past interactions (ADM model), as well as triadic closure and 
renewal of agents
\cite{karsai2014time,Laurent_2015,ubaldi2016asymptotic}. Here, through the co-evolution of the instantaneous network of interaction $G$ and the social bond network $B$, we modify 
the implementation of these mechanisms,
and integrate additional ones, namely: 
(i) the possible disappearance of a directed social bond when it becomes too weak; 
(ii) the influence of social context (e.g., two social agents belonging to the same group of discussion, having common neighbours, are more likely to interact with each other);
(iii) the distinction between intentional and casual interactions driven by the context.

To investigate which of the proposed mechanisms are relevant for the study of social systems, we test several variations of the resulting models. We put forward a systematic way to compare them with empirical data sets, by
computing the distance between model generated and empirical distributions for a given collection of observables.
We use this method to optimize the parameters for each model version, and then to rank versions according to their distance to empirical data.

\section{Framework}

\subsection{Interaction and social bond graphs}

Our framework consists in laws of evolution for two temporal networks, an interaction graph $G$ and a social bond graph $B$. We recall that a weighted temporal network 
 $g$ can be defined in discrete time as:
\begin{align*}
    g\colon\mathbb{N}\times V^{2} & \longrightarrow\mathbb{R}^{+}\\
    (t,i,j) & \longmapsto g_{ij}(t)
\end{align*}
where $V$ is the set of nodes and $g_{ij}(t)$ is the weight of the edge $(i,j)$ at time $t$. We denote by $N=|V|$ the number of nodes.

The interaction graph $G$ is an undirected and unweighted temporal network in discrete time, with finite duration $T$. The $N$ nodes of $G$ represent social agents, and $G_{ij}(t) = 1$ is interpreted as the fact that $i$ and $j$ are interacting at time $t$ (else, $G_{ij}(t) = 0$). We denote by $E(t)$  the set of such active edges of $G$ at $t$.
The social bond graph $B$ is a directed and weighted temporal network, on the same $N$ nodes and same timestamps as $G$: the weight $B_{ij}(t)$ stands for the social affinity of $i$ towards $j$ at $t$. The egonet of $i$ at time $t$ is defined as the set of neighbours of $i$ in $B$ at time $t$, i.e. 
$\gamma_{i}(B(t)) = \{j | B_{ij}(t) >0\}$.

We note here that we consider only positive interactions, for both $G$ and $B$. While negative (hostile) interactions do occur in social networks, and negative social
bonds exist as well, they are indeed typically difficult to observe concretely  \cite{thurner2018virtual}. In fact, negative social bonds are often 
deduced from an avoidance of interactions (i.e., two individuals interacting less than expected by chance)
\cite{ilany2013structural,thurner2018virtual,gelardi2019detecting,andres2022reconstructing}, an assumption that has been shown to be able to provide
support to social theories such as the social balance theory \cite{ilany2013structural,gelardi2019detecting}. Here therefore we do not distinguish between
an absence of interaction or of social bond and a negative one.

The evolutions of $G$ and $B$ are dependent on each other along the following lines.
First, interactions taking place at $t$ depend on interactions at the previous time: indeed, two agents belonging to the same group of discussion are more likely to interact in a close future. This can be formalized for instance by the existence of common neighbours in $G$ at the previous time step, giving rise to an influence of $G(t-1)$ on $G(t)$. 

Agents also choose their partners based on a long-term memory of their previous interactions. In particular, the more two nodes have interacted with each other in the past, the more likely they are to interact in the future. 
Hypothesizing that the edge weights of $B$ can encode this memory effect, it follows that 
the social bond weights at $t$ also influence $G(t)$ (a node will more likely choose a partner with whom it has a high affinity).

Reciprocally, the social bond graph is updated according to the interaction graph, following the reinforcement process of \cite{gelardi2021temporal}:
the weight $B_{ij}$ increases if $i$ and $j$ interact with each other, and stays the same or decreases if they do not: 
$$
\begin{cases}
G_{ij}(t)>0\implies B_{ij}(t+1) > B_{ij}(t)\\
G_{ij}(t)=0\implies B_{ij}(t+1)\leq B_{ij}(t) .
\end{cases}
$$
We initialize $B_{ij}$ as being $0$ for all times before the first interaction between $i$ and $j$ on $G$:
$\forall t\in\mathbb{N},\forall i,j\in V,
G_{ij}(\tau)=0,\forall\tau\leq t\implies B_{ij}(\tau)=0,\forall\tau\leq t$, thus assuming that no pre-existing social bonds exist between the nodes.

\begin{figure}
    \centering
    \includegraphics[scale=0.4]{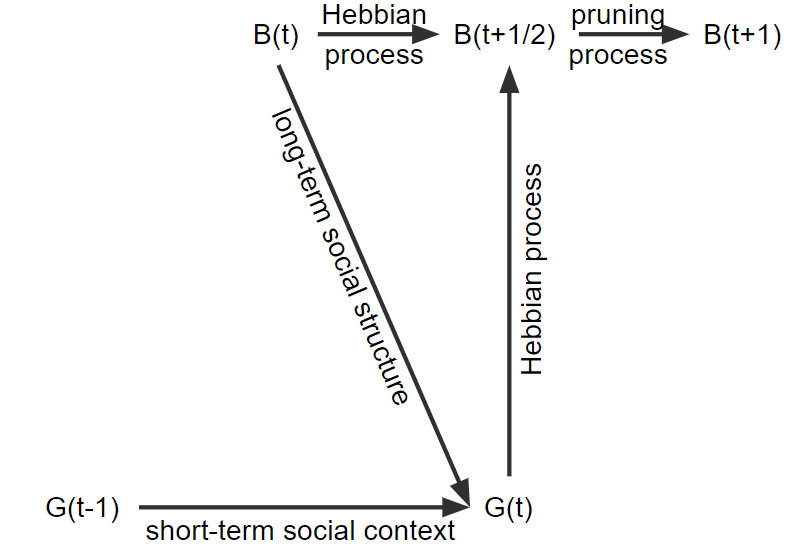}
    \caption{
    \textbf{Sketch of the dependencies between the interaction graph $G$ and the social bond graph $B$.} Edges having a higher weight in the social bond graph $B(t)$ are more likely to activate, i.e. to be part of the interaction graph $G(t)$. The computation of $B(t+1)$ is done in two steps: first $B$ is updated by the feedback of which edges were active in $G(t)$: unused ties decay while used ties strengthen. The output of this first step is denoted by $B(t+\frac{1}{2})$ because it refers to an intermediary step between $B(t)$ and $B(t+1)$, which is obtained from $B(t+\frac{1}{2})$ by a pruning process, consisting in removing weak unused social ties.
    The arrow from $G(t-1)$ to $G(t)$ is of a different nature than the arrow from $B(t)$ to $B(t+\frac{1}{2})$. The latter accounts for the inertia of the social bond graph, as $B$ can only encounter gradual change from one time to the next (implementing long-term memory). On the contrary, the arrow from $G(t-1)$ to $G(t)$ does not ensure that $G(t)$ will be similar to $G(t-1)$: it simply describes a short-term social context memory through the fact that the more two nodes share partners in $G(t-1)$, the more likely they are to be partners in $G(t)$.}
    \label{fig:01}
\end{figure}

In summary, $G(t)$ is determined both by $G(t-1)$ and $B(t)$, and in return, $B(t+1)$ is determined by $G(t)$ and $B(t)$  (see 
 Fig. \ref{fig:01}).

\subsection{Social mechanisms}

We use the framework described above to model several social mechanisms.

The first mechanism is a short-term reinforcement process with a long-term memory, through the co-evolution of $G$ and $B$: social agents remember with whom they have interacted
 and reinforce their social ties with their partners at each interaction, while unused ties weaken. In addition, we assume that weakened ties may vanish: at each timestep $B_{ij}$ has a certain probability to be reset to zero. 
 To  capture the realistic assumption that a node tends to shorten unfruitful partnerships to save time or energy, this probability increases as $B_{ij}$ decreases.

The second mechanism we consider is the cyclic closure in the social bond graph $B$. This mechanism captures the fact that, when a social agent initiates a new partnership, it may give the priority to the partners of its partners. 
Through this mechanism, the existing social bonds drive thus the interactions on $G$.

The third mechanism grasps the fact that two nodes belonging to the same group of discussion are more likely to start interacting together, whether or not they know each other \footnote{Note that this is different from
 focal closure, which suggests the formation of ties between individuals with common attributes or interests, and which is implemented by  links with randomly chosen individuals in \cite{Laurent_2015}}. 
This can be translated by an increased probability of interaction in $G(t)$ between nodes that were in the same connected component of $G(t-1)$ or, more simply, between nodes
that had common neighbors in $G(t-1)$.

The fourth mechanism is a dynamic triadic closure driven by the current context, accounting for the fact that if a node interacts simultaneously with two different nodes, these nodes are likely to also be interacting with each other. It is important to note that this mechanism leads to interactions that are contextual and may thus be of a fundamentally different social significance than intentional ones. In particular, we will take into account that contextual and intentional interactions on $G(t)$ might not influence the evolution of the social bonds in $B$ in the same way.

\begin{figure}
    \centering
    \includegraphics[width=\columnwidth]{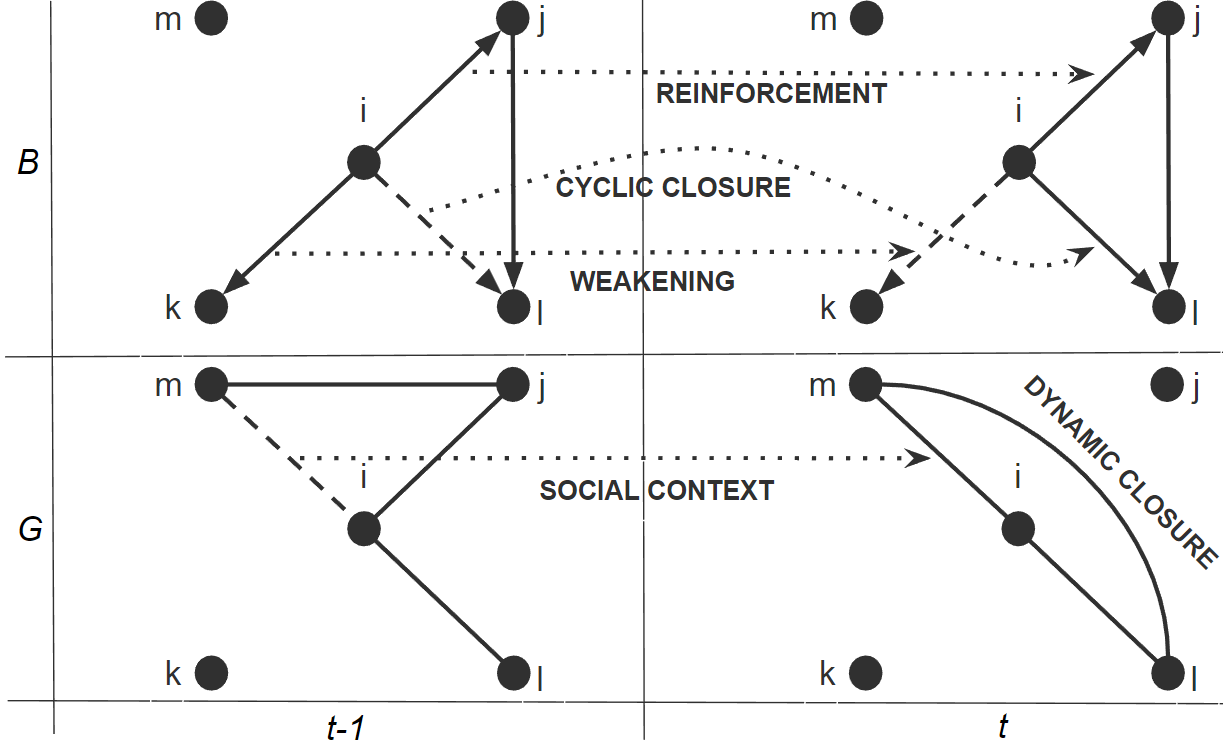}
    \caption{
    \textbf{Social mechanisms.} We focus on the node $i$ for two consecutive time steps $t-1$ and $t$, so that we do not represent all the ties of $B$.
    (i) \textit{reinforcement}: $i$ interacts with $j$ at $t-1$, leading to a reinforcement of the social affinity of $i$ towards $j$: $B_{ij}(t)>B_{ij}(t-1)$.
    (ii) \textit{cyclic closure}: at $t-1$, $i$ decides to interact with a new partner. First $i$ picks a known partner $j$ and then a known partner $l$ of $j$. The tie $B_{il}$ is created at $t$.
    (iii) \textit{weakening}: while $k$ is part of the egonet of $i$, $i$ does not interact with $k$ at $t-1$. This results in a weakening of the social affinity of $i$ towards $k$: $B_{ik}(t)<B_{ik}(t-1)$.
    (iv) \textit{social context}: the social affinity of $i$ towards $m$ at $t$ is temporarily increased by their common partners in $G(t-1)$.
    (v) \textit{dynamic closure}: once the intentional interactions have all been drawn at $t$,  $(m,i,l)$ is an open triangle in $G(t)$, which is likely to close because $m$ and $l$ are interacting with the same agent. 
    }
    \label{fig:02}
\end{figure}

The four mechanisms are summarized in Figure~\ref{fig:02}.

\subsection{Model implementation}
\label{subsec:1}
Let us now translate the mechanisms described into microscopic rules of evolution. To this aim, we focus on the AD model in discrete time
 \cite{perra2012activity,karsai2014time,Laurent_2015}: each node is endowed with an intrinsic activity parameter $a_{i}$, which gives its probability to be active at each time step. The difference between an active node and an inactive node is that only active nodes can emit intentional interactions.

\subsubsection{Creation of the temporal edges of $G(t)$}
 
At each time $t$, each active node $i$ makes $m_{i}$ attempts of {\em intentional} interactions, in a way depending on the interactions at the previous time step ($G(t-1)$) and of the current social bond graph ($B(t)$). At each such attempt: 
\begin{itemize}
    \item With probability $p_{g}$, $i$ will extend its egonet, i.e., create an interaction with a node $j$ with whom it has no social bond ($B_{ij}(t)=0$). In this case, $i$ chooses an interaction partner either uniformly at random (with probability $p_u$), or, with probability $1-p_{u}$, by triadic closure driven by the social bond graph $B$ (second mechanism above):  $i$ creates an interaction in $G(t)$ with a neighbour $j$ of a neighbour $k$ in $B(t)$.
    Moreover, the choice of $k$ and $j$ are driven by (i) the weights in the social bond graph, $B_{ik}(t)$ and $B_{kj}(t)$, and 
    (ii) the possible existence of a recent common social context (third mechanism). Specifically, 
    the first neighbour $k$ is chosen with probability $P(i\xrightarrow{}k)\propto B_{ik}(t)$, i.e., using the social affinity (independently from a social context). 
    The choice of $j$ as a neighbour of $k$ can be interpreted as a recommendation from $k$ to $i$; therefore, we include here the influence of a social context 
   recently shared by $k$ and $j$, and $j$ is chosen  among all neighbours of $k$ with probability:
    \begin{equation}
        \tilde{P}(k\xrightarrow{}j)\propto c_{kj}(t-1)B_{kj}(t) .
    \end{equation}
The coefficient $c_{kj}(t)$ is defined as:
    \begin{equation}
        c_{kj}(t)=1+\left|\gamma_{k}(G(t))\cap\gamma_{j}(G(t))\right| \ ,
    \end{equation}
    representing the boosting of the social affinity by the potential sharing of common neighbours in the previous time step ($\gamma_{\ell}(g)$ denotes the set of neighbours of a node $\ell$ in a graph $g$).
\item With probability $1-p_{g}$, $i$ does not extend its egonet, i.e., interacts with one of its neighbours in $B$. This neighbour  $j$ is chosen with probability 
$\tilde{P}(i\xrightarrow{}j)\propto c_{ij}(t-1)B_{ij}(t)$, i.e., proportionally to $i$'s affinity towards $j$, boosted by the potential existence of common neighbours in $G$ at the previous timestep (first and third mechanisms).
\end{itemize}

In addition to these intentional interactions, 
casual, {\em contextual} interactions can occur (fourth mechanism). To take this into account, we implement here a variation of the dynamic triadic closure. Namely, we consider that
for each open triangle in $G(t)$ made up of two intentional interactions e.g. $(i,k)$ and $(k,j)$, $i$ and $j$ interact with each other with probability $P_{c}(i,k,j)$ in a contextual, non intentional manner. For the open triangle $(i,k,j)$ to close, either $i$ or $j$ has to propose the contextual interaction. Denoting the probability that $i$ decides to close the triangle by $p_{ij}$, we have:
\begin{equation}
    P_{c}(i,k,j)=1-(1-p_{ij})(1-p_{ji})
\end{equation}

\begin{figure}
    \centering
    \includegraphics[scale=0.3]{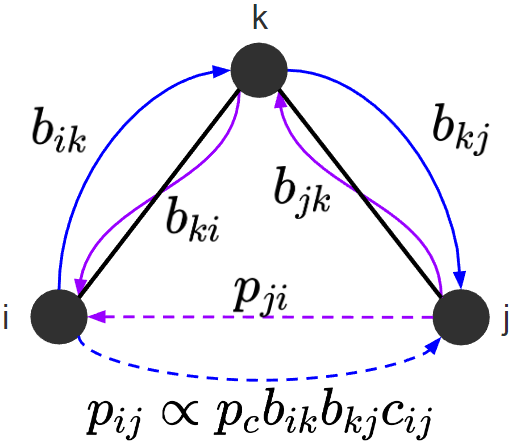}
    \caption{
    \textbf{Dynamic triadic closure.} After computation of the intentional interactions in $G(t)$, we identify its open triangles. Here $(i,k,j)$ is such an open triangle with edges $(i,k)$ and $(k,j)$ (black straight lines). Closing the triangle means either $i$ decides to interact with $j$ (probability $p_{ij}$), or vice-versa (probability $p_{ji}$). $p_{ij}$ depends both on how close are $i$ and $j$ in the current social context ($c_{ij}$) and how close they are relatively to their common partner of discussion ($b_{ik}$ and $b_{kj}$): if $i$ gives a lot of attention to $k$, and $k$ a lot of attention to $j$, then it is likely that $i$ and $j$ will interact.
    }
    \label{fig:015}
\end{figure}

In our implementation (Fig. \ref{fig:015}), $p_{ij}$ takes also into account whether or not $i$ is in the active state: as only active nodes can emit interactions, $p_{ij}=0$ if $i$ is inactive.
Moreover, we assume that it
depends both on the instantaneous social affinity 
$b_{ik}$ of $i$ towards $k$ and the instantaneous social affinity $b_{kj}$  of $k$ towards $j$. We define this instantaneous social affinity of a
node $\ell$ towards a node $m$ as follows: 
if $m$ is part of the egonet of $\ell$, then we simply define $b_{\ell m}$ as $\tilde{P}(\ell\xrightarrow{}m)$, i.e. 
$b_{\ell m}\propto c_{\ell m}(t-1)B_{\ell m}(t)$; 
if instead $B_{\ell m}$ is zero, we use $b_{\ell m}=p_{g}$ (probability that $\ell$ grows its egonet).
We thus use:
\begin{equation}
    p_{ij}(t)=\frac{p_{c}b_{ik}(t)b_{kj}(t)c_{ij}(t-1)}{1+p_{c}b_{ik}(t)b_{kj}(t)(c_{ij}(t-1)-1)}    
\end{equation}
where $0 \le p_{c} \le 1$ is a free parameter (we use $c_{ij}$ measured at $t-1$ as previously, as it is the social context of the previous time step that influences
the link creation at $t$). This form ensures that 
$p_{ij}$ grows with $p_{c}b_{ik}b_{kj}c_{ij}$ 
(i.e., is influenced by the social affinities and by the context) and remains between $0$ and $1$.

\subsubsection{Evolution of the social bonds of $B(t)$}

\begin{figure}
    \centering
    \includegraphics[width=\columnwidth]{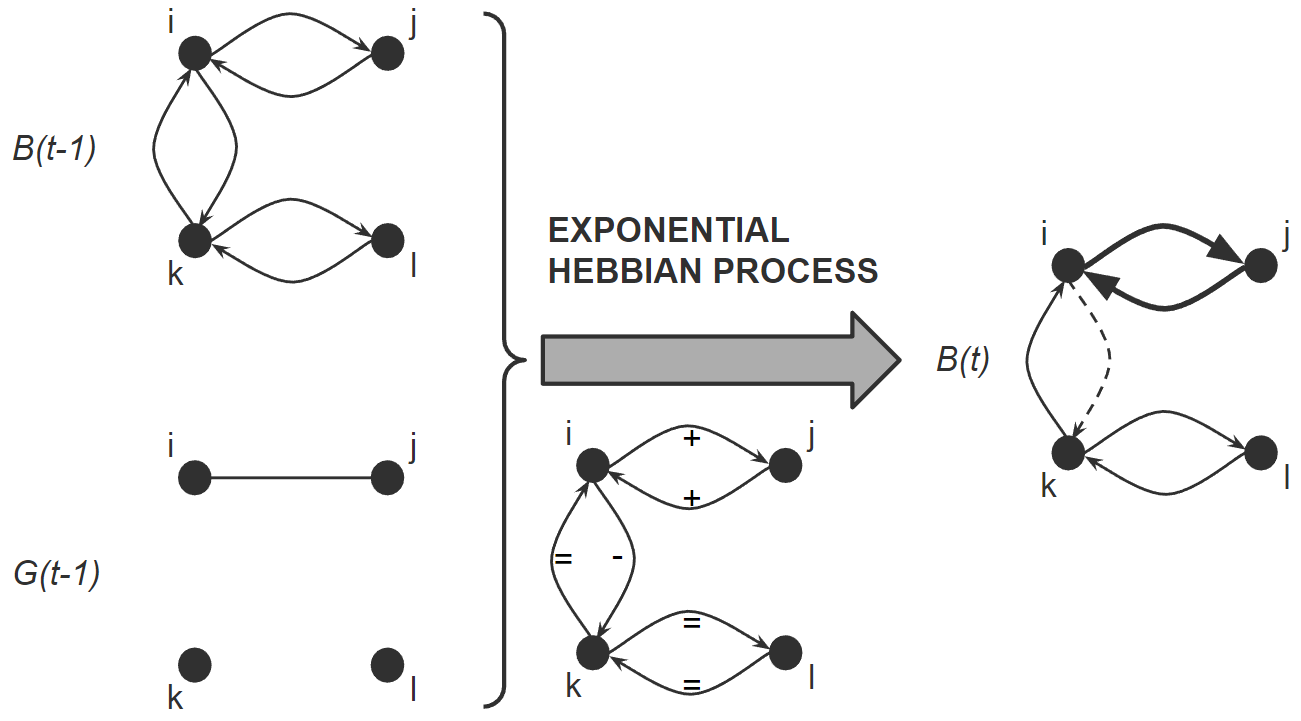}
    \caption{\textbf{Sketch of the Hebbian-like process} describing the evolution of the directed social bond weights from $B(t-1)$ to $B(t)$ due to the interactions in $G(t-1)$. Active edges are reinforced, inactive ties starting from an interacting node are weakened, and ties starting from a non-interacting node are unchanged.     
   }
    \label{fig:03}
\end{figure}

The interaction graph at $t$, $G(t)$, is thus composed of the intentional and contextual interactions of all active nodes at $t$. We denote the set of intentional interactions by $I(t)$, and the set of contextual ones by $C(t)$. These interactions determine the change in the social bond graph from time $t$ to the next time step $t+1$. The corresponding update (first mechanism) consists in two steps: a Hebbian-like process and a pruning process. During the Hebbian process, edges of $B(t)$ are either reinforced, weakened or let invariant, according to the rule introduced in \cite{gelardi2021temporal}: if a node 
$i$ interacts with $j$ but not $k$, then $B_{ij}$ and $B_{ji}$ may be reinforced, but $B_{ik}$ is weakened (see Figure \ref{fig:03}). If $i$ has no interaction at all, its social bonds are not changed.

As a refinement of the reinforcement rule \cite{gelardi2021temporal}, we introduce a distinction between contextual and intentional interactions.
To this aim, we denote by $R(t)$ the set of social ties that will be strengthened between $t$ and $t+1$, and by $W(t)$ the set of ties that cannot be weakened (among the ties starting from nodes that have an interaction in $G(t)$, as the nodes with no interaction at $t$ are not affected).

We choose $R$ and $W$ depending on the roles we give to intentional and contextual interactions.
A first possibility is to put all interactions on an equal footing: then all 
active edges are reinforced independently on whether they were intentional or contextual, i.e. $R=W=I\cup C$. 
If we consider only intentional interactions as relevant, and contextual interactions as noise, then edges from $C$ are not taken into account in the process: $R=W=I$. 
Finally, if we consider that contextual interactions are neutral, they should give rise neither to a reinforcement nor to a weakening, i.e. $R=I$ and $W=I\cup C$. These possible choices are summarized in Table \ref{tab:01}.

\begin{table}[!ht]
    \centering
    \begin{tabular}{c|c|c}
    Interpretation & $R$ &  $W$  \\
    \hline
{all interactions are equivalent } & $I\cup C$ &$I\cup C$\\
context interactions are neutral&  $I$ & $I\cup C$ \\
 context interactions are noise & $I$ & $I$ \\
    \end{tabular}
    \caption{\textbf{Choices for the update of $B$.}}
    \label{tab:01}
\end{table}

To precisely define the process, we need to specify at which rate a given tie strengthens or weakens. We denote strengthening rates by $\alpha$ and weakening rates by $\beta$.
In order to keep the weights of social ties bounded between $0$ and $1$ \cite{gelardi2021temporal}, we also consider rates in $[0,1]$, and we assume them constant.
While these rates are also uniform in \cite{gelardi2021temporal}, we consider here that they can be different for different individuals or different ties. 
We write the general evolution rules as:
\begin{equation}
\forall (i,j)\in R
\begin{cases}
    B_{ij}(t+1)=B_{ij}(t) + \alpha_{ij} (1 - B_{ij}(t) ) \\
    B_{ji}(t+1)= B_{ji}(t) + 
    \alpha_{ji} (1 - B_{ji}(t)) 
\end{cases}
\end{equation}
and:
\begin{equation}
\forall i\in \tilde{R},\forall k,(i,k)\notin R\cup W, B_{ik}(t+1)=(1-\beta_{ik})B_{ik}(t)
\end{equation}
where $\tilde{R}$ denotes the set of nodes involved in the links of $R$: $\tilde{R}=\{i|\exists j, (i,j)\in R\}$.

Note that in the original ADM model \cite{Laurent_2015}, the social bond weights are not bounded, and simply increase by $1$ at each interaction.

To obtain $B(t+1)$, we include an additional step, namely a pruning of the social bonds, to take into account the fact that weak social bonds might vanish
(in the original ADM instead, node disappearance is implemented uniformly at random \cite{Laurent_2015}, i.e., with no relation to the actual social bonds). 

To quantify how weak is a directed tie $(i,j)$, we compare the probability $P(i\xrightarrow{}j) \propto B_{ij}$ that $i$ selects $j$ among all its neighbours to interact with, with this same probability if all ties starting from $i$ had the same weight. Denoting by $d_{i}^{out}$ the number of out-links of $i$ in $B(t)$, a homogeneous partition of $i$'s interest towards its neighbours would correspond to 
%$B_{ij}^{hom}(t)=\sum_{k}B_{ik}(t) / d_{i}^{out}$, i.e., to
$P_{hom}(i\xrightarrow{}j) = 1/d_{i}^{out}$.
Therefore, we use as the probability to remove the directed tie $(i,j)$:
\begin{multline}
    \forall i\in \tilde{R},\forall j,(i,j)\notin R\cup W,\\
    P_{d}(ij)=\exp\left(-\lambda {d_{i}^{out}} P(i\xrightarrow{}j)\right)
\end{multline}
where $\lambda$ is a tunable parameter.
%$$\begin{cases}
%P(i\xrightarrow{}j)=\frac{B_{ij}(t)}{\sum_{k}B_{ik}(t)}\\
%d_{i}^{out}=\left|\{k|B_{ik}(t)>0\}\right|=|\gamma_{i}(B(t))|
%\end{cases}$$
$P_{d}(ij)$ is thus large if $P(i\xrightarrow{}j)$ is smaller than its homogeneous counterpart, and decreases exponentially when the importance of $j$ for $i$ increases.

\subsubsection{Model versions}

Even within the model implementation described in the previous paragraphs, we can define various versions of the model, with for instance different values or distributions of the parameters. Therefore, we first define  
a baseline version (version V1) with the following features:
\begin{itemize}
    \item $a_{i}$ is drawn from a power-law of exponent -1 with bounds $a^{\text{min}}$ and $a^{\text{max}}$;
    
    \item  $m_{i}$ is drawn from a uniform law in $\llbracket1,m^{\text{max}}\rrbracket$;
    
    \item $\alpha_{ij}=\beta_{ij} \equiv \alpha_i$ depends only on $i$, and $\alpha_{i}$ is drawn from a power-law of exponent -1 with bounds $0.001$ and $1$;

    \item the social context at the previous time step is taken into account through $c_{ij}(t-1)=1+\left|\gamma_{i}(G(t-1))\cap\gamma_{j}(G(t-1))\right|$;
    
    \item contextual interactions are neutral ($R=I$, $W=I\cup C$, see Table \ref{tab:01});
    
    \item the remaining free parameters are: $p_{g}$, $p_{u}$, $p_{c}$, $\lambda$.
\end{itemize}

We then implement variations with respect to the baseline, by changing in each case only one of the mechanism implementations, as summarized in Table \ref{tab:02}. We call these versions \textit{adjacent} versions, because they differ from the baseline in one aspect only. We tested 12 adjacent versions, numerated from 2 to 13. The version 14 
corresponds to the original ADM of \cite{Laurent_2015}, with the following properties:
\begin{itemize}
    \item the egonet growth rate is not constant. Instead of having a fixed probability $p_{g}$ of growing its egonet, each node $i$ grows it with a probability depending on its egonet size: $p_{g}(i)=\frac{c}{c+d_{i}^{out}}$, where $c\in\mathbb{N}$ is a model parameter;
    \item the recent social context is not taken into account: the direct influence of $G(t-1)$ on $G(t)$ is cut off, i.e. $c_{ij}(t)=1$;
    \item no contextual interactions are considered, i.e. $p_{c}=0$;
    \item $B$ has a linear reinforcement process  $B_{ij}(t+1)=B_{ij}(t)+1$ for each $(ij)$ in $G(t)$, and weakening of unused social bonds is not considered;
    \item a node pruning process: instead of removing social ties, we remove social agents with a constant probability $p_{d}$. After removing the social agent $i$, we reinsert it into the system to keep the number of agents constant, but with $B_{ij}=0$ $\forall j$.
\end{itemize}
This version is thus actually a \textit{composite} version (i.e., obtained by combining adjacent ones), because it differs from the baseline in more than one aspect. 

\begin{table*}
\hspace*{-1.5cm}
\renewcommand{\arraystretch}{1.3}
    \begin{tabular}{ccc|c|c|c|c|c|c|c|c|c|c|c|c|c|c|l}
        \cline{4-17}
        & & &  \multicolumn{14}{ c| }{version number} \\ \cline{4-17}
        & & & 1 & 2 & 3 & 4 & 5 & 6 & 7 & 8 & 9 & 10 & 11 & 12 & 13 & 14\\
        \cline{2-17}
        \multirow{9}{*}{\rotatebox{90}{\parbox[c]{4cm}{\centering social mechanisms}}} &
        \multicolumn{1}{ |c  }{social context} &
        \multicolumn{1}{ c| }{} & Yes & - & - & - & No & - & - & - & - & - & - & - & - & No\\
        \cline{2-17}
        & \multicolumn{1}{ |c  }{additional interactions} &
        \multicolumn{1}{ c| }{} & neutral & - & - & - & - & equivalent & noise & None & - & - & - & - & -  & None\\
        \cline{2-17}
        & \multicolumn{1}{ |c  }{egonet growth} &
        \multicolumn{1}{ c| }{} & constant & - & - & variable & - & - & - & - & - & - & - & - & -  & variable\\ \cline{2-17}
        & \multicolumn{1}{ |c  }{\textit{social bond graph update}} &
        \multicolumn{1}{ c| }{} & &  &  &  &  &  & &  &  &  &  & & &  &\\
        & \multicolumn{1}{ |c  }{Hebbian process} &
        \multicolumn{1}{ c| }{} & $\alpha_{i}$ & linear & - & - & - & - & - & - & $\alpha$ & $\alpha_{i}$, $\beta_{i}$ & $\alpha_{ij}$, $\beta_{ij}$ & - & - & linear\\
        & \multicolumn{1}{ |c  }{$(R,W)$} &
        \multicolumn{1}{ c| }{} & $(I,I\cup C)$ & - & - & - & - & $(I\cup C,I\cup C)$ & $(I,I)$ & $(I,I)$ & - & - & - & - &  - & $(I,I)$\\
        & \multicolumn{1}{ |c  }{Pruning process} &
        \multicolumn{1}{ c| }{} & social tie & - & social agent & - & - & - & - & - & - & - & - & - &  - & social agent\\
        \cline{2-17}
        & \multicolumn{1}{ |c  }{$m_{i}$} &
        \multicolumn{1}{ c| }{} & $\mathcal{U}(\llbracket1,m^{\text{max}}\rrbracket)$ & - & - & - & - & - & - & - & - & - & - & - & constant & constant\\ \cline{2-17}
        & \multicolumn{1}{ |c  }{node activity} &
        \multicolumn{1}{ c| }{} & $a_{i}$ & - & - & - & - & - & - & - & - & - & - & $a$ &  - & -\\
        \cline{2-17}
    \end{tabular}
    \caption{\textbf{Model versions.} The version 1 is the baseline version, while the version 14 is the original ADM.
    The symbol - means identical to the baseline version. In the row entitled ``additional interactions'', we precise the role of the interactions obtained through the dynamic triadic closure mechanism. The indication ``None'' means that this mechanism does not exist, i.e. $p_{c}=0$.
    In the row entitled ``Hebbian process'', a symbol $\alpha_{i}$ alone means three things. First, the Hebbian process used is an exponential process. Second $\alpha_{ij}=\beta_{ij}=\alpha_{i}$, and third $\alpha_{i}$ is drawn independently for each $i$ from the same power-law of exponent $-1$.
    Similarly in the row entitled ``node activity'', a symbol $a_{i}$ means that $a_{i}$ is drawn from a power-law of exponent -1 independently for each node.
    On the contrary, an unscripted symbol, like $a$ or $\alpha$, means that the same value is assigned to every node.
    We put an additional symbol $\beta_{i}$ or $\beta_{ij}$ when the decay rate is drawn independently from the strengthening rate. However, $\alpha_{i}$, $\beta_{i}$, $\alpha_{ij}$, $\beta_{ij}$ are all drawn from power-law distributions with the same exponent -1.
    In the row titled ``$m_{i}$'', the symbol $\mathcal{U}(\llbracket1,m^{\text{max}}\rrbracket)$ means that $m_{i}$ is drawn independently for each $i$ from the uniform law on the set of integers $\{1,2,\ldots,m^{\text{max}}\}$.
    }
    \label{tab:02}
\end{table*}

\section{Comparison with empirical data sets}

\begin{table}
    \centering
    \begin{tabular}{|c|c|c|c|c|}
        \hline
        & name & nodes & timestamps & temporal edges\\
        \hline
        \rotatebox{0}{Conferences}&
        \makecell{conf16\\
        conf17\\
        }&
        \makecell{138\\274
        }&
        \makecell{3 635\\7 250
        }&
        \makecell{153 371\\
        229 536
        }\\
        \hline
        \rotatebox{0}{Schools}&
        \makecell{utah\\highschool3
        }&
        \makecell{630\\327
        }&
        \makecell{1 250\\7 375}&
        \makecell{353 708\\188 508}\\
        \hline
        \rotatebox{0}{Workplace}&
        \makecell{work2
        }&
        \makecell{217
        }&
        \makecell{18 488}&
        \makecell{78 249}\\
        \hline
    \end{tabular}
    \caption{\textbf{Sizes of the empirical data sets considered in this paper.}}
    \label{tab:03}
\end{table}

We consider as references several publicly available empirical data sets describing face-to-face interactions in different contexts, namely two scientific conferences, two schools and a workplace (See Table  \ref{tab:03} and Supplemental Material, SM). As our aim is to evaluate which hypotheses made on some social mechanisms yield realistic temporal networks, we will thus evaluate how close are the temporal networks $G$ generated by each model version to each reference empirical data set. Note that we compare $G$ and not $B$, as the empirical data sets correspond to instantaneous interactions.

The properties of the temporal networks generated by each model version naturally depend on the version parameters. Some 
can be extracted or estimated directly from the reference data set: the number of nodes $N$, the duration $T$, and the observed minimum and maximum node activities, $a^{\text{min}}_{\text{obs}}$ and $a^{\text{max}}_{\text{obs}}$. The other parameters are however a priori unknown and tunable. To limit the number of free parameters, we fix the 
bounds for the power-law followed by the strengthening and weakening rates of the social bonds, $\alpha_{ij}$ and $\beta_{ij}$ (with $\alpha^{\text{min}}=0.001$ and $\alpha^{\text{max}}=1$). The list of remaining free parameters for each model version is given in Table \ref{tab:05}.

\begin{table*}
%\hspace*{-1.5cm}
\renewcommand{\arraystretch}{1.3}
    \begin{tabular}{ccc|c|c|c|c|}
        \cline{4-7}
        & & & related versions & parameter nature & parameter bounds & related mechanisms\\
        \cline{2-7}
        \multirow{6}{*}{\rotatebox{90}{\parbox[c]{5cm}{\centering free parameters}}} &
        \multicolumn{1}{ |c  }{$p_{u}$} &
        \multicolumn{1}{ c| }{} & all except $4$ and $14$
        %$\widehat{4},\widehat{14}$ 
        & probability & $0.001-1$ & egonet growth\\
        \cline{2-7}
        & \multicolumn{1}{ |c  }{$p_{g}$} &
        \multicolumn{1}{ c| }{} & 
        all except $4$ and $14$
        %$\widehat{4},\widehat{14}$
        & probability & $0.001-1$ & egonet growth\\
        \cline{2-7}
        & \multicolumn{1}{ |c  }{$p_{c}$} &
        \multicolumn{1}{ c| }{} & all except $4$ and $14$
        %$\widehat{8},\widehat{14}$ 
        & probability & $0.001-1$ & dynamic triadic closure\\
        \cline{2-7}
        & \multicolumn{1}{ |c  }{$p_{d}$} &
        \multicolumn{1}{ c| }{} & 3,14 & probability & $0.001-1$ & node pruning\\
        \cline{2-7}
        & \multicolumn{1}{ |c  }{$a$} &  \multicolumn{1}{ c| }{} & 12 & probability & $a^{\text{min}}_{\text{obs}}-a^{\text{max}}_{\text{obs}}$ & interaction proposal\\
        \cline{2-7}
        & \multicolumn{1}{ |c  }{$a^{\text{min}}$} &  \multicolumn{1}{ c| }{} & all except 12 & probability & $a^{\text{min}}_{\text{obs}}-a^{\text{max}}_{\text{obs}}$ & interaction proposal\\
        \cline{2-7}
        & \multicolumn{1}{ |c  }{$a^{\text{max}}$} &  \multicolumn{1}{ c| }{} & all except 12 & probability & $a^{\text{min}}_{\text{obs}}-1$ & interaction proposal\\
        \cline{2-7}
        & \multicolumn{1}{ |c  }{$\alpha$} &  \multicolumn{1}{ c| }{} & 9 & rate & $0.001-1$ & Hebbian process\\
        \cline{2-7}
        & \multicolumn{1}{ |c  }{$\lambda$} &  \multicolumn{1}{ c| }{} & all except 3 and 14 & intensity & $0.01-10$ & edge pruning\\
        \cline{2-7}
        & \multicolumn{1}{ |c  }{$c$} &
        \multicolumn{1}{ c| }{} & 4,14 & integer & $1-4$ & egonet growth\\
        \cline{2-7}
        & \multicolumn{1}{ |c  }{$m$} &
        \multicolumn{1}{ c| }{} & 13,14 & integer & $1-4$ & interaction proposal \\
        \cline{2-7}
        & \multicolumn{1}{ |c  }{$m^{\text{max}}$} &
        \multicolumn{1}{ c| }{} & all except 13 and 14 & integer & $1-4$ & interaction proposal \\
        \cline{2-7}
    \end{tabular}
    \caption{\textbf{Free parameters of the models.} 
    }
    \label{tab:05}
\end{table*}

Our procedure is thus the following: we first define a set of observables of interest, and a comparison method (a distance) between the outcome of each model version and each reference data set.
For each reference and version, we then use a genetic algorithm to find the parameter values minimizing their difference. Note that these optimal parameter values can be different for different references. 

For each observable $O$, we can then gather the comparison between each model $M$ and each empirical reference $E$ into 
a distance tensor $D[\mathcal{O}]_{M,E}$, and subsequently 
rank all model versions by giving them a score for each observable: the higher the score, the closer the model observable with respect to the empirical ones. Combining the ranks for all observables yields then a 
global ranking of models.

%%%%%%%%%%%%%%%

\subsection{Observables}\label{para:1}

As face-to-face interactions are local in space and time it seems natural to study observables related to small spatio-temporal scales, like nodes, edges or small subgraphs. The simplest observables related to such an object $ob$ are:
\begin{itemize}
    \item its activity duration: number of consecutive time steps $ob$ exists in the temporal graph;
    \item its interactivity duration: number of consecutive time steps $ob$ is absent from the temporal graph;
    \item its aggregated weight: number of times $ob$ has been present in total in the temporal graph;
    \item its newborn activity: number of consecutive time steps $ob$ exists just after its first occurrence in the temporal graph.
\end{itemize}
If $ob$ is not a trivial sub-graph like nodes or edges, its size can also be an observable of interest.

\begin{figure}
    \centering
    \includegraphics[width=\columnwidth]{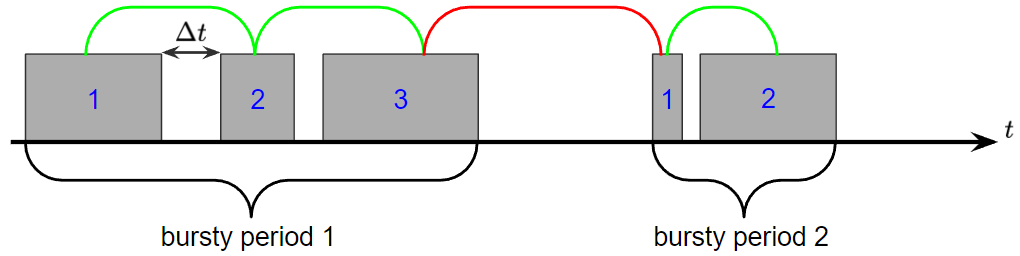}
    \caption{\label{fig:014}\textbf{Number of events per bursty period.} The sketch represents the activation timeline of a pair of nodes $(i,j)$. A grey rectangle represents an event for $(i,j)$, i.e. a maximal period of uninterrupted interaction between $i$ and $j$. A bursty period is a maximal collection of events that follow each other in time by a delay less than a given threshold, taken here to be 3 time steps. We draw a green junction between two consecutive events if they belong to the same bursty period, i.e. if they are separated by $\Delta t\leq3$ time steps, and a red junction if $\Delta t > 3$. In the example shown, we obtain two distinct bursty periods, with $3$ events in the first and $2$ in the second.
    }
\end{figure}

Let us now recall some useful definitions:
\paragraph{event:} (see also Figure \ref{fig:014})
An event is the combination of an edge $(i,j)$, a starting time $t_{0}$ and a stopping time $t_{f}$ such that $(i,j)$ is inactive at $t_{0}-1$ and $t_{f}+1$, and is active  $\forall t$ such that $t_{0}\leq t\leq t_{f}$.

\paragraph{bursty period \cite{Karsai2012}:} 
Two events are defined as adjacent if they are defined on the same edge and if the delay between them is less than a given time lapse $\Delta t$. A bursty period is a maximal collection of adjacent events (see Fig. \ref{fig:014}). 

\paragraph{aggregated network:}
The aggregated network on the whole temporal interval $\llbracket1,T\rrbracket$ is the weighted undirected graph $A$ such that $A_{ij}$ is the aggregated weight of the edge $(i,j)$, i.e., the number of time steps such that $G_{ij}(t)=1$. 

\paragraph{aggregation level:}
We define the interaction graph aggregated at level $n$, $G^{(n)}$, as follows: 
$(i,j)\in G^{(n)}(t)\iff(i,j)\in G(\llbracket nt,n(t+1)\llbracket)$. 
Note that $G^{(n)}$ is unweighted and undirected. We have $G^{(1)}=G$, and $G^{(T)}$ is an unweighted aggregated network on the all temporal interval. Observables of $G^{(n)}$ are called the observables at aggregation level $n$.

\begin{figure}
\subfigure[conf16]{\includegraphics[width=\columnwidth]{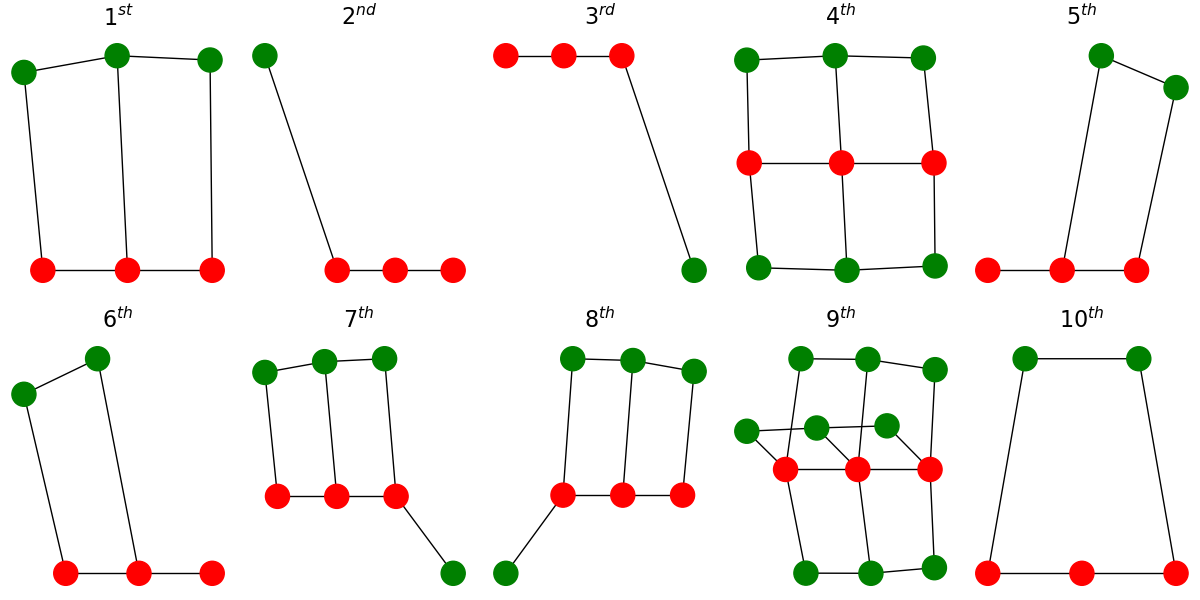}}
\subfigure[utah]{\includegraphics[width=\columnwidth]{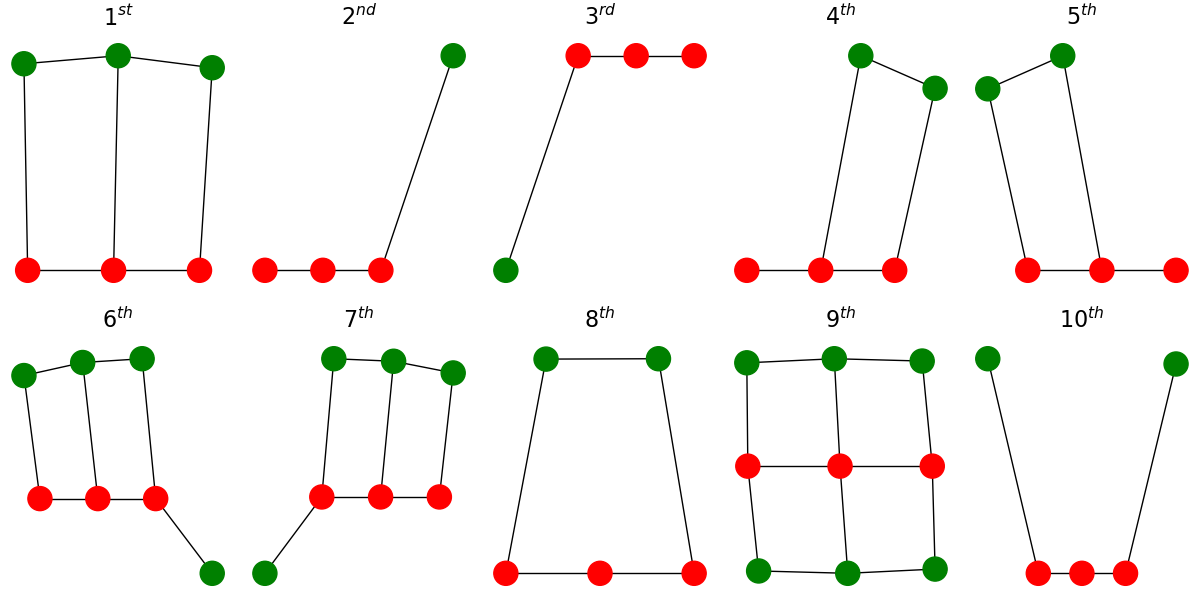}}
    \caption{\label{fig:04}\textbf{10 most frequent (3,1)-ETN in two empirical data sets.}
    (a) data set ``conf16'' (scientific conference), (b) data set ``utah''
    (primary school in Utah).
    Although those two social contexts are very different their ETNs are quite similar.
    }
\end{figure}

\paragraph{Egocentric Temporal Network (ETN) \cite{Longa2022,longa2022neighbourhood}:}
An ETN (see Fig. \ref{fig:04}) corresponds to a representation of the diversity of the interaction partners of a given node (the ego, in red in Fig. \ref{fig:04}) at $d$ consecutive times. In Fig. \ref{fig:04}, each ETN reads from left to right (time flow direction). Green circles represent neighbors of the red node. An horizontal edge is drawn between two circles iff they correspond to the same node at different times. The duration of an ETN is called its depth. A $(d,n)-$ETN is an ETN of depth $d$ and aggregation level $n$.

\paragraph{ETN vector:}
An ETN vector is a vector $V$ where the component $V_{i}$ is the aggregated weight of the ETN $i$.

We can now define the set of observables we will use to characterize and compare the temporal networks. The observables related to (temporal) subgraphs are (see table \ref{tab:04}):
\begin{itemize}
    \item aggregated weights for edges, (2,1)-ETN and (3,1)-ETN;
    \item size of connected components of the interaction graph;
    \item activity and interactivity duration for nodes and edges;
    \item newborn activity for edges;
    \item number of events per bursty period (Fig. \ref{fig:014}).
\end{itemize}

\begin{table*}
    \renewcommand{\arraystretch}{1.3}
    \centering
    \begin{tabular}{ccc|c|c|c|c|c|}
        & & \multicolumn{6}{c}{object}
        \\
        \cline{3-8}
        & \multicolumn{1}{c|}{} & nodes & edges & events & (2,1)-ETN & (3,1)-ETN & connected components of $G(t)$
        \\
        \cline{2-8}
        \multirow{5}{*}{\rotatebox{90}{\parbox[c]{2.5cm}{\centering observable type}}} &
        \multicolumn{1}{|c|}{aggregated weight}
        & & $\times$ &  & $\times$ & $\times$ &
        \\
        \cline{2-8}
        &
        \multicolumn{1}{|c|}{activity duration}
        &
        $\times$ & $\times$ & $\times^{\ast}$ & & &
        \\
        \cline{2-8}
        &
        \multicolumn{1}{|c|}{activity interduration}
        &
        $\times$ & $\times$ & & & &
        \\
        \cline{2-8}
        &
        \multicolumn{1}{|c|}{newborn activity}
        &
        & $\times$ & & & &
        \\
        \cline{2-8}
        &
        \multicolumn{1}{|c|}{size}
        &
        & & & & & $\times$
        \\
        \cline{2-8}
    \end{tabular}
\caption{\textbf{Sub-temporal graph observables.} A $(d,n)$-ETN is an ETN of depth $d$ and aggregation level $n$.
At the crossing between the row entitled ``activity duration'' and the column entitled ``events'', we wrote an asterisk because the observable we used with events as objects and activity duration as type is called the number of events per bursty period. 
}
\label{tab:04}
\end{table*}

In addition, we also consider:
\begin{itemize}
    \item the clustering coefficient of the aggregated network;
    \item the degree assortativity in the aggregated network; 
    \item the $(3,n)$-ETN vector including the weights of ETNs computed in the aggregation levels $n$ from 1 to 10: this allows to take into account various timescales in a single observable.
\end{itemize}

\subsection{Comparison method}

\subsubsection{Distance tensor}
\label{subsubsec:0}
We want to quantify how close are a synthetic temporal network and a reference empirical data set, with respect to a given observable.
In order to be able to aggregate across observables and obtain a global distance and score, we consider for each observable 
a distance bounded between 0 and 1. Moreover, we need to consider different metrics for 
 observables for which we have either
 (i) a distribution (e.g. activity durations or aggregated weights), or (ii) only one numerical value for each network (e.g. the clustering coefficient) or (iii) only one vectorial realization (e.g., ETN vectors).
 
\paragraph{Point observables.}
Let us first consider an observable $\mathcal{O}$ for which we have only one realization per data set $\mathcal{O}(\mathcal{D})\in\mathbb{R}$, where $\mathcal{D}$ is the data set. Then we take as metric:
\begin{equation}
D[\mathcal{O}]_{\mathcal{D},\mathcal{D'}}=\frac{|\mathcal{O}(\mathcal{D})-\mathcal{O}(\mathcal{D'})|}{2\max(|\mathcal{O}(\mathcal{D})|,|\mathcal{O}(\mathcal{D'})|)}    
\end{equation}

This metric is bounded between 0 and 1, and reaches its maximum value only when $\mathcal{O}(\mathcal{D})=-\mathcal{O}(\mathcal{D'})$.

\paragraph{Observables with multiple realizations per data set}
If $\mathcal{O}(\mathcal{D})$ is a variable whose distribution $P$ can be sampled, we need to consider a distance between the synthetic and empirical distributions. However $\mathcal{O}(\mathcal{D})$ and $\mathcal{O}(\mathcal{D'})$ may yield distributions not equally sampled, possibly on different supports. We choose here to obtain distributions of equal size, by 
completing the least sampled distribution with zeros, and compare them with the Jensen-Shannon divergence (JSD), which is bounded between 0 and 1. For two discrete distributions $p$ and $q$:
\begin{equation}
\begin{cases}
\text{JSD}(p,q)=\frac{1}{2}(\text{KL}(p||m)+\text{KL}(q||m))\\
m=\frac{1}{2}(p+q)\\
\text{KL}(p||m)=\sum_{i}p_{i}\log_{2}(\frac{p_{i}}{m_{i}})
\end{cases}
\end{equation}
We thus consider the metric:
\begin{equation}
D[\mathcal{O}]_{\mathcal{D},\mathcal{D'}}=\text{JSD}(P[\mathcal{O}(\mathcal{D})],P[\mathcal{O}(\mathcal{D'})]).
\end{equation}

\paragraph{Vector observables.}
\label{para:2}
Now let us consider that $\mathcal{O}(\mathcal{D})\in\mathbb{R}^{d}$, with some $d\in\mathbb{N}$. In our context, this corresponds to the ETN vectors. As we are interested in the relative frequencies of the ETNs, we use the cosine similarity:
%$$D^{2}[\mathcal{O}]_{\mathcal{D},\mathcal{D'}}=1-
\begin{equation}
\text{sim}(\mathcal{O}(\mathcal{D}),\mathcal{O}(\mathcal{D'}))=
\frac{\mathcal{O}(\mathcal{D})\cdot \mathcal{O}(\mathcal{D'})}{||\mathcal{O}(\mathcal{D})||~||\mathcal{O}(\mathcal{D'})||}.
\end{equation}

In fact, in the case of $(3,n)$-ETN, we have a family of $n$ vectors $\{v_{p}\}_{p=1,\cdots,n}$: the $i^{th}$ component of $v_{p}$ is the ratio between the number of occurrences of the motif $i$ at aggregation level $p$ and the number of occurrences of the most frequent motif at aggregation level $p$.

We thus define the similarity between the two families of $n$ vectors as the product of cosine similarities between each pair of vectors at the same level of aggregation:
\begin{equation}
\text{Sim}(v_{1},\ldots,v_{n} ; v'_{1},\ldots,v'_{n})=\prod_{p=1}^{n}\text{sim}(v_{p},v'_{p}),
\end{equation}
and the distance between the two families is $1-\text{Sim}$.

\subsubsection{From distances to a score}

\begin{figure}
    \centering
    \includegraphics[width=\columnwidth]{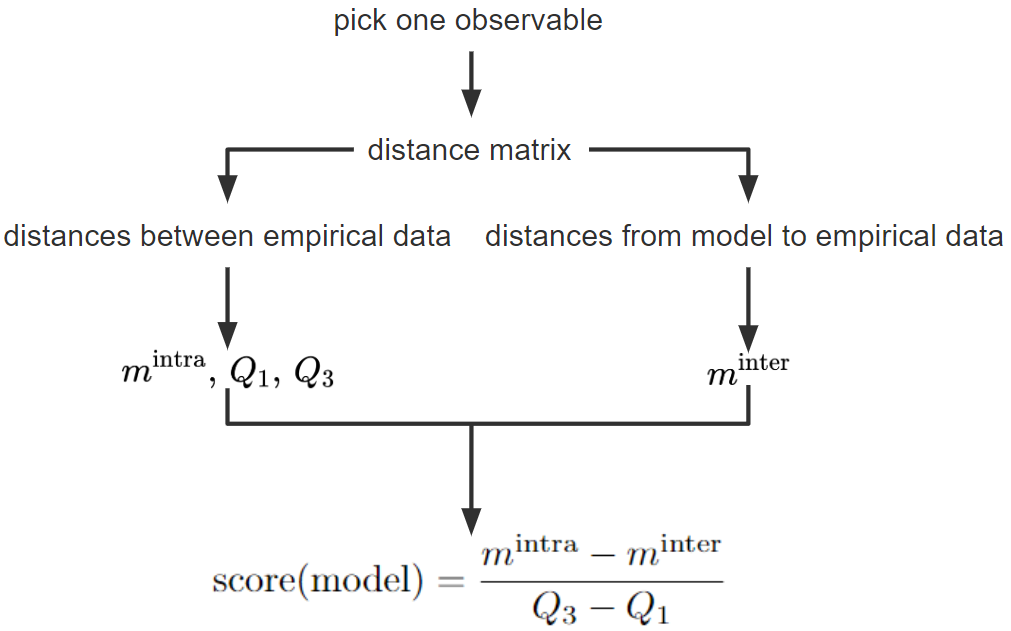}
    \caption{\label{fig:05}\textbf{Computing the score of a model.} The evaluated data sets consist in all versions of the models presented in Table \ref{tab:02} and the empirical data sets presented in Table \ref{tab:03}. For each data set, a different score is computed for each observable. The higher the score of a model, the closer the distribution of the associated observable is from the distributions in the empirical data sets. Said otherwise, a higher score means more realistic statistical properties for the associated observable.
    }
\end{figure}

Our goal is to understand which model versions 
are best able to reproduce empirical properties observed in a series of data sets.
For each model version, and each observable, we thus define a score by comparing the minimal distances between synthetic and empirical data with the distance between empirical data sets themselves.
To this aim, given an observable $\mathcal{O}$ and a model version $\mathcal{V}$, we:
\begin{enumerate}
    \item compute for each empirical data set $\mathcal{E}$ its distance
    $\delta[\mathcal{O}]_{\mathcal{V},\mathcal{E}}$
    to the set of model instances, i.e., the minimal distance 
    $\min_{\mathcal{D}} D[\mathcal{O}]_{\mathcal{D},\mathcal{E}}$ 
    over all instances $\mathcal{D}$ of the model $\mathcal{V}$;
    \item  
    compute the median of the distances between $\mathcal{V}$ and all empirical data sets:
    \begin{equation}
    m^{\text{inter}}_\mathcal{O}(\mathcal{V})=\text{median}(\delta[\mathcal{O}]_{\mathcal{V},\mathcal{E}})\end{equation}
    where the index $\mathcal{E}$ runs over the empirical data sets;
    \item compute the characteristic distance between empirical data sets themselves:
    \begin{equation}
    m^{\text{intra}}_\mathcal{O}=\text{median}(D[\mathcal{O}]_{\mathcal{E},\mathcal{E'}})
    \end{equation}
    where the indices $\mathcal{E},\mathcal{E'}$  both run over the empirical data sets ($\mathcal{E}\neq\mathcal{E'}$);
    \item compute the interquartile range of distances between empirical data sets, $Q_3 - Q_1$;
    \item deduce the score of the model version $\mathcal{V}$ for observable $\mathcal{O}$:
    \begin{equation} \label{eq:score}
    \text{score$_\mathcal{O}$($\mathcal{V}$)}=\frac{m^{\text{intra}}_\mathcal{O}
    -m^{\text{inter}}_\mathcal{O}(\mathcal{V})}{Q_{3}-Q_{1}} .
    \end{equation}
\end{enumerate}
This procedure is illustrated in Fig.~\ref{fig:05}. A higher score corresponds to 
the fact that the model version has instances with
statistical properties closer to the empirical ones, for the chosen observable.

Note that, while this procedure is intended to provide a score to models, we can also apply it also to each empirical data set. The interpretation is then not a ``score'', but quantifies how close a data set is to the other ones.

\subsection{Results}

We first illustrate that our approach providing a score using the proximity tensor is compatible with a qualitative direct appreciation of the distributions. We then detail the genetic tuning of the free parameters. This allows us to identify the best model belonging to the class investigated here. We then investigate in more details the interplay between observables and the role of each mechanism in our model, i.e., which observables change when a given mechanism or hypothesis is changed. 

\subsubsection{Illustration}

\begin{figure*}
    \includegraphics[width=0.9\columnwidth]{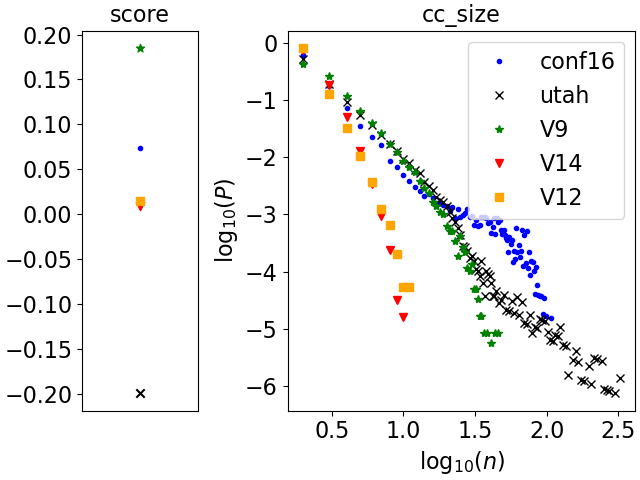}
%    \label{fig:6a}
    \includegraphics[width=0.9\columnwidth]{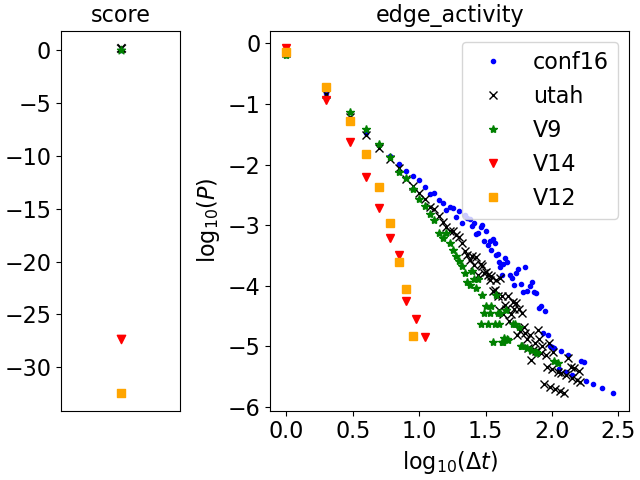}
%    \label{fig:6b}
    \includegraphics[width=0.9\columnwidth]{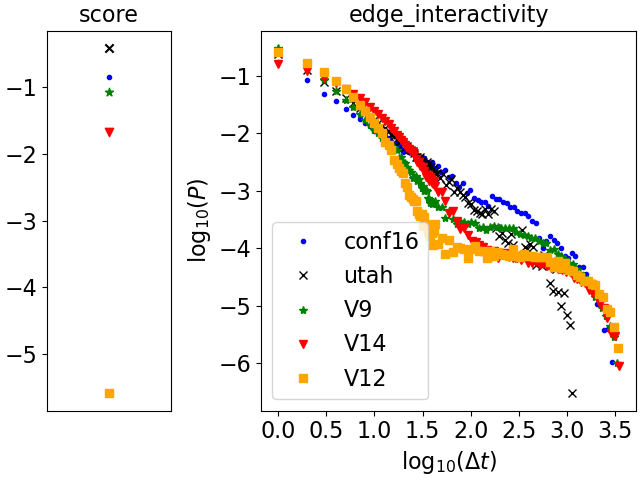}
%    \label{fig:6c}
    \includegraphics[width=0.9\columnwidth]{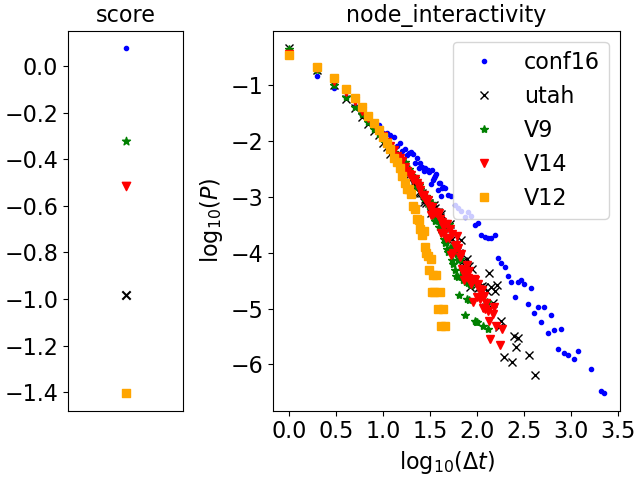}
%    \label{fig:6d}
\caption{\label{fig:06}\textbf{Illustration of the scores for several observables and models.} 
The figure displays in each panel the distribution of an observable ("cc\_size": size of connected components) for two data sets (the conference ``conf16'' and the school ``utah'') and three model versions: the original ADM (V14), the adjacent version with the highest average score (V9) and the version with the lowest average score (V12). The three models were optimized with respect to the ``utah'' reference. The score is computed in each case by Eq. \ref{eq:score}: a higher score is associated with a distribution closer to the empirical one. In the top left panel, ``utah'' has a low score because the ``cc\_size'' observable is the only distribution which is not similar for all empirical references.
}
\end{figure*}

Figure \ref{fig:06} displays the distribution of several observables for two empirical data sets corresponding to different contexts and three model versions. This illustrates how, for each observable that can be sampled, a higher score is associated with 
a distribution closer to the empirical ones.

For point observables (the clustering coefficient and degree assortativity of the fully aggregated network), the score associated with a point observable does not necessarily reflect the degree of proximity with an empirical reference (not shown):
this is due to the fact that these point observables are highly variable from one empirical data set to another.

\begin{figure}[h]
    \subfigure[best adjacent version\label{fig:16a}]{
    \includegraphics[width=0.8\columnwidth]{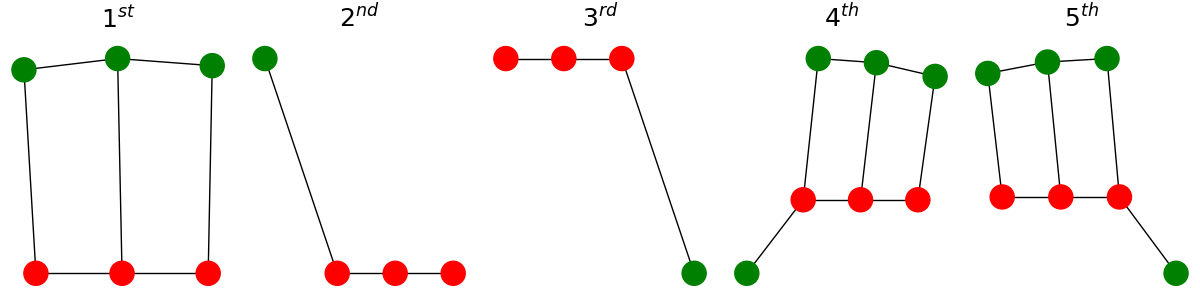}}
    \subfigure[empirical reference 'utah'\label{fig:16b}]{                                \includegraphics[width=.8\columnwidth]{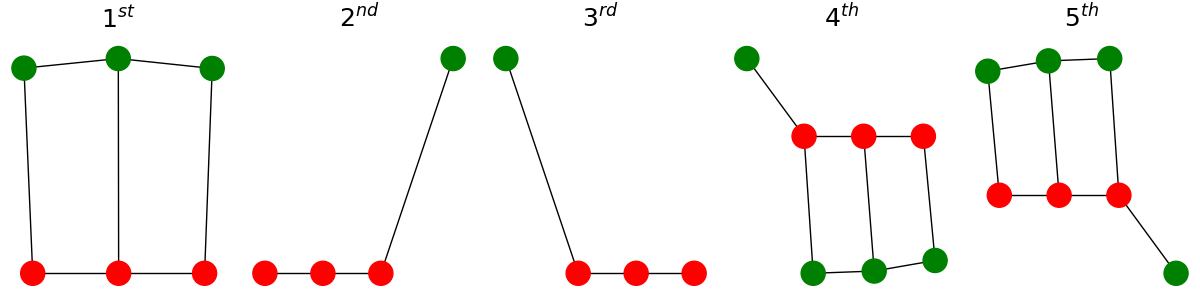}}
    \subfigure[worst adjacent version\label{fig:16c}]{
    \includegraphics[width=0.8\columnwidth]{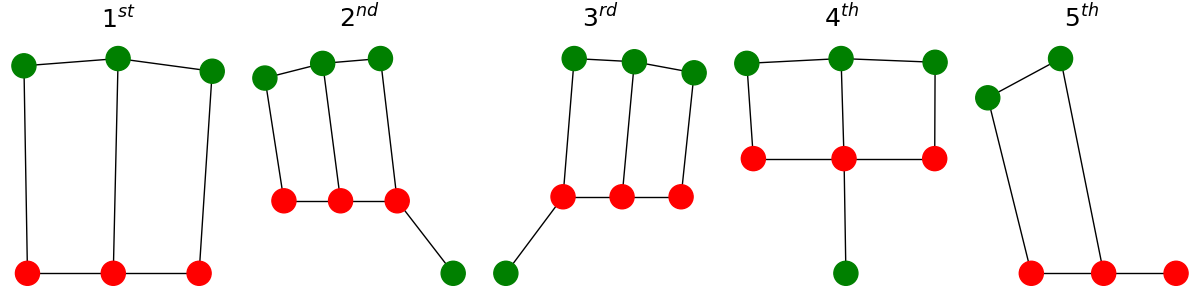}}
    \subfigure[ETN autosimilarity\label{fig:16d}]{
    \includegraphics[width=0.8\columnwidth]{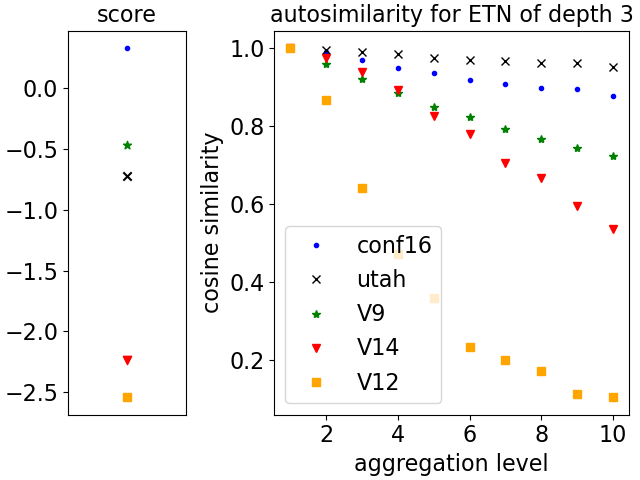}}
\caption{\label{fig:016}\textbf{Interpretation of the score of the ETN vector observable.}
Fig \ref{fig:16a}, \ref{fig:16b}, \ref{fig:16c}: Five most frequent ETN of depth 3 at aggregation level 5 for two models tuned w.r.t. the empirical reference ``utah''. The model with the highest score (V9) has the same motifs as the reference, but not the worst version (V12).
Fig \ref{fig:16d}: ETN autosimilarity for three models (the best V9, the worst V12 and the original ADM V14) and two empirical references. The ETN autosimilarity of a given data set at aggregation level $n$ is the cosine similarity between the motifs observed in this data set at level $n$ and the motifs observed in this data set at level $1$.
Empirical references are highly autosimilar with respect to this measure. For models, a higher score for the ETN vector corresponds to a higher ETN autosimilarity.
}
\end{figure}

It is more difficult to check the accordance between a high score for the ETN vector observable and realistic motifs because we can visualize only a few motifs. As an illustration however, we display in Fig. \ref{fig:16a} the five most frequent motifs at aggregation level $5$ of the ``utah'' data set and the instances associated with this reference of the models with highest and lowest ETN scores.
The ``utah'' instance of the version with the highest ETN score has exactly the same 5 most frequent (3,5)-ETN as the ``utah'' reference, while this is not the case of the instance of the version with the lowest score.

Figure \ref{fig:16d} moreover shows the ETN autosimilarity for three model versions and two references. We define the ETN autosimilarity of a data set at a given depth $d$ and aggregation level $n$ as the ETN similarity (defined in \ref{para:2}) between the $(d,n)$- and $(d,1)$-ETN vectors of this data set.
The empirical references are highly autosimilar, i.e. their ETN autosimilarity is close to $1$ for various levels of aggregation. We also display in the figure 
 the ETN autosimilarity of three model versions (V9, V12 and V14), using in each case the instance tuned to be as close as possible to the reference ``utah''.
The higher the score of a model version, the closer its ETN autosimilarity curve to the ``utah'' reference.

\subsubsection{Tuning the models' parameters by a genetic algorithm}
\label{subsubsec:2}

For each model version and each reference data set,
we want to obtain the parameter values that yield temporal networks instances as close as possible to the reference.
Recall that given a reference data set and a model version, there are three types of parameters: (i) frozen parameters that depend only on the version, like the bounds for the power-laws of strengthening and decay rates $\alpha_{ij}$, $\beta_{ij}$; (ii) readable parameters that depend only on the reference, like $N$, $T$, $a^{\text{min}}_{\text{obs}}$ and $a^{\text{max}}_{\text{obs}}$; (iii) free parameters, that depend both on the version and the reference, that we tune to get as close as possible to the reference data set, e.g., $p_{c}$ or $m^{\text{max}}$ (see Table \ref{tab:05} for the  list of parameters).

To tune the free parameters, we use a genetic algorithm (described in 
%Appendix \ref{sec:3})
the SM), with a fitness set to the distance between the reference data set and the instance of the temporal network generated by the model. 
However, computing the distances for all observables is computationally costly while,
in a genetic algorithm, the fitness computation should be fast as it is computed at each iteration and for each genetic sequence. 
Therefore, we choose here to use as fitness only the distance relative to the ETN vector with the first ten levels of aggregation, i.e. the $(3,n)$-ETN for $n=1,\ldots,10$.
This observable is indeed computationally efficient and covers various time and spatial scales.

We find that this is enough for the model to improve on other observables too: we illustrate this point in the SM by comparing random instances with tuned instances along several
observables. Some distributions remain different from their empirical counterparts, in particular the distributions of sizes of connected components
(``cc\_size"), which however differ also between data sets. 
A better agreement and better scores might be obtained at the cost of an increased computational effort, by including additional features in the genetic
algorithm fitness. Overall, how to keep the computational effort of the genetic algorithm low while obtaining a good similarity between model and data statistics on a large range of observables remains an open interesting question.
We have also checked that the fitness is positively correlated with the score of every observable, 
which means that, despite these limitations, the genetic tuning does what it was intended to: obtain instances with closer statistical properties from empirical references than 
random instances in \textit{every} observable. In the SM, 
we also investigate how the values of the tuned parameters are distributed across versions and references. 

\subsubsection{Most realistic model within the ADM class}\label{subsubsec:1}

\begin{figure}
\subfigure[averaging the ranks]{
        \includegraphics[width=\columnwidth]{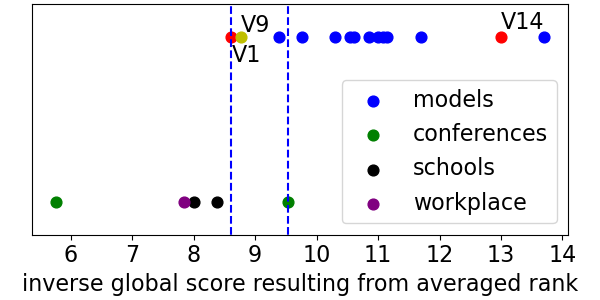}}
  %      \label{fig:10a}
\subfigure[averaging the scores]{
        \includegraphics[width=\columnwidth]{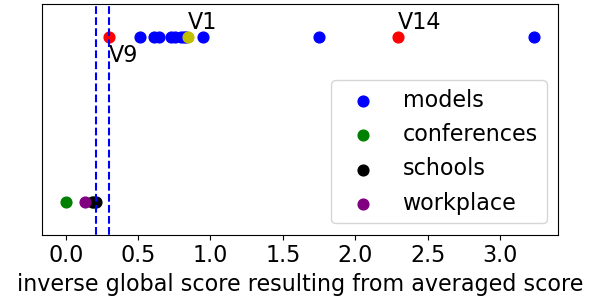}}
    \caption{\label{fig:010}\textbf{Global ranking of the models 1 to 14 according to two possible ranking strategies.} 
    In each panel, the $y$ coordinate is arbitrary: empirical data sets have been placed at $y=0$ while artificial data sets are aligned at $y=1$. The global ranking is obtained by sorting the models by increasing order of the $x$ coordinate (hence ``inverse global score''). 
    In each panel, the best version as well as the original ADM are highlighted as red dots. The best version in the other ranking strategy is displayed as a yellow dot. Blue vertical dashed lines indicate either the crossover or the gap between empirical and artificial data sets.
    (a): The $x$-coordinate is given by the model rank averaged over all observables. The best version is the baseline V1, almost \textit{ex aequo} with version 9.
    (b) the $x$-coordinate is given by the opposite of the averaged score, shifted by the maximum averaged score to take positive values. Version 9 ($\alpha_{ij}=\beta_{ij}=\alpha$) is here 
    the best one. In both panels, the original ADM is ranked rather low.
    There is either an overlap (a) or a small gap (b) between models and empirical data sets. Thus the  class of models considered here is able to generate synthetic data sets with statistical properties close from real data sets.
    }
\end{figure}

To compare the models, 
we first compute for each observable a ranking of the model versions using their score, computed 
using the distances between each instance obtained by the genetic tuning and the corresponding reference data set.
To then determine the best model among the 14 versions presented above, it is necessary to define a global score for each model version. 
We consider two possible strategies:
\begin{itemize}
    \item the global score of a model (or data set) is given by its rank averaged over all observables;
    \item the global score of a model (or data set) is given by its score summed over all observables and the global rank is just the rank according to the global score.
\end{itemize}
Note that other global ranks could be obtained by attributing different weights to the score or rank for different observables. We choose here however not to favor an observable over another.
The resulting rankings are shown on Figure \ref{fig:010}. The original ADM performs very low in both rankings, and the two best versions are the baseline version and the version 9, i.e. with $\alpha_{ij}=\beta_{ij}=\alpha$.
We also show in the SM the rankings of all model versions for each observable separately: despite a rather large variability between rankings, 
the baseline version remains within the five first ranks for 6 observables, and the version 9 for 8 observables.

In the next subsections, we investigate this global result in more details, to understand in particular how each model performs with respect to each observable, and the impact of the
various mechanisms on the models' performances.

\subsubsection{Similarity between observables}

\begin{figure}[h]
  \subfigure[Similarity matrix between observables\label{fig:9a}]{
        \includegraphics[width=0.9\columnwidth]{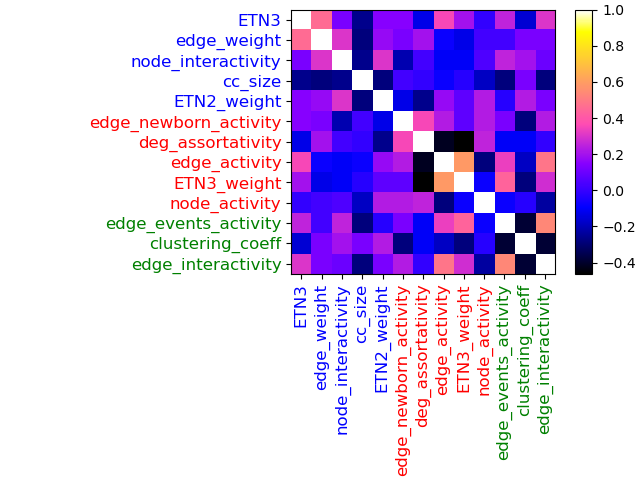}}
       \subfigure[Similarity network of observables \label{fig:9b}]{
        \includegraphics[width=0.8\columnwidth]{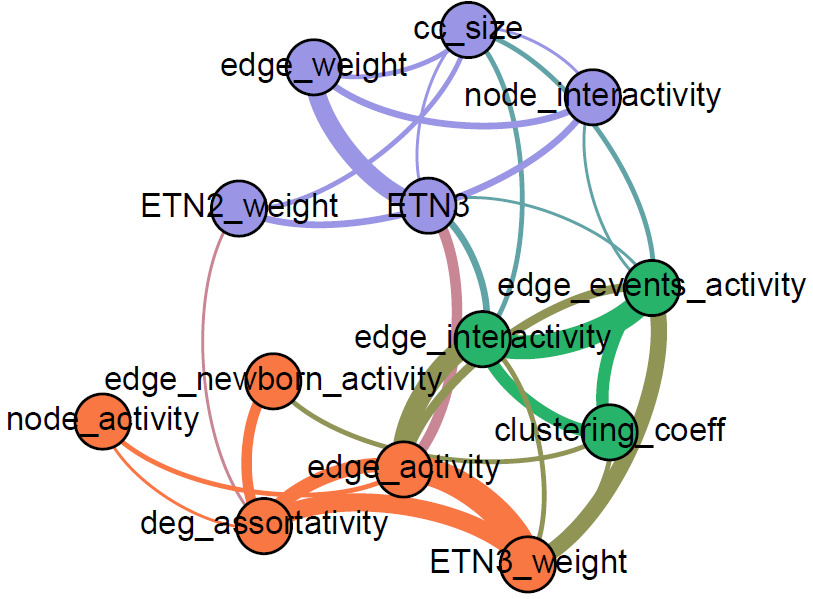}}
%        \label{fig:7b}
    \caption{\label{fig:07}\textbf{Similarity between observables.}
    (a) Similarity matrix obtained by computing the Kendall similarity between the associated rankings of our model versions
    (note that the results might be different if considering a different ensemble of models).
    The observables are quite independent from each other (low absolute values for the similarity for almost all observables pairs). Strong negative couplings are observed only between the point observables and some distribution observables. The ETN vector is either positively or weakly coupled to every other observable, in accordance with the conclusion of \ref{subsubsec:2}: improving on motifs generally means improving on other observables.
     (b) The matrix is turned into a weighted network by taking as edge weight the absolute value of the Kendall similarity. A community detection algorithm based on modularity optimization detects three groups of observables (colored according to the group they belong to). The thickness of an edge is proportional to its weight and we filter out small weights for
     visualization purposes.
    }
\end{figure}

First, we need to investigate the fact that the observables we have chosen to characterize our social temporal networks are not independent. In particular, when modifying a modeling hypothesis or a parameter value, several observables may be modified in a correlated way. Understanding these correlations can help better interpret the effect of varying the modeling hypotheses. 
We thus define a similarity between two observables as the Kendall tau between the rankings of the models using these observables.
The resulting  similarity matrix between observables is shown in Fig.~\ref{fig:07}. We then extract groups of correlated observables by converting this matrix into a weighted undirected network: the nodes of this network are the observables and the weight $w_{\mathcal{O}\mathcal{O'}}$ is the absolute value of the Kendall similarity between rankings of observables $\mathcal{O}$ and $\mathcal{O'}$. 
The network is shown in Fig.~\ref{fig:07}, on which we use the community detection algorithm of the software Gephi \cite{bastian2009gephi}, based on modularity maximization, to obtain the three following groups:
\begin{itemize}
    \item  group I (blue): node activity interduration, edge weight, size of connected components, ETN vector and (2,1)-ETN weight;
    \item group II (orange): node, edge and newborn edge activity duration, degree assortativity and (3,1)-ETN weight;
    \item group III (green): edge activity interduration, events activity duration and clustering coefficient.
\end{itemize}

\subsubsection{Impact of hypotheses on model performances}

\begin{figure}[h!]
\subfigure[Score variation for each group of observables\label{fig:5a}]{
        \includegraphics[width=0.9\columnwidth]{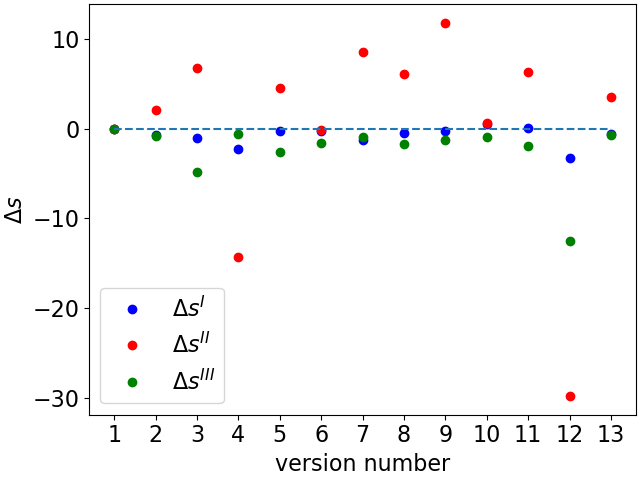}}
\subfigure[Contribution of each observable to each group score\label{fig:5c}]{
        \includegraphics[width=0.9\columnwidth]{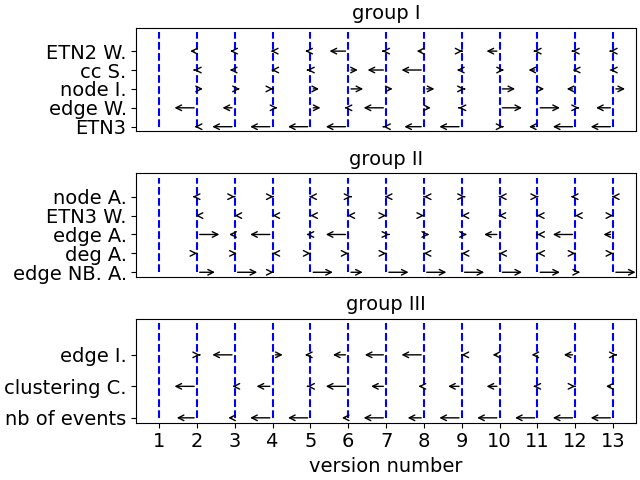}}
    \caption{\label{fig:08}\textbf{Score variation resulting from the change of hypothesis with respect to the baseline model.} For each version, we compute the difference between its score and the score of the baseline (V1). In panel \ref{fig:5a}, these differences are summed over all observables of each of the groups identified in Fig.~\ref{fig:07}. 
    The sum over group I (resp. II, III) is called the score I (resp. II, III)
    and is denoted by $\Delta s^{\text{I}}$ (resp. $\Delta s^{\text{II}}$, $\Delta s^{\text{III}}$).
    (a): Score differences for each adjacent model version. 
    Blue dots correspond to scores I, red dots to scores II and green dots to scores III. The horizontal dashed line indicates the line of no changes in score.
    (b): Differences in score for each observable, represented as horizontal arrows. A difference is positive when the arrow points to the right and negative when it points to the left. Vertical blue dashed lines correspond to $\Delta s=0$. 
    The arrow length is proportional to the difference in score, rescaled so that the longest arrow spans an arbitrary length of $0.5$ unit. %Rescaling factors are different for each group.
    %Versions show no variation for many observables from all groups. 
    The edge weight and the size of connected components (cc\_size) are the only observables on which a significant improvement is observed with respect to the baseline version. 
    }
\end{figure}

In order to have more precise information about how hypotheses impact each observable, depending on the group it belongs to, we define for each model version its score relative to a group of observables as follows:
\begin{enumerate}
    \item for each observable in the group (I, II or III), we compute the score of the model version as well the score of the baseline version V1;
    \item we compute the difference between the score of the version and the score of the baseline version;
    \item we sum the differences obtained for each observable in the group.
\end{enumerate}
Figure \ref{fig:08} shows the resulting group scores for the various versions. We also indicate the relative contribution of each observable inside the group to the group score. Finally, we summarize in Table  \ref{tab:5b} which hypotheses lead to an improvement or a worsening with respect to the baseline version.

\begin{table}
        \begin{tabular}{c|c|c|c|c}
        n & difference with version 1 & $\Delta s^{\text{I}}$ & $\Delta s^{\text{II}}$ & $\Delta s^{\text{III}}$\\
        \hline
        2 & linear process & 0 & + & 0\\
        3 & node pruning & 0 & + & -\\
        4 & varying egonet growth & - & - & 0\\
        5 & $c_{ij}(t)=1$ & 0 & + & -\\
        6 &$R=W=I\cup C$ & 0 & 0 & -\\
        7 &$R=W=I$ & 0 & + & 0\\
        8 &$p_{c}=0$ & 0 & + & 0\\
        9 &$\alpha$ & 0 & + & 0\\
        10 &$\alpha_{i}$, $\beta_{i}$ & 0 & 0 & 0\\
        11 &$\alpha_{ij}$, $\beta_{ij}$ & 0 & + & -\\
        12 &$a$ & - & - & -\\
        13 &$m$ & 0 & + & 0
    \end{tabular}
\caption{\textbf{Outcome of hypotheses on the sign of the three group scores.}\label{tab:5b}
 For each version, we indicate the sign of scores I, II and III. Many versions have a score similar to the basis version. Improvement is only observed for group II, and comes often with a loss over the two other groups.}
\end{table}

Figure \ref{fig:5a} indicates that the baseline version seems to be optimal for observables from group I and III, since no version exhibits improvement on either group. However, 8 out of 12 adjacent versions show an improvement for group II. The most common signature is $0,+,0$: 5 versions show no change on groups I and III and an improvement on group II.

In terms of mechanisms, updating the social bond graph with a linear Hebbian process with no decay (V2) improves over the exponential Hebbian process of the baseline version, but if we use an exponential Hebbian process with a uniform value $\alpha_{ij}=\beta_{ij}=\alpha$ (V9), then we get still better results.
Thus, in order to recover a more realistic social system, agents should all update their social ties in the same way, i.e. with the same homogeneous parameter $\alpha$.
The observation for the intrinsic activity $a_{i}$ is the opposite: imposing a uniform value $a_{i}=a$ (V12) leads to a drastic loss in score for all groups. Heterogeneity in the intrinsic activities seems to be necessary to recover a realistic social system.
On the other hand, a uniform number of emitted interactions (V13) leads to an improvement. Actually (see SM), %(cf Appendix \ref{sec:2}), 
the value for $m$ or $m^{\text{max}}$ returned by the genetic tuning is $1$ in most cases: a higher value probably causes the nodes to have a too large instantaneous degree, i.e. agents interact with more other agents than what is realistic, leading to unrealistic ETN motifs.

Figure \ref{fig:5a} also yields interesting insights concerning the update of the social bond graph and the contextual interactions. A uniform node pruning (V3) leads to poor performance on group III observables quite as equivalent as the gain over group II. 
Not taking into account the social context, i.e. putting $c_{ij}=1$ (V5) also leads to opposite changes: we gain over group II and loose over group III. Regarding contextual interactions, considering them leads to a significant improvement under the condition that they are treated as pure noise (V7). Having no contextual interaction at all (V8) also leads to an improvement, but of smaller amplitude.
Thus adding noise in our system makes it more realistic, which can be understood by the fact that 
many interactions have in fact little social significance and occur only due to context.

Figure \ref{fig:5c} gives more detailed information by indicating the relative contribution of each observable to the group score. In particular, some observables always give a negligible contribution to the score of the group they belong to.
This is in part due to the fact that some observables are shared across all versions, i.e. their realizations are similar in all versions. This is the case of ``ETN2 weight'' and ``ETN3 weight'', whose distribution always match almost perfectly the empirical case (after genetic tuning).
%Hence those observables are not relevant to consider, because they are too easy to recover.

Other observables are shared across almost all versions, like ``node activity'' and ``node interactivity'', which are similar for all versions except version 12, characterized by $a_{i}=a$ (however, for this version the loss in score relatively to those two observables is negligible compared to the loss relative to the other ones).

Overall, the 8 major observables, which are mainly responsible for the observed group scores, are:
\begin{itemize}
    \item in the group I: ``size of connected components'', ``edge weight'', ``ETN3''
    \item in the group II: ``edge activity'', ``edge newborn activity''
    \item in the group III: ``edge interactivity'', ``clustering coeff'', ``edge events activity''
\end{itemize}
All observables relative to edges are major observables. However, the fact that an observable contributes a lot to the score of its group does not mean that it is necessarily relevant: as the point observables are not shared across empirical references, we must be careful when we score a model relatively to them. For instance, if we considered that only relevant observables should be robust over empirical social systems, then the clustering coefficient and the degree assortativity should not be used to score and rank models.
Why some observables contribute more than others might also depend on how shared they are between references: if an observable has almost the exact same realizations in every empirical reference, then the associated interquartile range will be almost zero, which can lead to high variations in the score for models (cf. Eq. \ref{eq:score}).

It is finally important to note that, except for the original ADM (V14), the model versions considered differ from the baseline version by one hypothesis only.
The question arising naturally is the following: if we accumulate modifications with respect to the baseline version, do variations in score accumulate accordingly? 
If so, Table \ref{tab:5b} could be used to design even more realistic models by combining the hypotheses that lead to improvements:  for each mechanism, we can check whether
the variation from the baseline leads to an improvement or not, and combine the variations that do. We explore this avenue in the SM
for several composite versions. The relation between the score of a composite version and the scores of its adjacent components is however non trivial, and the
best version remains V9 even when taking into account the composite versions.

\section{Discussion}
\label{sec:0}

In this paper, we have presented a general framework allowing to design various models by controlling their qualitative aspects. 
We have considered a modeling framework based on the idea of a co-evolution of an observed interaction network and an underlying and unobserved social bond network. Within the overall framework of the activity driven model with memory \cite{perra2012activity,Laurent_2015}, we hypothesized that social bonds partially drive the observed interactions, together with an influence of the current social context, and that interactions impact social bonds \cite{gelardi2021temporal}: the corresponding strengthening and weakening of social bonds take into account the fact that an interaction reinforces a social bond, and that resources (time, energy) are needed to maintain a social bond, so that the absence of interaction weakens it.
Instead of the usual exploration of a parameter space for a given set of mechanisms, we have then considered, within this framework, an exploration of a hypotheses space, corresponding to representations of several possible social mechanisms. 
Parameters corresponding to each hypothesis are then tuned by a genetic algorithm to maximize the similarity between model instances and a given empirical data set. While such similarity can be defined a priori in many ways, we find that using only the ETN vector to quantify it and tune the parameters leads to an improvement for many other observables, indicating that many statistical properties of a social temporal network are related to its ETN motifs \cite{longa2022neighbourhood}. We recall that the ETN vector is given by the list and frequencies of ETN motifs at various levels of aggregation (1 to 10 in our case), which thus encodes several spatiotemporal scales. This procedure allows us to define a score for each model, relative to each observable considered and globally, and to deduce which mechanisms lead to more realistic artificial temporal networks. In particular, many of the model versions considered perform better than the original Activity Driven model with memory.
Once tuned, each model version can produce synthetic data sets of arbitrary sizes and durations and with realistic properties, which can be used for instance as support for numerical simulations of 
dynamical processes on temporal networks.

Our work entails a number of limitations that are worth discussing. First, 
the list of observables we consider to rank  models is somewhat arbitrary: we investigated observables of different types (point, with multiple realizations, vector) and dealing with various scales, but other observables could be thought of, while some might 
be removed from the list because of their variability among the empirical references (e.g., clustering coefficient). Second, the scoring mechanism may also be improved. Indeed, a higher score is not always clearly associated with a value of the observable closer to the empirical value. Future work will thus address the issue of building another ad hoc score measure with a clearer interpretation.

The use of a series of statistical properties to determine whether a model is producing realistic temporal networks can also be discussed. Indeed, empirical data sets show large activity variations, i.e., in the number of interactions per timestamp. These variations can be driven by changes in population size or in intrinsic activity \cite{kobayashi2020two}, either due to imposed schedules or to spontaneous bursts. Such patterns
 cannot be recovered in the class of models we have explored, for which the number of interactions per timestamp is stationary with small fluctuations. 
Exploring other classes of models would be necessary to account for the large empirical variations.  
The methodology considered in this paper could however then still be used to cover such extended classes. In particular, our results suggest that the full exploration of the hypotheses space is not necessary, as properties of composite models could be predicted from their adjacent components. 
%Of course it remains to determine in which extent this could be true.

Despite these limitations, the partial exploration we performed allowed to determine models with a much higher degree of realism than the original ADM, and also to show the interest of modeling several social mechanisms such as taking into account the social context, considering casual interactions (dynamic triadic closure) and updating the underlying social bond ties through an exponential Hebbian process with both strengthening and weakening mechanisms.
The class of models we have considered could also be extended, e.g. by adding group memberships, or by considering various types of Hebbian processes: delayed or anti-Hebbian process, or allowing negative interactions and possibly negative social bonds \cite{gelardi2019detecting,andres2022reconstructing}.

%
% ****** End of file apssamp.tex ******
\providecommand{\noopsort}[1]{}\providecommand{\singleletter}[1]{#1}%

\begin{thebibliography}{38}%
\makeatletter
\providecommand \@ifxundefined [1]{%
 \@ifx{#1\undefined}
}%
\providecommand \@ifnum [1]{%
 \ifnum #1\expandafter \@firstoftwo
 \else \expandafter \@secondoftwo
 \fi
}%
\providecommand \@ifx [1]{%
 \ifx #1\expandafter \@firstoftwo
 \else \expandafter \@secondoftwo
 \fi
}%
\providecommand \natexlab [1]{#1}%
\providecommand \enquote  [1]{``#1''}%
\providecommand \bibnamefont  [1]{#1}%
\providecommand \bibfnamefont [1]{#1}%
\providecommand \citenamefont [1]{#1}%
\providecommand \href@noop [0]{\@secondoftwo}%
\providecommand \href [0]{\begingroup \@sanitize@url \@href}%
\providecommand \@href[1]{\@@startlink{#1}\@@href}%
\providecommand \@@href[1]{\endgroup#1\@@endlink}%
\providecommand \@sanitize@url [0]{\catcode `\\12\catcode `\$12\catcode
  `\&12\catcode `\#12\catcode `\^12\catcode `\_12\catcode `\%12\relax}%
\providecommand \@@startlink[1]{}%
\providecommand \@@endlink[0]{}%
\providecommand \url  [0]{\begingroup\@sanitize@url \@url }%
\providecommand \@url [1]{\endgroup\@href {#1}{\urlprefix }}%
\providecommand \urlprefix  [0]{URL }%
\providecommand \Eprint [0]{\href }%
\providecommand \doibase [0]{https://doi.org/}%
\providecommand \selectlanguage [0]{\@gobble}%
\providecommand \bibinfo  [0]{\@secondoftwo}%
\providecommand \bibfield  [0]{\@secondoftwo}%
\providecommand \translation [1]{[#1]}%
\providecommand \BibitemOpen [0]{}%
\providecommand \bibitemStop [0]{}%
\providecommand \bibitemNoStop [0]{.\EOS\space}%
\providecommand \EOS [0]{\spacefactor3000\relax}%
\providecommand \BibitemShut  [1]{\csname bibitem#1\endcsname}%
\let\auto@bib@innerbib\@empty
%</preamble>
\bibitem [{\citenamefont {Granovetter}(1973)}]{Granovetter:1973}%
  \BibitemOpen
  \bibfield  {author} {\bibinfo {author} {\bibfnamefont {M.~S.}\ \bibnamefont
  {Granovetter}},\ }\bibfield  {title} {\bibinfo {title} {The strength of weak
  ties},\ }\href {http://www.jstor.org/stable/2776392} {\bibfield  {journal}
  {\bibinfo  {journal} {American Journal of Sociology}\ }\textbf {\bibinfo
  {volume} {78}},\ \bibinfo {pages} {1360} (\bibinfo {year}
  {1973})}\BibitemShut {NoStop}%
\bibitem [{\citenamefont {Hinde}(1976)}]{Hinde:1976}%
  \BibitemOpen
  \bibfield  {author} {\bibinfo {author} {\bibfnamefont {R.~A.}\ \bibnamefont
  {Hinde}},\ }\bibfield  {title} {\bibinfo {title} {Interactions, relationships
  and social structure},\ }\href {http://www.jstor.org/stable/2800384}
  {\bibfield  {journal} {\bibinfo  {journal} {Man}\ }\textbf {\bibinfo {volume}
  {11}},\ \bibinfo {pages} {1} (\bibinfo {year} {1976})}\BibitemShut {NoStop}%
\bibitem [{\citenamefont {Wasserman}\ and\ \citenamefont
  {Faust}(1994)}]{wasserman1994social}%
  \BibitemOpen
  \bibfield  {author} {\bibinfo {author} {\bibfnamefont {S.}~\bibnamefont
  {Wasserman}}\ and\ \bibinfo {author} {\bibfnamefont {K.}~\bibnamefont
  {Faust}},\ }\href@noop {} {\emph {\bibinfo {title} {Social network analysis:
  Methods and applications}}},\ Vol.~\bibinfo {volume} {8}\ (\bibinfo
  {publisher} {Cambridge university press},\ \bibinfo {year}
  {1994})\BibitemShut {NoStop}%
\bibitem [{\citenamefont {Mossong}\ \emph {et~al.}(2008)\citenamefont
  {Mossong}, \citenamefont {Hens}, \citenamefont {Jit}, \citenamefont
  {Beutels}, \citenamefont {Auranen}, \citenamefont {Mikolajczyk},
  \citenamefont {Massari}, \citenamefont {Salmaso}, \citenamefont {Tomba},
  \citenamefont {Wallinga}, \citenamefont {Heijne}, \citenamefont
  {Sadkowska-Todys}, \citenamefont {Rosinska},\ and\ \citenamefont
  {Edmunds}}]{Mossong:2008}%
  \BibitemOpen
  \bibfield  {author} {\bibinfo {author} {\bibfnamefont {J.}~\bibnamefont
  {Mossong}}, \bibinfo {author} {\bibfnamefont {N.}~\bibnamefont {Hens}},
  \bibinfo {author} {\bibfnamefont {M.}~\bibnamefont {Jit}}, \bibinfo {author}
  {\bibfnamefont {P.}~\bibnamefont {Beutels}}, \bibinfo {author} {\bibfnamefont
  {K.}~\bibnamefont {Auranen}}, \bibinfo {author} {\bibfnamefont
  {R.}~\bibnamefont {Mikolajczyk}}, \bibinfo {author} {\bibfnamefont
  {M.}~\bibnamefont {Massari}}, \bibinfo {author} {\bibfnamefont
  {S.}~\bibnamefont {Salmaso}}, \bibinfo {author} {\bibfnamefont {G.~S.}\
  \bibnamefont {Tomba}}, \bibinfo {author} {\bibfnamefont {J.}~\bibnamefont
  {Wallinga}}, \bibinfo {author} {\bibfnamefont {J.}~\bibnamefont {Heijne}},
  \bibinfo {author} {\bibfnamefont {M.}~\bibnamefont {Sadkowska-Todys}},
  \bibinfo {author} {\bibfnamefont {M.}~\bibnamefont {Rosinska}},\ and\
  \bibinfo {author} {\bibfnamefont {W.~J.}\ \bibnamefont {Edmunds}},\
  }\bibfield  {title} {\bibinfo {title} {Social contacts and mixing patterns
  relevant to the spread of infectious diseases},\ }\href@noop {} {\bibfield
  {journal} {\bibinfo  {journal} {PLoS Med}\ }\textbf {\bibinfo {volume} {5}},\
  \bibinfo {pages} {e74} (\bibinfo {year} {2008})}\BibitemShut {NoStop}%
\bibitem [{\citenamefont {Conlan}\ \emph {et~al.}(2011)\citenamefont {Conlan},
  \citenamefont {Eames}, \citenamefont {Gage}, \citenamefont {von Kirchbach},
  \citenamefont {Ross}, \citenamefont {Saenz},\ and\ \citenamefont
  {Gog}}]{Conlan:2011}%
  \BibitemOpen
  \bibfield  {author} {\bibinfo {author} {\bibfnamefont {A.~J.~K.}\
  \bibnamefont {Conlan}}, \bibinfo {author} {\bibfnamefont {K.~T.~D.}\
  \bibnamefont {Eames}}, \bibinfo {author} {\bibfnamefont {J.~A.}\ \bibnamefont
  {Gage}}, \bibinfo {author} {\bibfnamefont {J.~C.}\ \bibnamefont {von
  Kirchbach}}, \bibinfo {author} {\bibfnamefont {J.~V.}\ \bibnamefont {Ross}},
  \bibinfo {author} {\bibfnamefont {R.~A.}\ \bibnamefont {Saenz}},\ and\
  \bibinfo {author} {\bibfnamefont {J.~R.}\ \bibnamefont {Gog}},\ }\bibfield
  {title} {\bibinfo {title} {Measuring social networks in british primary
  schools through scientific engagement},\ }\href@noop {} {\bibfield  {journal}
  {\bibinfo  {journal} {Proceedings of the Royal Society B: Biological
  Sciences}\ }\textbf {\bibinfo {volume} {278}},\ \bibinfo {pages} {1467}
  (\bibinfo {year} {2011})}\BibitemShut {NoStop}%
\bibitem [{\citenamefont {Malik}(2018)}]{malik2018bias}%
  \BibitemOpen
  \bibfield  {author} {\bibinfo {author} {\bibfnamefont {M.~M.}\ \bibnamefont
  {Malik}},\ }\emph {\bibinfo {title} {Bias and beyond in digital trace
  data}},\ \href@noop {} {Ph.D. thesis},\ \bibinfo  {school} {Doctoral
  dissertation, Carnegie Mellon University} (\bibinfo {year}
  {2018})\BibitemShut {NoStop}%
\bibitem [{\citenamefont {Schaible}\ \emph {et~al.}(2021)\citenamefont
  {Schaible}, \citenamefont {Oliveira}, \citenamefont {Zens},\ and\
  \citenamefont {G{\'e}nois}}]{schaible2021sensing}%
  \BibitemOpen
  \bibfield  {author} {\bibinfo {author} {\bibfnamefont {J.}~\bibnamefont
  {Schaible}}, \bibinfo {author} {\bibfnamefont {M.}~\bibnamefont {Oliveira}},
  \bibinfo {author} {\bibfnamefont {M.}~\bibnamefont {Zens}},\ and\ \bibinfo
  {author} {\bibfnamefont {M.}~\bibnamefont {G{\'e}nois}},\ }\bibfield  {title}
  {\bibinfo {title} {Sensing close-range proximity for studying face-to-face
  interaction},\ }in\ \href@noop {} {\emph {\bibinfo {booktitle} {Handbook of
  Computational Social Science, Volume 1}}}\ (\bibinfo  {publisher}
  {Routledge},\ \bibinfo {year} {2021})\ pp.\ \bibinfo {pages}
  {219--239}\BibitemShut {NoStop}%
\bibitem [{\citenamefont {Cattuto}\ \emph {et~al.}(2010)\citenamefont
  {Cattuto}, \citenamefont {Van~den Broeck}, \citenamefont {Barrat},
  \citenamefont {Colizza}, \citenamefont {Pinton},\ and\ \citenamefont
  {Vespignani}}]{cattuto2010dynamics}%
  \BibitemOpen
  \bibfield  {author} {\bibinfo {author} {\bibfnamefont {C.}~\bibnamefont
  {Cattuto}}, \bibinfo {author} {\bibfnamefont {W.}~\bibnamefont {Van~den
  Broeck}}, \bibinfo {author} {\bibfnamefont {A.}~\bibnamefont {Barrat}},
  \bibinfo {author} {\bibfnamefont {V.}~\bibnamefont {Colizza}}, \bibinfo
  {author} {\bibfnamefont {J.-F.}\ \bibnamefont {Pinton}},\ and\ \bibinfo
  {author} {\bibfnamefont {A.}~\bibnamefont {Vespignani}},\ }\bibfield  {title}
  {\bibinfo {title} {Dynamics of person-to-person interactions from distributed
  rfid sensor networks},\ }\href {https://doi.org/10.1371/journal.pone.0011596}
  {\bibfield  {journal} {\bibinfo  {journal} {PLoS ONE}\ }\textbf {\bibinfo
  {volume} {5}},\ \bibinfo {pages} {e11596} (\bibinfo {year}
  {2010})}\BibitemShut {NoStop}%
\bibitem [{\citenamefont {Salath\'e}\ \emph {et~al.}(2010)\citenamefont
  {Salath\'e}, \citenamefont {Kazandjieva}, \citenamefont {Lee}, \citenamefont
  {Levis}, \citenamefont {Feldman},\ and\ \citenamefont
  {Jones}}]{salathe2010high}%
  \BibitemOpen
  \bibfield  {author} {\bibinfo {author} {\bibfnamefont {M.}~\bibnamefont
  {Salath\'e}}, \bibinfo {author} {\bibfnamefont {M.}~\bibnamefont
  {Kazandjieva}}, \bibinfo {author} {\bibfnamefont {J.~W.}\ \bibnamefont
  {Lee}}, \bibinfo {author} {\bibfnamefont {P.}~\bibnamefont {Levis}}, \bibinfo
  {author} {\bibfnamefont {M.~W.}\ \bibnamefont {Feldman}},\ and\ \bibinfo
  {author} {\bibfnamefont {J.~H.}\ \bibnamefont {Jones}},\ }\bibfield  {title}
  {\bibinfo {title} {A high-resolution human contact network for infectious
  disease transmission},\ }\href {https://doi.org/10.1073/pnas.1009094108}
  {\bibfield  {journal} {\bibinfo  {journal} {Proceedings of the National
  Academy of Sciences}\ }\textbf {\bibinfo {volume} {107}},\ \bibinfo {pages}
  {22020} (\bibinfo {year} {2010})}\BibitemShut {NoStop}%
\bibitem [{\citenamefont {Stehl{\'e}}\ \emph {et~al.}(2011)\citenamefont
  {Stehl{\'e}}, \citenamefont {Voirin}, \citenamefont {Barrat}, \citenamefont
  {Cattuto}, \citenamefont {Isella}, \citenamefont {Pinton}, \citenamefont
  {Quaggiotto}, \citenamefont {Van~den Broeck}, \citenamefont {R{\'e}gis},
  \citenamefont {Lina} \emph {et~al.}}]{stehle2011high}%
  \BibitemOpen
  \bibfield  {author} {\bibinfo {author} {\bibfnamefont {J.}~\bibnamefont
  {Stehl{\'e}}}, \bibinfo {author} {\bibfnamefont {N.}~\bibnamefont {Voirin}},
  \bibinfo {author} {\bibfnamefont {A.}~\bibnamefont {Barrat}}, \bibinfo
  {author} {\bibfnamefont {C.}~\bibnamefont {Cattuto}}, \bibinfo {author}
  {\bibfnamefont {L.}~\bibnamefont {Isella}}, \bibinfo {author} {\bibfnamefont
  {J.-F.}\ \bibnamefont {Pinton}}, \bibinfo {author} {\bibfnamefont
  {M.}~\bibnamefont {Quaggiotto}}, \bibinfo {author} {\bibfnamefont
  {W.}~\bibnamefont {Van~den Broeck}}, \bibinfo {author} {\bibfnamefont
  {C.}~\bibnamefont {R{\'e}gis}}, \bibinfo {author} {\bibfnamefont
  {B.}~\bibnamefont {Lina}}, \emph {et~al.},\ }\bibfield  {title} {\bibinfo
  {title} {High-resolution measurements of face-to-face contact patterns in a
  primary school},\ }\href@noop {} {\bibfield  {journal} {\bibinfo  {journal}
  {PloS one}\ }\textbf {\bibinfo {volume} {6}},\ \bibinfo {pages} {e23176}
  (\bibinfo {year} {2011})}\BibitemShut {NoStop}%
\bibitem [{\citenamefont {Barrat}\ \emph {et~al.}(2013)\citenamefont {Barrat},
  \citenamefont {Cattuto}, \citenamefont {Colizza}, \citenamefont {Gesualdo},
  \citenamefont {Isella}, \citenamefont {Pandolfi}, \citenamefont {Pinton},
  \citenamefont {Rav{\`a}}, \citenamefont {Rizzo}, \citenamefont {Romano} \emph
  {et~al.}}]{barrat2013empirical}%
  \BibitemOpen
  \bibfield  {author} {\bibinfo {author} {\bibfnamefont {A.}~\bibnamefont
  {Barrat}}, \bibinfo {author} {\bibfnamefont {C.}~\bibnamefont {Cattuto}},
  \bibinfo {author} {\bibfnamefont {V.}~\bibnamefont {Colizza}}, \bibinfo
  {author} {\bibfnamefont {F.}~\bibnamefont {Gesualdo}}, \bibinfo {author}
  {\bibfnamefont {L.}~\bibnamefont {Isella}}, \bibinfo {author} {\bibfnamefont
  {E.}~\bibnamefont {Pandolfi}}, \bibinfo {author} {\bibfnamefont {J.-F.}\
  \bibnamefont {Pinton}}, \bibinfo {author} {\bibfnamefont {L.}~\bibnamefont
  {Rav{\`a}}}, \bibinfo {author} {\bibfnamefont {C.}~\bibnamefont {Rizzo}},
  \bibinfo {author} {\bibfnamefont {M.}~\bibnamefont {Romano}}, \emph
  {et~al.},\ }\bibfield  {title} {\bibinfo {title} {Empirical temporal networks
  of face-to-face human interactions},\ }\href@noop {} {\bibfield  {journal}
  {\bibinfo  {journal} {The European Physical Journal Special Topics}\ }\textbf
  {\bibinfo {volume} {222}},\ \bibinfo {pages} {1295} (\bibinfo {year}
  {2013})}\BibitemShut {NoStop}%
\bibitem [{\citenamefont {Stopczynski}\ \emph {et~al.}(2014)\citenamefont
  {Stopczynski}, \citenamefont {Sekara}, \citenamefont {Sapiezynski},
  \citenamefont {Cuttone}, \citenamefont {Madsen}, \citenamefont {Larsen},\
  and\ \citenamefont {Lehmann}}]{stopczynski2014measuring}%
  \BibitemOpen
  \bibfield  {author} {\bibinfo {author} {\bibfnamefont {A.}~\bibnamefont
  {Stopczynski}}, \bibinfo {author} {\bibfnamefont {V.}~\bibnamefont {Sekara}},
  \bibinfo {author} {\bibfnamefont {P.}~\bibnamefont {Sapiezynski}}, \bibinfo
  {author} {\bibfnamefont {A.}~\bibnamefont {Cuttone}}, \bibinfo {author}
  {\bibfnamefont {M.~M.}\ \bibnamefont {Madsen}}, \bibinfo {author}
  {\bibfnamefont {J.~E.}\ \bibnamefont {Larsen}},\ and\ \bibinfo {author}
  {\bibfnamefont {S.}~\bibnamefont {Lehmann}},\ }\bibfield  {title} {\bibinfo
  {title} {Measuring large-scale social networks with high resolution},\ }\href
  {https://doi.org/10.1371/journal.pone.0095978} {\bibfield  {journal}
  {\bibinfo  {journal} {PLoS ONE}\ }\textbf {\bibinfo {volume} {9}},\ \bibinfo
  {pages} {e95978} (\bibinfo {year} {2014})}\BibitemShut {NoStop}%
\bibitem [{\citenamefont {Toth}\ \emph {et~al.}(2015)\citenamefont {Toth},
  \citenamefont {Leecaster}, \citenamefont {Pettey}, \citenamefont
  {Gundlapalli}, \citenamefont {Gao}, \citenamefont {Rainey}, \citenamefont
  {Uzicanin},\ and\ \citenamefont {Samore}}]{toth2015role}%
  \BibitemOpen
  \bibfield  {author} {\bibinfo {author} {\bibfnamefont {D.~J.}\ \bibnamefont
  {Toth}}, \bibinfo {author} {\bibfnamefont {M.}~\bibnamefont {Leecaster}},
  \bibinfo {author} {\bibfnamefont {W.~B.}\ \bibnamefont {Pettey}}, \bibinfo
  {author} {\bibfnamefont {A.~V.}\ \bibnamefont {Gundlapalli}}, \bibinfo
  {author} {\bibfnamefont {H.}~\bibnamefont {Gao}}, \bibinfo {author}
  {\bibfnamefont {J.~J.}\ \bibnamefont {Rainey}}, \bibinfo {author}
  {\bibfnamefont {A.}~\bibnamefont {Uzicanin}},\ and\ \bibinfo {author}
  {\bibfnamefont {M.~H.}\ \bibnamefont {Samore}},\ }\bibfield  {title}
  {\bibinfo {title} {The role of heterogeneity in contact timing and duration
  in network models of influenza spread in schools},\ }\href@noop {} {\bibfield
   {journal} {\bibinfo  {journal} {Journal of The Royal Society Interface}\
  }\textbf {\bibinfo {volume} {12}},\ \bibinfo {pages} {20150279} (\bibinfo
  {year} {2015})}\BibitemShut {NoStop}%
\bibitem [{\citenamefont {Sapiezynski}\ \emph {et~al.}(2019)\citenamefont
  {Sapiezynski}, \citenamefont {Stopczynski}, \citenamefont {Lassen},\ and\
  \citenamefont {Lehmann}}]{sapiezynski2019interaction}%
  \BibitemOpen
  \bibfield  {author} {\bibinfo {author} {\bibfnamefont {P.}~\bibnamefont
  {Sapiezynski}}, \bibinfo {author} {\bibfnamefont {A.}~\bibnamefont
  {Stopczynski}}, \bibinfo {author} {\bibfnamefont {D.~D.}\ \bibnamefont
  {Lassen}},\ and\ \bibinfo {author} {\bibfnamefont {S.}~\bibnamefont
  {Lehmann}},\ }\bibfield  {title} {\bibinfo {title} {Interaction data from the
  copenhagen networks study},\ }\href@noop {} {\bibfield  {journal} {\bibinfo
  {journal} {Scientific Data}\ }\textbf {\bibinfo {volume} {6}},\ \bibinfo
  {pages} {315} (\bibinfo {year} {2019})}\BibitemShut {NoStop}%
\bibitem [{\citenamefont {Holme}(2015)}]{holme2015modern}%
  \BibitemOpen
  \bibfield  {author} {\bibinfo {author} {\bibfnamefont {P.}~\bibnamefont
  {Holme}},\ }\bibfield  {title} {\bibinfo {title} {Modern temporal network
  theory: a colloquium},\ }\href@noop {} {\bibfield  {journal} {\bibinfo
  {journal} {The European Physical Journal B}\ }\textbf {\bibinfo {volume}
  {88}},\ \bibinfo {pages} {1} (\bibinfo {year} {2015})}\BibitemShut {NoStop}%
\bibitem [{\citenamefont {Holme}\ and\ \citenamefont
  {Saram{\"a}ki}(2012)}]{holme2012temporal}%
  \BibitemOpen
  \bibfield  {author} {\bibinfo {author} {\bibfnamefont {P.}~\bibnamefont
  {Holme}}\ and\ \bibinfo {author} {\bibfnamefont {J.}~\bibnamefont
  {Saram{\"a}ki}},\ }\bibfield  {title} {\bibinfo {title} {Temporal networks},\
  }\href@noop {} {\bibfield  {journal} {\bibinfo  {journal} {Physics reports}\
  }\textbf {\bibinfo {volume} {519}},\ \bibinfo {pages} {97} (\bibinfo {year}
  {2012})}\BibitemShut {NoStop}%
\bibitem [{\citenamefont {Stehl{\'e}}\ \emph {et~al.}(2010)\citenamefont
  {Stehl{\'e}}, \citenamefont {Barrat},\ and\ \citenamefont
  {Bianconi}}]{stehle2010dynamical}%
  \BibitemOpen
  \bibfield  {author} {\bibinfo {author} {\bibfnamefont {J.}~\bibnamefont
  {Stehl{\'e}}}, \bibinfo {author} {\bibfnamefont {A.}~\bibnamefont {Barrat}},\
  and\ \bibinfo {author} {\bibfnamefont {G.}~\bibnamefont {Bianconi}},\
  }\bibfield  {title} {\bibinfo {title} {Dynamical and bursty interactions in
  social networks},\ }\href@noop {} {\bibfield  {journal} {\bibinfo  {journal}
  {Physical review E}\ }\textbf {\bibinfo {volume} {81}},\ \bibinfo {pages}
  {035101} (\bibinfo {year} {2010})}\BibitemShut {NoStop}%
\bibitem [{\citenamefont {Karsai}\ \emph {et~al.}(2012)\citenamefont {Karsai},
  \citenamefont {Kaski}, \citenamefont {Barab{\'a}si},\ and\ \citenamefont
  {Kert{\'e}sz}}]{Karsai2012}%
  \BibitemOpen
  \bibfield  {author} {\bibinfo {author} {\bibfnamefont {M.}~\bibnamefont
  {Karsai}}, \bibinfo {author} {\bibfnamefont {K.}~\bibnamefont {Kaski}},
  \bibinfo {author} {\bibfnamefont {A.-L.}\ \bibnamefont {Barab{\'a}si}},\ and\
  \bibinfo {author} {\bibfnamefont {J.}~\bibnamefont {Kert{\'e}sz}},\
  }\bibfield  {title} {\bibinfo {title} {Universal features of correlated
  bursty behaviour},\ }\href {https://doi.org/10.1038/srep00397} {\bibfield
  {journal} {\bibinfo  {journal} {Scientific Reports}\ }\textbf {\bibinfo
  {volume} {2}},\ \bibinfo {pages} {397} (\bibinfo {year} {2012})}\BibitemShut
  {NoStop}%
\bibitem [{\citenamefont {Vestergaard}\ \emph {et~al.}(2014)\citenamefont
  {Vestergaard}, \citenamefont {G\'enois},\ and\ \citenamefont
  {Barrat}}]{vestergaard2014how}%
  \BibitemOpen
  \bibfield  {author} {\bibinfo {author} {\bibfnamefont {C.~L.}\ \bibnamefont
  {Vestergaard}}, \bibinfo {author} {\bibfnamefont {M.}~\bibnamefont
  {G\'enois}},\ and\ \bibinfo {author} {\bibfnamefont {A.}~\bibnamefont
  {Barrat}},\ }\bibfield  {title} {\bibinfo {title} {How memory generates
  heterogeneous dynamics in temporal networks},\ }\href
  {https://doi.org/10.1103/PhysRevE.90.042805} {\bibfield  {journal} {\bibinfo
  {journal} {Phys. Rev. E}\ }\textbf {\bibinfo {volume} {90}},\ \bibinfo
  {pages} {042805} (\bibinfo {year} {2014})}\BibitemShut {NoStop}%
\bibitem [{\citenamefont {Karsai}\ \emph {et~al.}(2014)\citenamefont {Karsai},
  \citenamefont {Perra},\ and\ \citenamefont {Vespignani}}]{karsai2014time}%
  \BibitemOpen
  \bibfield  {author} {\bibinfo {author} {\bibfnamefont {M.}~\bibnamefont
  {Karsai}}, \bibinfo {author} {\bibfnamefont {N.}~\bibnamefont {Perra}},\ and\
  \bibinfo {author} {\bibfnamefont {A.}~\bibnamefont {Vespignani}},\ }\bibfield
   {title} {\bibinfo {title} {Time varying networks and the weakness of strong
  ties.},\ }\href@noop {} {\bibfield  {journal} {\bibinfo  {journal} {Sci Rep}\
  }\textbf {\bibinfo {volume} {4}},\ \bibinfo {pages} {4001} (\bibinfo {year}
  {2014})}\BibitemShut {NoStop}%
\bibitem [{\citenamefont {Laurent}\ \emph {et~al.}(2015)\citenamefont
  {Laurent}, \citenamefont {Saramäki},\ and\ \citenamefont
  {Karsai}}]{Laurent_2015}%
  \BibitemOpen
  \bibfield  {author} {\bibinfo {author} {\bibfnamefont {G.}~\bibnamefont
  {Laurent}}, \bibinfo {author} {\bibfnamefont {J.}~\bibnamefont {Saramäki}},\
  and\ \bibinfo {author} {\bibfnamefont {M.}~\bibnamefont {Karsai}},\
  }\bibfield  {title} {\bibinfo {title} {From calls to communities: a model for
  time-varying social networks},\ }\bibfield  {journal} {\bibinfo  {journal}
  {The European Physical Journal B}\ }\textbf {\bibinfo {volume} {88}},\ \href
  {https://doi.org/10.1140/epjb/e2015-60481-x} {10.1140/epjb/e2015-60481-x}
  (\bibinfo {year} {2015})\BibitemShut {NoStop}%
\bibitem [{\citenamefont {Gelardi}\ \emph {et~al.}(2021)\citenamefont
  {Gelardi}, \citenamefont {Le~Bail}, \citenamefont {Barrat},\ and\
  \citenamefont {Claidiere}}]{gelardi2021temporal}%
  \BibitemOpen
  \bibfield  {author} {\bibinfo {author} {\bibfnamefont {V.}~\bibnamefont
  {Gelardi}}, \bibinfo {author} {\bibfnamefont {D.}~\bibnamefont {Le~Bail}},
  \bibinfo {author} {\bibfnamefont {A.}~\bibnamefont {Barrat}},\ and\ \bibinfo
  {author} {\bibfnamefont {N.}~\bibnamefont {Claidiere}},\ }\bibfield  {title}
  {\bibinfo {title} {From temporal network data to the dynamics of social
  relationships},\ }\href@noop {} {\bibfield  {journal} {\bibinfo  {journal}
  {Proceedings of the Royal Society B}\ }\textbf {\bibinfo {volume} {288}},\
  \bibinfo {pages} {20211164} (\bibinfo {year} {2021})}\BibitemShut {NoStop}%
\bibitem [{\citenamefont {Jo}\ \emph {et~al.}(2011)\citenamefont {Jo},
  \citenamefont {Pan},\ and\ \citenamefont {Kaski}}]{jo2011emergence}%
  \BibitemOpen
  \bibfield  {author} {\bibinfo {author} {\bibfnamefont {H.-H.}\ \bibnamefont
  {Jo}}, \bibinfo {author} {\bibfnamefont {R.~K.}\ \bibnamefont {Pan}},\ and\
  \bibinfo {author} {\bibfnamefont {K.}~\bibnamefont {Kaski}},\ }\bibfield
  {title} {\bibinfo {title} {Emergence of bursts and communities in evolving
  weighted networks},\ }\href@noop {} {\bibfield  {journal} {\bibinfo
  {journal} {PloS one}\ }\textbf {\bibinfo {volume} {6}},\ \bibinfo {pages}
  {e22687} (\bibinfo {year} {2011})}\BibitemShut {NoStop}%
\bibitem [{\citenamefont {Kumpula}\ \emph {et~al.}(2007)\citenamefont
  {Kumpula}, \citenamefont {Onnela}, \citenamefont {Saram{\"a}ki},
  \citenamefont {Kaski},\ and\ \citenamefont
  {Kert{\'e}sz}}]{kumpula2007emergence}%
  \BibitemOpen
  \bibfield  {author} {\bibinfo {author} {\bibfnamefont {J.~M.}\ \bibnamefont
  {Kumpula}}, \bibinfo {author} {\bibfnamefont {J.-P.}\ \bibnamefont {Onnela}},
  \bibinfo {author} {\bibfnamefont {J.}~\bibnamefont {Saram{\"a}ki}}, \bibinfo
  {author} {\bibfnamefont {K.}~\bibnamefont {Kaski}},\ and\ \bibinfo {author}
  {\bibfnamefont {J.}~\bibnamefont {Kert{\'e}sz}},\ }\bibfield  {title}
  {\bibinfo {title} {Emergence of communities in weighted networks},\
  }\href@noop {} {\bibfield  {journal} {\bibinfo  {journal} {Physical review
  letters}\ }\textbf {\bibinfo {volume} {99}},\ \bibinfo {pages} {228701}
  (\bibinfo {year} {2007})}\BibitemShut {NoStop}%
\bibitem [{\citenamefont {Orsini}\ \emph {et~al.}(2015)\citenamefont {Orsini},
  \citenamefont {Dankulov}, \citenamefont {Colomer-de Sim{\'o}n}, \citenamefont
  {Jamakovic}, \citenamefont {Mahadevan}, \citenamefont {Vahdat}, \citenamefont
  {Bassler}, \citenamefont {Toroczkai}, \citenamefont {Bogu{\~n}{\'a}},
  \citenamefont {Caldarelli}, \citenamefont {Fortunato},\ and\ \citenamefont
  {Krioukov}}]{orsini2015quantifying}%
  \BibitemOpen
  \bibfield  {author} {\bibinfo {author} {\bibfnamefont {C.}~\bibnamefont
  {Orsini}}, \bibinfo {author} {\bibfnamefont {M.~M.}\ \bibnamefont
  {Dankulov}}, \bibinfo {author} {\bibfnamefont {P.}~\bibnamefont {Colomer-de
  Sim{\'o}n}}, \bibinfo {author} {\bibfnamefont {A.}~\bibnamefont {Jamakovic}},
  \bibinfo {author} {\bibfnamefont {P.}~\bibnamefont {Mahadevan}}, \bibinfo
  {author} {\bibfnamefont {A.}~\bibnamefont {Vahdat}}, \bibinfo {author}
  {\bibfnamefont {K.~E.}\ \bibnamefont {Bassler}}, \bibinfo {author}
  {\bibfnamefont {Z.}~\bibnamefont {Toroczkai}}, \bibinfo {author}
  {\bibfnamefont {M.}~\bibnamefont {Bogu{\~n}{\'a}}}, \bibinfo {author}
  {\bibfnamefont {G.}~\bibnamefont {Caldarelli}}, \bibinfo {author}
  {\bibfnamefont {S.}~\bibnamefont {Fortunato}},\ and\ \bibinfo {author}
  {\bibfnamefont {D.}~\bibnamefont {Krioukov}},\ }\bibfield  {title} {\bibinfo
  {title} {Quantifying randomness in real networks},\ }\href@noop {} {\bibfield
   {journal} {\bibinfo  {journal} {Nature Communications}\ }\textbf {\bibinfo
  {volume} {6}},\ \bibinfo {pages} {8627} (\bibinfo {year} {2015})}\BibitemShut
  {NoStop}%
\bibitem [{\citenamefont {Longa}\ \emph
  {et~al.}(2022{\natexlab{a}})\citenamefont {Longa}, \citenamefont {Cencetti},
  \citenamefont {Lepri},\ and\ \citenamefont {Passerini}}]{Longa2022}%
  \BibitemOpen
  \bibfield  {author} {\bibinfo {author} {\bibfnamefont {A.}~\bibnamefont
  {Longa}}, \bibinfo {author} {\bibfnamefont {G.}~\bibnamefont {Cencetti}},
  \bibinfo {author} {\bibfnamefont {B.}~\bibnamefont {Lepri}},\ and\ \bibinfo
  {author} {\bibfnamefont {A.}~\bibnamefont {Passerini}},\ }\bibfield  {title}
  {\bibinfo {title} {An efficient procedure for mining egocentric temporal
  motifs},\ }\href {https://doi.org/10.1007/s10618-021-00803-2} {\bibfield
  {journal} {\bibinfo  {journal} {Data Mining and Knowledge Discovery}\
  }\textbf {\bibinfo {volume} {36}},\ \bibinfo {pages} {355} (\bibinfo {year}
  {2022}{\natexlab{a}})}\BibitemShut {NoStop}%
\bibitem [{\citenamefont {Longa}\ \emph
  {et~al.}(2022{\natexlab{b}})\citenamefont {Longa}, \citenamefont {Cencetti},
  \citenamefont {Lehmann}, \citenamefont {Passerini},\ and\ \citenamefont
  {Lepri}}]{longa2022neighbourhood}%
  \BibitemOpen
  \bibfield  {author} {\bibinfo {author} {\bibfnamefont {A.}~\bibnamefont
  {Longa}}, \bibinfo {author} {\bibfnamefont {G.}~\bibnamefont {Cencetti}},
  \bibinfo {author} {\bibfnamefont {S.}~\bibnamefont {Lehmann}}, \bibinfo
  {author} {\bibfnamefont {A.}~\bibnamefont {Passerini}},\ and\ \bibinfo
  {author} {\bibfnamefont {B.}~\bibnamefont {Lepri}},\ }\bibfield  {title}
  {\bibinfo {title} {Neighbourhood matching creates realistic surrogate
  temporal networks},\ }\href@noop {} {\bibfield  {journal} {\bibinfo
  {journal} {arXiv}\ ,\ \bibinfo {pages} {arXiv:2205.08820}} (\bibinfo {year}
  {2022}{\natexlab{b}})}\BibitemShut {NoStop}%
\bibitem [{\citenamefont {Dunbar}\ \emph {et~al.}(2009)\citenamefont {Dunbar},
  \citenamefont {Korstjens}, \citenamefont {Lehmann},\ and\ \citenamefont
  {Project}}]{dunbar2009time}%
  \BibitemOpen
  \bibfield  {author} {\bibinfo {author} {\bibfnamefont {R.~I.}\ \bibnamefont
  {Dunbar}}, \bibinfo {author} {\bibfnamefont {A.~H.}\ \bibnamefont
  {Korstjens}}, \bibinfo {author} {\bibfnamefont {J.}~\bibnamefont {Lehmann}},\
  and\ \bibinfo {author} {\bibfnamefont {B.~A. C.~R.}\ \bibnamefont
  {Project}},\ }\bibfield  {title} {\bibinfo {title} {Time as an ecological
  constraint},\ }\href@noop {} {\bibfield  {journal} {\bibinfo  {journal}
  {Biological Reviews}\ }\textbf {\bibinfo {volume} {84}},\ \bibinfo {pages}
  {413} (\bibinfo {year} {2009})}\BibitemShut {NoStop}%
\bibitem [{\citenamefont {Miritello}\ \emph {et~al.}(2013)\citenamefont
  {Miritello}, \citenamefont {Lara}, \citenamefont {Cebrian},\ and\
  \citenamefont {Moro}}]{miritello2013limited}%
  \BibitemOpen
  \bibfield  {author} {\bibinfo {author} {\bibfnamefont {G.}~\bibnamefont
  {Miritello}}, \bibinfo {author} {\bibfnamefont {R.}~\bibnamefont {Lara}},
  \bibinfo {author} {\bibfnamefont {M.}~\bibnamefont {Cebrian}},\ and\ \bibinfo
  {author} {\bibfnamefont {E.}~\bibnamefont {Moro}},\ }\bibfield  {title}
  {\bibinfo {title} {Limited communication capacity unveils strategies for
  human interaction},\ }\href@noop {} {\bibfield  {journal} {\bibinfo
  {journal} {Scientific reports}\ }\textbf {\bibinfo {volume} {3}},\ \bibinfo
  {pages} {1} (\bibinfo {year} {2013})}\BibitemShut {NoStop}%
\bibitem [{\citenamefont {Perra}\ \emph {et~al.}(2012)\citenamefont {Perra},
  \citenamefont {Gon\c{c}alves}, \citenamefont {Pastor-Satorras},\ and\
  \citenamefont {Vespignani}}]{perra2012activity}%
  \BibitemOpen
  \bibfield  {author} {\bibinfo {author} {\bibfnamefont {N.}~\bibnamefont
  {Perra}}, \bibinfo {author} {\bibfnamefont {B.}~\bibnamefont
  {Gon\c{c}alves}}, \bibinfo {author} {\bibfnamefont {R.}~\bibnamefont
  {Pastor-Satorras}},\ and\ \bibinfo {author} {\bibfnamefont {A.}~\bibnamefont
  {Vespignani}},\ }\bibfield  {title} {\bibinfo {title} {{Activity driven
  modeling of time varying networks.}},\ }\href
  {https://doi.org/10.1038/srep00469} {\bibfield  {journal} {\bibinfo
  {journal} {Scientific reports}\ }\textbf {\bibinfo {volume} {2}},\ \bibinfo
  {pages} {469} (\bibinfo {year} {2012})}\BibitemShut {NoStop}%
\bibitem [{\citenamefont {Ubaldi}\ \emph {et~al.}(2016)\citenamefont {Ubaldi},
  \citenamefont {Perra}, \citenamefont {Karsai}, \citenamefont {Vezzani},
  \citenamefont {Burioni},\ and\ \citenamefont
  {Vespignani}}]{ubaldi2016asymptotic}%
  \BibitemOpen
  \bibfield  {author} {\bibinfo {author} {\bibfnamefont {E.}~\bibnamefont
  {Ubaldi}}, \bibinfo {author} {\bibfnamefont {N.}~\bibnamefont {Perra}},
  \bibinfo {author} {\bibfnamefont {M.}~\bibnamefont {Karsai}}, \bibinfo
  {author} {\bibfnamefont {A.}~\bibnamefont {Vezzani}}, \bibinfo {author}
  {\bibfnamefont {R.}~\bibnamefont {Burioni}},\ and\ \bibinfo {author}
  {\bibfnamefont {A.}~\bibnamefont {Vespignani}},\ }\bibfield  {title}
  {\bibinfo {title} {Asymptotic theory of time-varying social networks with
  heterogeneous activity and tie allocation},\ }\href@noop {} {\bibfield
  {journal} {\bibinfo  {journal} {Scientific Reports}\ }\textbf {\bibinfo
  {volume} {6}},\ \bibinfo {pages} {35724} (\bibinfo {year}
  {2016})}\BibitemShut {NoStop}%
\bibitem [{\citenamefont {Thurner}(2018)}]{thurner2018virtual}%
  \BibitemOpen
  \bibfield  {author} {\bibinfo {author} {\bibfnamefont {S.}~\bibnamefont
  {Thurner}},\ }\bibfield  {title} {\bibinfo {title} {Virtual social science},\
  }\href@noop {} {\bibfield  {journal} {\bibinfo  {journal} {arXiv preprint
  arXiv:1811.08156}\ } (\bibinfo {year} {2018})}\BibitemShut {NoStop}%
\bibitem [{\citenamefont {Ilany}\ \emph {et~al.}(2013)\citenamefont {Ilany},
  \citenamefont {Barocas}, \citenamefont {Koren}, \citenamefont {Kam},\ and\
  \citenamefont {Geffen}}]{ilany2013structural}%
  \BibitemOpen
  \bibfield  {author} {\bibinfo {author} {\bibfnamefont {A.}~\bibnamefont
  {Ilany}}, \bibinfo {author} {\bibfnamefont {A.}~\bibnamefont {Barocas}},
  \bibinfo {author} {\bibfnamefont {L.}~\bibnamefont {Koren}}, \bibinfo
  {author} {\bibfnamefont {M.}~\bibnamefont {Kam}},\ and\ \bibinfo {author}
  {\bibfnamefont {E.}~\bibnamefont {Geffen}},\ }\bibfield  {title} {\bibinfo
  {title} {Structural balance in the social networks of a wild mammal},\
  }\href@noop {} {\bibfield  {journal} {\bibinfo  {journal} {Animal Behaviour}\
  }\textbf {\bibinfo {volume} {85}},\ \bibinfo {pages} {1397} (\bibinfo {year}
  {2013})}\BibitemShut {NoStop}%
\bibitem [{\citenamefont {Gelardi}\ \emph {et~al.}(2019)\citenamefont
  {Gelardi}, \citenamefont {Fagot}, \citenamefont {Barrat},\ and\ \citenamefont
  {Claidière}}]{gelardi2019detecting}%
  \BibitemOpen
  \bibfield  {author} {\bibinfo {author} {\bibfnamefont {V.}~\bibnamefont
  {Gelardi}}, \bibinfo {author} {\bibfnamefont {J.}~\bibnamefont {Fagot}},
  \bibinfo {author} {\bibfnamefont {A.}~\bibnamefont {Barrat}},\ and\ \bibinfo
  {author} {\bibfnamefont {N.}~\bibnamefont {Claidière}},\ }\bibfield  {title}
  {\bibinfo {title} {Detecting social (in)stability in primates from their
  temporal co-presence network},\ }\href
  {https://doi.org/https://doi.org/10.1016/j.anbehav.2019.09.011} {\bibfield
  {journal} {\bibinfo  {journal} {Animal Behaviour}\ }\textbf {\bibinfo
  {volume} {157}},\ \bibinfo {pages} {239} (\bibinfo {year}
  {2019})}\BibitemShut {NoStop}%
\bibitem [{\citenamefont {Andres}\ \emph {et~al.}(2022)\citenamefont {Andres},
  \citenamefont {Casiraghi}, \citenamefont {Vaccario},\ and\ \citenamefont
  {Schweitzer}}]{andres2022reconstructing}%
  \BibitemOpen
  \bibfield  {author} {\bibinfo {author} {\bibfnamefont {G.}~\bibnamefont
  {Andres}}, \bibinfo {author} {\bibfnamefont {G.}~\bibnamefont {Casiraghi}},
  \bibinfo {author} {\bibfnamefont {G.}~\bibnamefont {Vaccario}},\ and\
  \bibinfo {author} {\bibfnamefont {F.}~\bibnamefont {Schweitzer}},\ }\bibfield
   {title} {\bibinfo {title} {Reconstructing signed relations from interaction
  data},\ }\href@noop {} {\bibfield  {journal} {\bibinfo  {journal} {arXiv}\ ,\
  \bibinfo {pages} {arXiv:2209.03219}} (\bibinfo {year} {2022})}\BibitemShut
  {NoStop}%
\bibitem [{Note1()}]{Note1}%
  \BibitemOpen
  \bibinfo {note} {Note that this is different from focal closure, which
  suggests the formation of ties between individuals with common attributes or
  interests, and which is implemented by links with randomly chosen individuals
  in \cite {Laurent_2015}}\BibitemShut {NoStop}%
\bibitem [{\citenamefont {Bastian}\ \emph {et~al.}(2009)\citenamefont
  {Bastian}, \citenamefont {Heymann},\ and\ \citenamefont
  {Jacomy}}]{bastian2009gephi}%
  \BibitemOpen
  \bibfield  {author} {\bibinfo {author} {\bibfnamefont {M.}~\bibnamefont
  {Bastian}}, \bibinfo {author} {\bibfnamefont {S.}~\bibnamefont {Heymann}},\
  and\ \bibinfo {author} {\bibfnamefont {M.}~\bibnamefont {Jacomy}},\
  }\bibfield  {title} {\bibinfo {title} {Gephi: an open source software for
  exploring and manipulating networks},\ }in\ \href@noop {} {\emph {\bibinfo
  {booktitle} {Proceedings of the international AAAI conference on web and
  social media}}},\ Vol.~\bibinfo {volume} {3}\ (\bibinfo {year} {2009})\ pp.\
  \bibinfo {pages} {361--362}\BibitemShut {NoStop}%
\bibitem [{\citenamefont {Kobayashi}\ and\ \citenamefont
  {G\'enois}(2020)}]{kobayashi2020two}%
  \BibitemOpen
  \bibfield  {author} {\bibinfo {author} {\bibfnamefont {T.}~\bibnamefont
  {Kobayashi}}\ and\ \bibinfo {author} {\bibfnamefont {M.}~\bibnamefont
  {G\'enois}},\ }\bibfield  {title} {\bibinfo {title} {Two types of
  densification scaling in the evolution of temporal networks},\ }\href
  {https://doi.org/10.1103/PhysRevE.102.052302} {\bibfield  {journal} {\bibinfo
   {journal} {Phys. Rev. E}\ }\textbf {\bibinfo {volume} {102}},\ \bibinfo
  {pages} {052302} (\bibinfo {year} {2020})}\BibitemShut {NoStop}%
\end{thebibliography}
\end{document}

\title{
Modeling framework unifying contact and social networks: \\
Supplemental Material
}

\author{Didier Le Bail}
\affiliation{Aix Marseille Univ, Universit\'e de Toulon, CNRS, CPT, Marseille, France}
\author{Mathieu G\'enois}%
\affiliation{Aix Marseille Univ, Universit\'e de Toulon, CNRS, CPT, Marseille, France}
\author{Alain Barrat}
\affiliation{Aix Marseille Univ, Universit\'e de Toulon, CNRS, CPT, Marseille, France}
\email{alain.barrat@cpt.univ-mrs.fr}

\date{\today}% It is always \today, today,

\maketitle

%\tableofcontents
\setcounter{section}{0}
\setcounter{table}{0}
\setcounter{equation}{0}
\setcounter{figure}{0}
\renewcommand{\figurename}{Supplemental Figure}
\renewcommand{\tablename}{Supplemental Table}
\renewcommand{\thesection}{Supplemental Section \arabic{section}}

\section{Empirical data}
\label{sec:1}
\subsection{Data sets}

The empirical temporal networks we use represent face-to-face interaction data. Individuals are represented as nodes, and an edge is drawn between two nodes each time the associated individuals are interacting with each other. In practice, 
interactions are detected by wearable sensors
that exchange low-power radio signals
\cite{sociopatterns,toth2015role}.
This typically allows to detect face-to-face close 
proximity ($\sim$ 1 meter) with a temporal resolution of about 20 seconds.
%two persons are considered as interacting if they are face-to-face for at least 20 seconds at distance of communication ($\sim$ 1 meter) \cite{cattuto2010dynamics,sociopatterns}. Hence the temporal resolution is of 20 seconds and the shortest possible interaction lasts 20 seconds. These interactions are collected by equipping consenting people with sensors. Two strategies exist: radiofrequency identification devices (RFIDs), employed by SocioPatterns, or wireless ranging enabled nodes (WRENs), employed by Toth et al. \cite{toth2015role} for schools in the USA. 
The data sets we consider here are publicly available thanks to two independent collaborations.
%The specific data sets we use are the following.
The first is provided by Toth et al. \cite{toth2015role} and the others by the SocioPatterns collaboration \cite{sociopatterns}:
\begin{itemize}
    \item the ``utah'' data set describes the proximity interactions which occurred on November 28 and 29, 2012 in an urban public middle school in Utah (USA) \cite{toth2015role};
    \item the ``highschool3'' data set gives the interactions between 327 students of nine classes within a high school in Marseille, during 5 days in December 2013 \cite{mastrandrea2015contact};
    \item the ``conf16'' data set was collected during the 3rd GESIS Computational Social Science Winter Symposium, held on November 30 and December 1, 2016. It is described and called ``WS16'' in \cite{genois2022combining};
    \item the ``conf17'' data set was collected during the International Conference on Computational Social Science, held from July 10 to 13, 2017. It is described and called ``ICCSS17'' in \cite{genois2022combining};
    %\item the 'conf18' data set was collected during the Eurosymposium on Computational Social Science, held from December 5 to 7, 2018. It is described and called 'ECSS18' in \cite{genois2022combining};
    %\item the 'conf19' data set was collected during the 41st European Conference on Information Retrieval, held from April 14 to 18, 2019. It is described and called 'ECIR19' in \cite{genois2022combining};
    \item the ``work2'' data set contains the temporal network of contacts between individuals recorded in an office building in France during two weeks in 2015 \cite{Genois2018}.
\end{itemize}

In practice, as timestamps in the data are multiple of the temporal resolution, we divide the times by the temporal resolution in order to relabel them as successive integers, removing moreover timestamps in which no interactions are observed.

\subsection{Statistical properties}
Empirical temporal networks show broad distributions for many observables \cite{barrat2013empirical}, which indicates strong fluctuations as well as the absence of any characteristic scale. These distributions are very similar for temporal networks obtained in various social contexts, which indicates that their statistical properties may be emergent properties that do not depend strongly on the microscopic dynamics of human relationships.
Instead they may depend on a small number of features of these relationships, which would make it possible to obtain 
realistic temporal networks without a deep understanding of human behaviour.

We compare the statistical properties of the empirical data sets with respect to the  observables we have considered in this paper.  
For the point observables studied, conferences show different properties from the two schools and the workplace. Conferences have a high clustering coefficient  ($\sim0.7$) and a negative degree assortativity ($\sim-0.1$). The three other data sets have a lower clustering coefficient ($0.3$-$0.5$) and a positive degree assortativity ($0.02$-$0.07$).

\begin{figure*}
    \subfigure[size of connected components\label{fig:13a}]{
        \includegraphics[width=0.6\columnwidth]{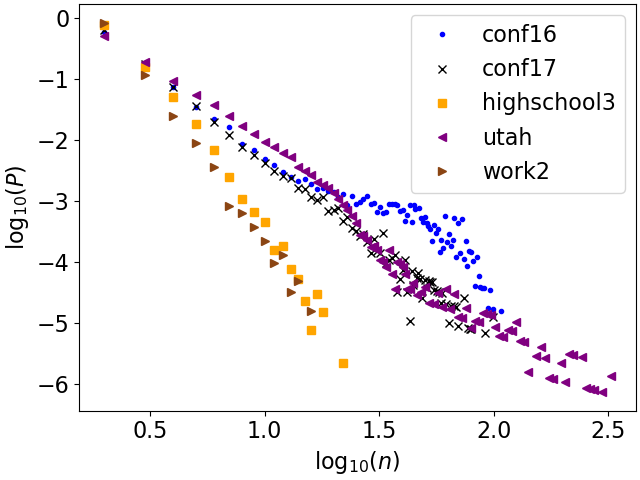}}
    %    \label{fig:13a}
    \subfigure[edge activity duration\label{fig:13b}]{
        \includegraphics[width=0.6\columnwidth]{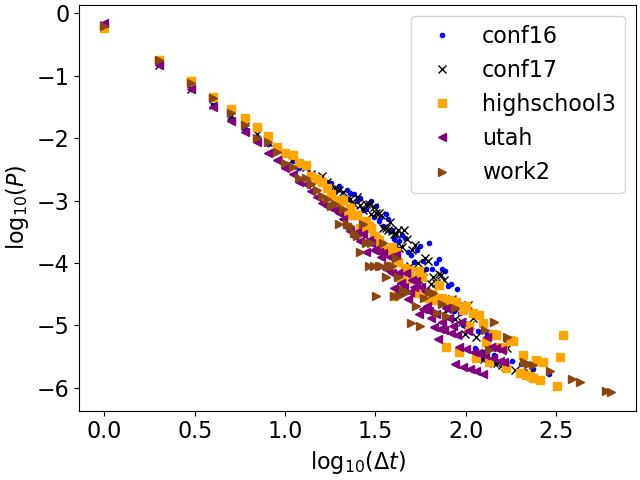}}
    \subfigure[node interactivity duration\label{fig:13c}]{
        \includegraphics[width=0.6\columnwidth]{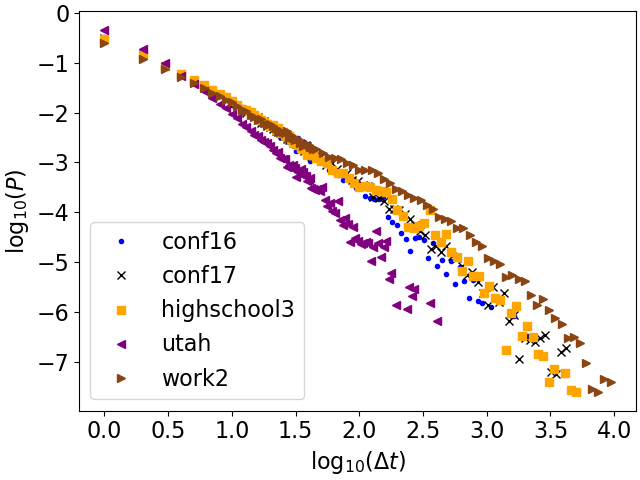}}
    \subfigure[similarity matrix between empirical data sets relative to point observables\label{fig:13d}]{
        \includegraphics[width=0.6\columnwidth]{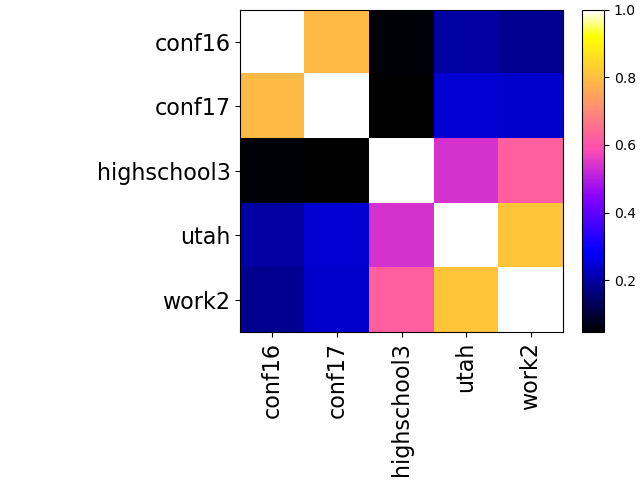}}
    \subfigure[similarity matrix between empirical data sets relative to distribution observables\label{fig:13e}]{
        \includegraphics[width=0.6\columnwidth]{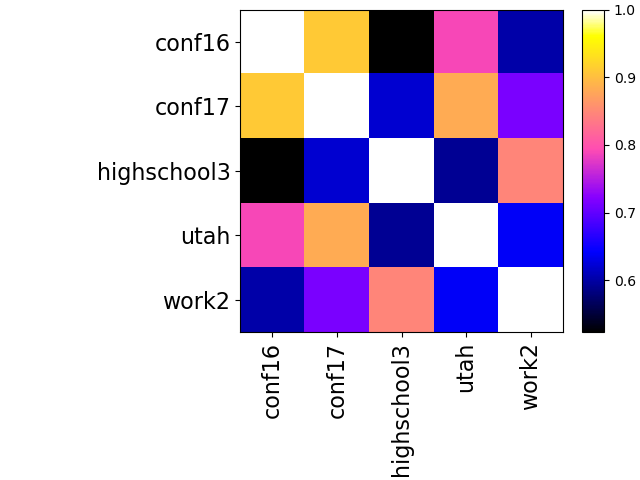}}
    \subfigure[similarity matrix between empirical data sets relative to the $(3,n)$-ETN vector observables for $n=1,\ldots,10$ \label{fig:13f}]{
        \includegraphics[width=0.6\columnwidth]{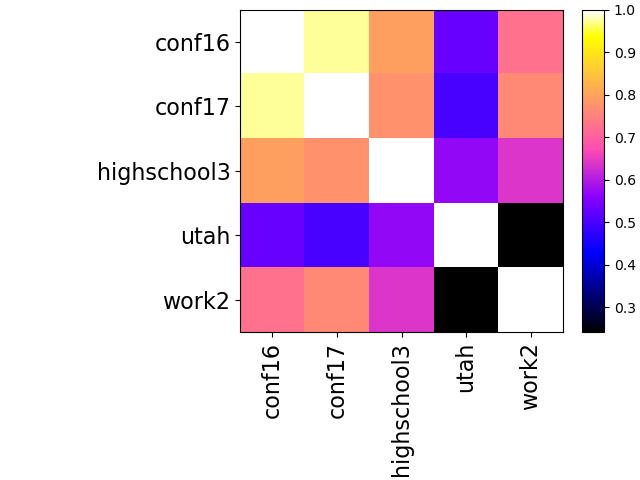}}
    \caption{\label{fig:013}\textbf{Comparison of empirical data sets.}
    Panels \ref{fig:13a} to \ref{fig:13c} show the distributions of the sizes of connected components, of the edge activity durations and of the node interactivity durations measured on the five empirical data sets.   These distribution observables are similar across empirical data sets. The size of connected components (panel \ref{fig:13a}) is an expected exception since it depends strongly on the edge activity, which is not shared across data sets.
    Panels \ref{fig:13d} to \ref{fig:13f} give the similarity matrices between empirical data sets for the various types of observables (\ref{fig:13d}: point observables; 
    \ref{fig:13e}: distribution observables; \ref{fig:13f}: ETN vector observable): the element at row $x$ and column $y$ of a matrix gives the similarity between the observables measured in 
    data sets $x$ and $y$ (the similarity for an ensemble of observables is defined as the product of the similarities measured for each observable). 
    For the point observables, the conferences are clearly separated from the other data sets.
    Panel \ref{fig:13f} shows instead that the motifs are quite robust: different contexts, like a workplace, conferences or a highschool exhibit similar motifs. 
    Only the primary school ``utah'' seems to distinguish itself by its motifs.
    A possible explanation is the age of the social agents at play: in the ``utah'' data set, they are mainly children while in all other data sets, social agents are young adults or older.}
\end{figure*}

On the other hand, distribution and vector observables show similar properties across data sets (as illustrated in Supplemental Figure \ref{fig:013})
despite their differences in number of nodes and edge activity  (the mean number of interactions per time step). In particular the duration of an interaction follows the same statistics in the conferences data sets, where the edge activity is high, as in the workplace data set, where the edge activity is low.
Thus it seems that the duration of a face-to-face interaction is not influenced by how rare those interactions are.

As the number of observables is high, it is useful to introduce a global similarity matrix relative to each type of observable.
We define the similarity between two data sets $\mathcal{D}$ and $\mathcal{D'}$ relative to a given type T of observable as the product of the similarities between those data sets with respect to each observable $\mathcal{O}$ of the chosen type:
\begin{equation}
\text{Sim}_{\text{T}}(\mathcal{D},\mathcal{D'})=\prod_{\mathcal{O},\text{Type}(\mathcal{O})=\text{T}}(1-D[\mathcal{O}]_{\mathcal{D},\mathcal{D'}})
\end{equation}
where $D$ is the distance tensor introduced in the main text.
As we have three types of observables, this gives rise to three similarity matrices visible on Supplemental Figures \ref{fig:13d}, \ref{fig:13e} and \ref{fig:13f}.

\begin{figure*}
    \subfigure[point similarity histogram\label{fig:21a}]{
        \includegraphics[width=0.6\columnwidth]{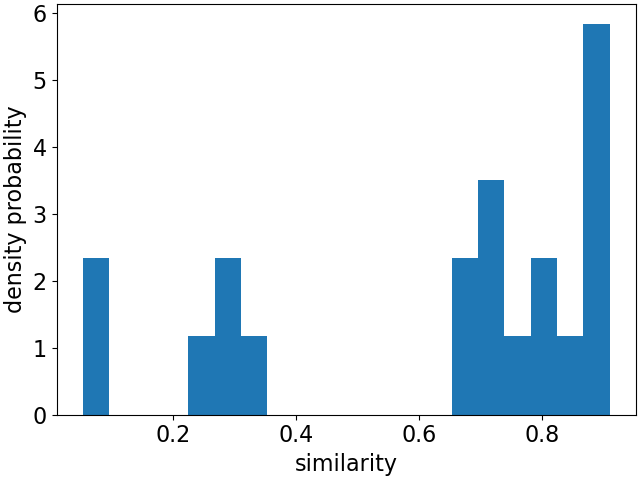}}
    \subfigure[distribution similarity histogram\label{fig:21b}]{
        \includegraphics[width=0.6\columnwidth]{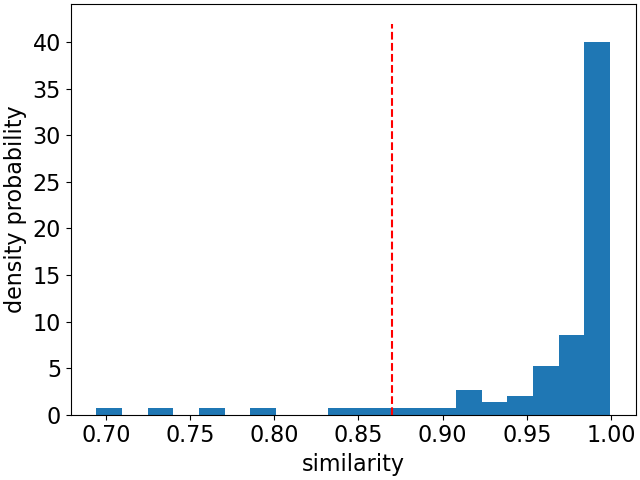}}
    \subfigure[vector similarity histogram\label{fig:21c}]{
        \includegraphics[width=0.6\columnwidth]{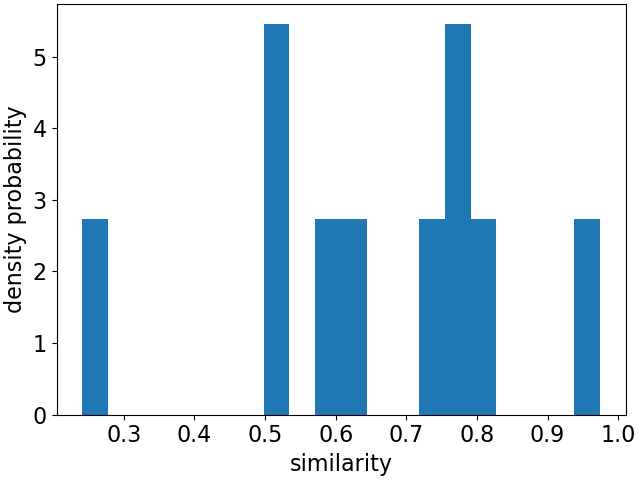}}
    \subfigure[weight of (2,1)-ETN in empirical data sets\label{fig:21d}]{
        \includegraphics[width=0.6\columnwidth]{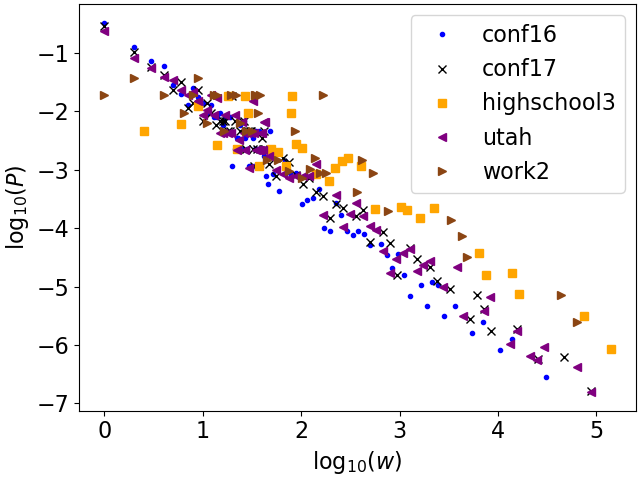}}
    \subfigure[size of connected components in empirical data sets\label{fig:21e}]{
        \includegraphics[width=0.6\columnwidth]{cc_size_XP.png}}
    \subfigure[weight of (3,1)-ETN in empirical data sets\label{fig:21f}]{
        \includegraphics[width=0.6\columnwidth]{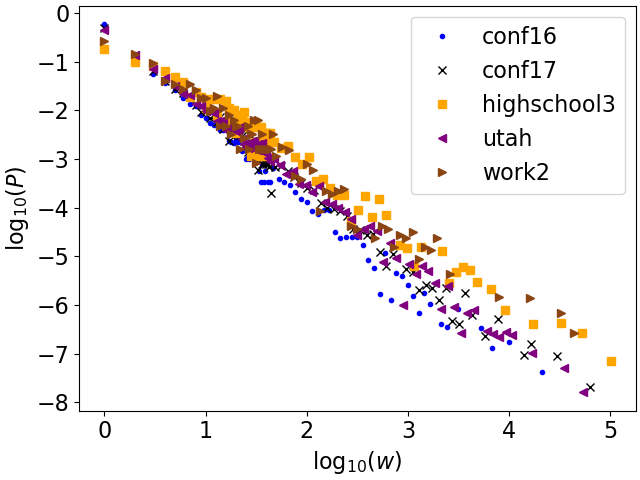}}
    \caption{\label{fig:021}\textbf{Similarity histograms with respect to each type of observable.}
    Panels \ref{fig:21a} to (\ref{fig:21c}: Histograms of similarities between empirical data sets with respect to each observable of a given type.
    In panel \ref{fig:21b}, the vertical red dotted line indicates the arbitrarily chosen threshold value separating lower similarities from higher ones (cf main text).
    Panels \ref{fig:21d} to \ref{fig:21f}: Distribution observables.
    }
\end{figure*}

Supplemental Figure \ref{fig:013} shows that differences exist between empirical data sets for all types of observables. Can we identify the observables responsible for the low similarity observed between some data sets?
To do so, we first collect every similarity value between each pair of data sets with respect to each observable of each type, and build an histogram from it. 
This way, we obtain one similarity histogram per type of observable, as displayed on Supplemental Figure \ref{fig:021}.
We can then choose a threshold and collect all the observables responsible for similarity values below this threshold.
As there are only two point observables and one vector observable, we consider this issue only for the distribution observables. 
It turns out that all similarities below 0.87 are only due to the weights of (2,1) and (3,1)-ETN (``ETN2\_weight'' and ``ETN3\_weight'').
However Supplemental Figures \ref{fig:21d} and \ref{fig:21f} seem to indicate a robustness of these observables, while 
the distribution of connected components sizes varies more  (Supplemental Fig.~\ref{fig:21e}).
We have used a similarity measure based on the well-known
Jensen-Shannon divergence (JSD), a well-known distance. In fact, Supplemental Figure panels \ref{fig:21d}, \ref{fig:21e} and \ref{fig:21f}
indicate that the heads of the distributions are similar for the observable ``cc\_size'' while they are more dissimilar for ``ETN2\_weight'' and ``ETN3\_weight''.
It might thus be interesting to define another measure of similarity than the JSD that would coincide more with the visual inspection of these broad distributions.

\clearpage
\section{Description and illustration of the genetic algorithm}
\label{sec:3}
Recall that a model instance has several free parameters, and that we want to select the instance closest to a given reference data set.
The tuning of the free parameters is performed using a genetic algorithm. In the language of those algorithms, it can be described as follows:
\begin{enumerate}
    \item Decide of a genetic code (sequence of binary digits) for the free parameters.
    In our case, probabilities or floats were coded on ten digits while integers were coded on four.
    \item Initialize randomly a population of genetic sequences.
    We took 40 sequences.
    \item Compute the fitness of each sequence. To minimize computation, we stored in a dictionary the fitness of already evaluated sequences. To evaluate the fitness of a new sequence, first we translate the sequence into parameter values, second we generate the artificial data set using those parameters and our model, third we define the fitness of our sequence as the ETN vector similarity between the generated data set and the associated reference.
    \item Take from the population the sequences with the greatest fitness, discard the others and refill the population with combinations between the selected sequences.
    In practice we took the 7 best sequences, to which we added a sequence generated at random to avoid premature convergence to a local maximum. We combined sequences by pairs, using single-locus and double-loci crossover as well as random mutations. We also imposed the best sequence to survive in the next generation, to ensure the fitness of the best individual is monotonically increasing across generations.
    \item Define a stopping criterion.
    The algorithm stops after 20 generations and returns the parameters which yield the highest fitness. We have checked that this is enough for the fitness to saturate in all cases.
\end{enumerate}

In Supplemental Figure \ref{fig:022}, we illustrate the evolution of the fitness across generations of sequences.
\begin{figure}
    \subfigure[fitness of instances of model V3 w.r.t. reference 'utah' across generations\label{fig:22a}]{
        \includegraphics[width=0.8\columnwidth]{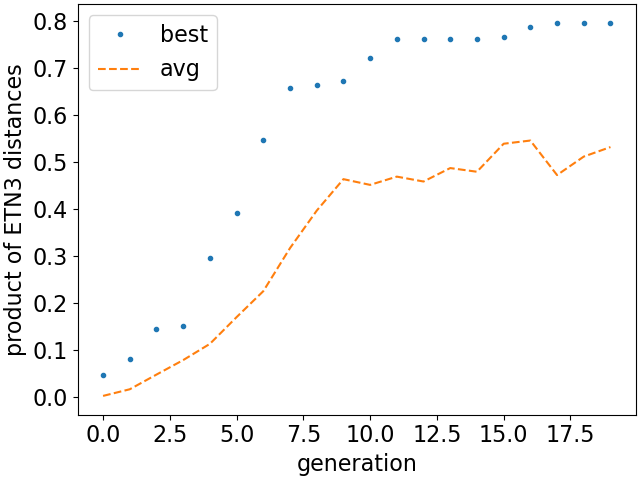}}
    \subfigure[fitness distribution in a random and a tuned population\label{fig:22b}]{
        \includegraphics[width=0.8\columnwidth]{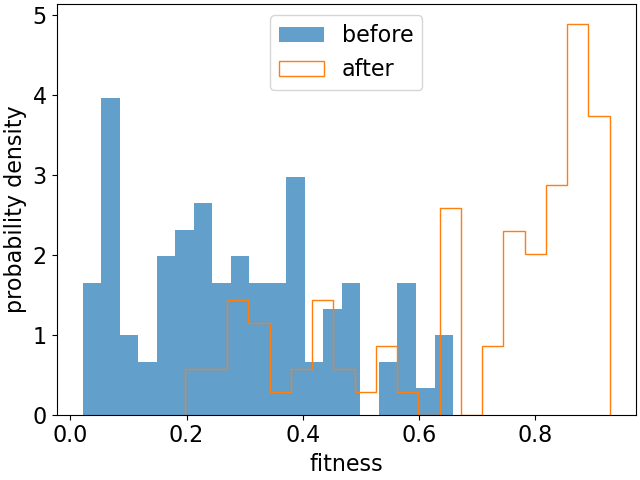}}
    \subfigure[fitness improvement under tuning\label{fig:22c}]{
        \includegraphics[width=0.8\columnwidth]{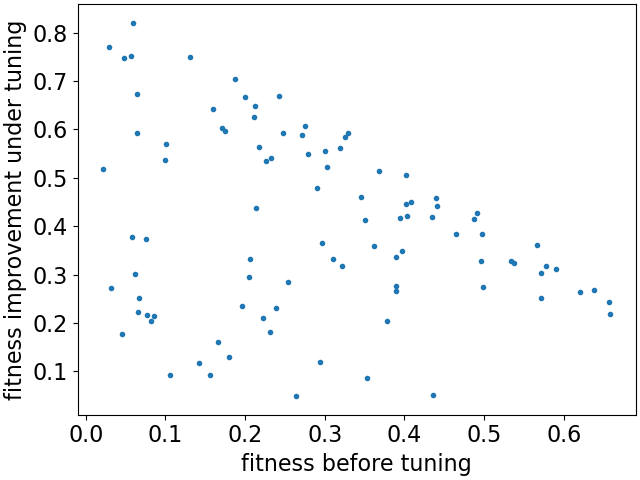}}
    \caption{\label{fig:022}\textbf{Fitness evolution under genetic tuning.}
    Panel \ref{fig:22a}: A typical evolution of fitness during the genetic algorithm. ``best" refers to the sequence with highest fitness at each generation, and ``average"
    to the average population fitness.
    Panel \ref{fig:22b}: For every model version and every reference we collect the best fitness in the population of sequences before and after the genetic tuning. In each case we build the fitness histogram. Before tuning, the most probable value for the fitness is close to zero while after tuning, it lies around 0.8-0.9.
    Panel \ref{fig:22c}: The fitness improvement, defined as the difference between the fitness after tuning and the fitness before, seems to decrease linearly with the initial fitness, bounding the final fitness below a maximum value of 0.8-0.9, independent of the initial fitness.
    }
\end{figure}
Panel \ref{fig:22c} shows the existence of a maximum fitness, around 0.8-0.9 and independent of the initial fitness. This  
indicates the existence of a limit on the ETN vector similarity, intrinsic either to the algorithm used for tuning, or the class of models considered, or both.
Further investigation would be needed to understand the origin of this limit, and how to go beyond.

\begin{figure*}[thb]
\subfigure[cc\_size, V1]{
        \includegraphics[width=0.9\columnwidth]{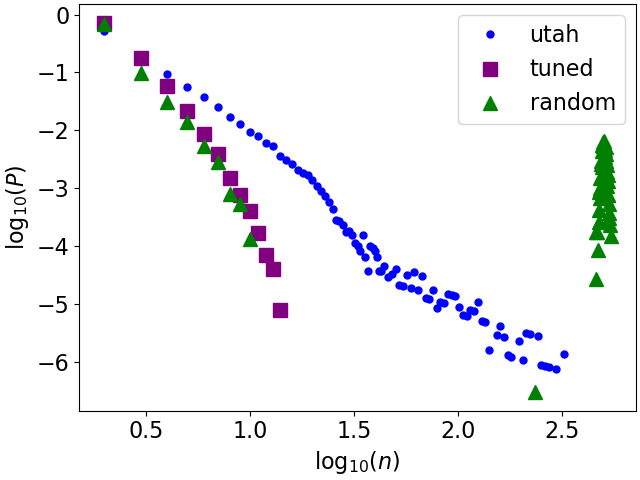}}
\subfigure[cc\_size, V9]{
        \includegraphics[width=0.9\columnwidth]{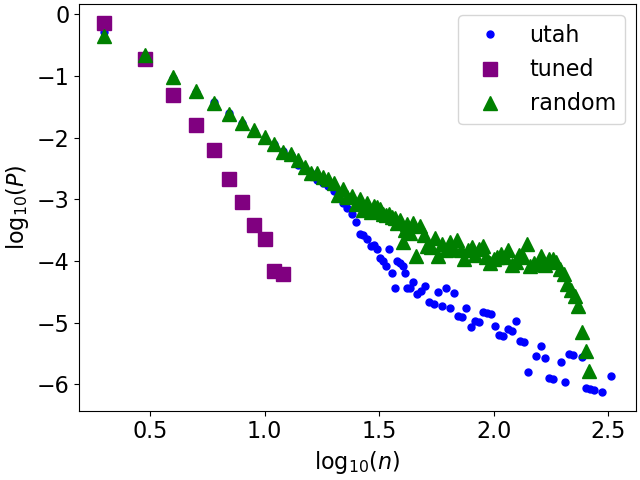}}
\subfigure[edge\_activity, V1]{
        \includegraphics[width=0.9\columnwidth]{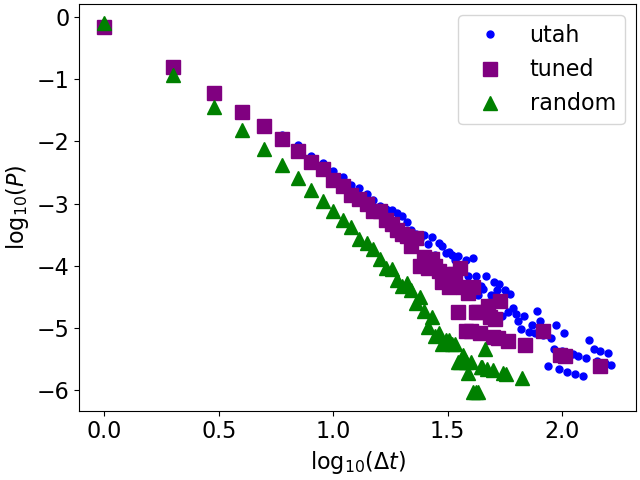}}
\subfigure[edge\_activity, V9]{
        \includegraphics[width=0.9\columnwidth]{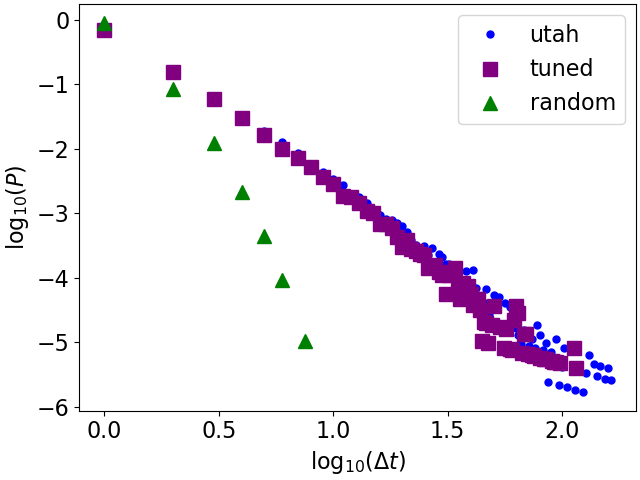}}
\subfigure[ETN3\_weight, V1]{
        \includegraphics[width=0.9\columnwidth]{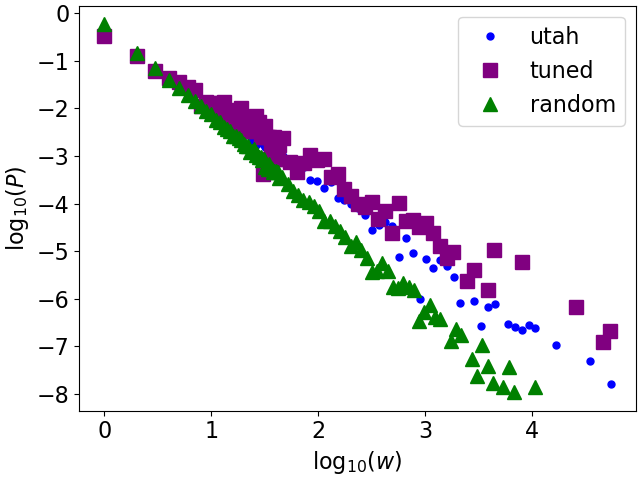}}
\subfigure[ETN3\_weight, V9]{
        \includegraphics[width=0.9\columnwidth]{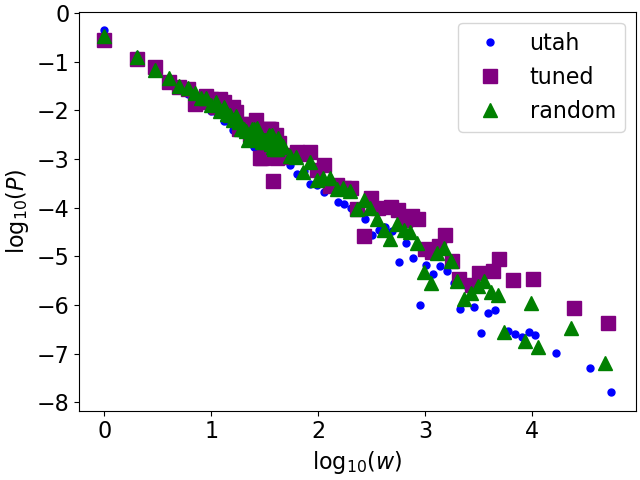}}
    \caption{\label{fig:011}\textbf{Effect of genetic tuning on the observables with multiple realizations.} Each row is associated 
    with a different observable ("cc\_size": size of connected components in $G(t)$) and each column to a model version (V1 and V9). The empirical reference ``utah" is used
    in all cases.    Each panel displays the observable distribution in the reference data set and in two instances of the given model version: one instance with parameter values chosen at random, and one instance with parameters genetically tuned.
    Although the genetic tuning takes only the ETN vector observable into account to compare instances with references, it also impacts the distributions of other observables. Results are similar for other versions and other observables.
%    It seems that the tuning has negatively impacted the "cc\_size" distribution ; however it is expected because in the ADM class, the total number of social interactions at time $t$ does not depend on $t$. This prevents the model distribution to match the empirical case, in which the number of social interactions at a given time follows a large distribution.
    %(see SM). (cf Appendix \ref{sec:3}).
   % Weights for (2,1) and (3,1)-ETN are the observables that undergo the smallest change.
    %Note that even the worst adjacent model (V12) has a realistic distribution for the weights of (3,1)-ETN.
    }
\end{figure*}

Supplemental Figure \ref{fig:011} finally compares several observables measured on random model instances with tuned instances: 
a clear improvement in the similarity of distributions between the model and the empirical data is obtained after tuning: even if the genetic algorithm
takes into account only the similarity with the ETN vector, optimizing this feature allows to improve also on other observables,
even if the agreement is not always perfect (Note that the worst agreement is obtained for the distributions of connected component sizes; however, this observable
also varies substantially from one data set to another).

\clearpage
\section{Variability of the genetic tuning across references}
\label{sec:2}
Each model version is tuned successively to the various empirical data sets taken as reference.
Here we detail how the tuned parameters of a version change with the reference. In particular, 
for each parameter, we want to check whether it is distributed at random after tuning.
If yes it means that this parameter has no impact on ETN motifs, and thus should be removed from the list of free parameters.
If its distribution is a Dirac peaked at a given value, this parameter should also be removed from the list of free parameters, and become a parameter frozen at this value.

\subsection{Integer parameters}

There are three integer parameters:
\begin{itemize}
    \item $c$: sets the egonet growth probability when it is variable according to $p_{g}(i)=\frac{c}{c+d_{i}}$, where $i$ is a social agent and $d_{i}$ its number of neighbours in the social bond graph;
    \item $m$: number of intentional interactions emitted by a social agent when it is constant;
    \item $m^{\text{max}}$: maximum number of intentional interactions when it is variable across social agents.
\end{itemize}
\begin{figure}
    \subfigure[distribution of $c$ \label{fig:23a}]{
        \includegraphics[width=0.9\columnwidth]{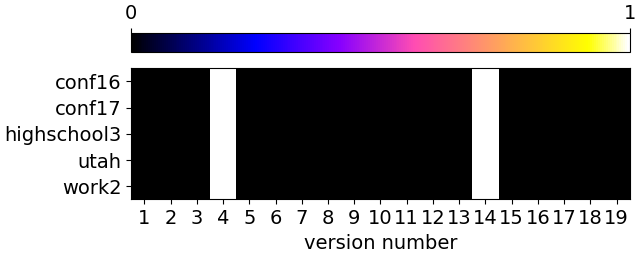}}
    \subfigure[distribution of $m$\label{fig:23b}]{
        \includegraphics[width=0.9\columnwidth]{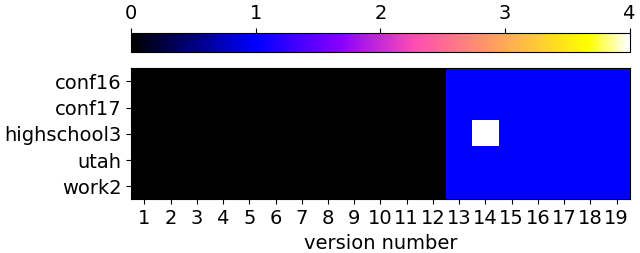}}
 \subfigure[distribution of $m^{\text{max}}$\label{fig:23c}]{
        \includegraphics[width=0.9\columnwidth]{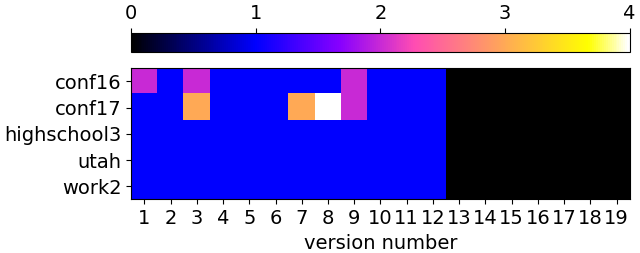}}
    \caption{\label{fig:023}\textbf{Distribution of the integer parameters among the tuned instances.}
    For each integer parameter, we indicate the value it takes for each instance after tuning. We put this value to zero when the parameter is not defined.
    Panel \ref{fig:23a}: the only value selected by the genetic algorithm is $c=1$.
    Panel \ref{fig:23b}: $m$ after tuning equals 1 for all instances where it is defined, except the instance (V14, 'highschool3') for which  $m=4$.
    Panel \ref{fig:23c}: For instances tuned w.r.t. the school or workplace references, $m^{\text{max}}$  equals 1.
    For instances tuned w.r.t. the conference references, $m^{\text{max}}$ varies depending on the model version.
    }
\end{figure}

We show in Supplemental Figure \ref{fig:023} how $c$, $m$ and $m^{\text{max}}$ are distributed among the tuned instances.
The only value selected by the genetic algorithm for $c$ is $c=1$. This makes sense because $c$ can be interpreted as the minimum size of the egonet of a social agent.
Similarly, nodes tend to interact with a few contacts at each time step so it makes sense that the value $m=1$ is preferred by the algorithm (cf panel \ref{fig:23b}). Only for the instance (V14, ``highschool3'') we observe $m\neq1$. A different behaviour is observed for $m^{\text{max}}$. For most instances it equals 1 but for instances tuned w.r.t. the conference references, it takes higher values depending on the model version.
This makes sense because the conference data sets have the highest density, i.e. number of active links per time step.

\subsection{Float parameters}

There are 9 float parameters:
\begin{itemize}
    \item $a$: probability that a node emits intentional interactions;
    \item $\lambda$: sets the probability of removing a directed tie in the social bond graph, based on the tie's weight;
    \item $\alpha$: sets the rate at which the weight of a tie changes in the social bond graph upon activation or inactivation;
    \item $a^{\text{min}}$: lower bound for the probability of emitting intentional interactions;
    \item $a^{\text{max}}$: upper bound for the probability of emitting intentional interactions;
    \item $p_{c}$: sets the probability of dynamic triadic closure;
    \item $p_{u}$: when a node chooses to grow its egonet, probability of growing it by choosing a new partner at random instead of triadic closure;
    \item $p_{g}$: probability that a node grows its egonet; 
    \item $p_{d}$: probability that a node resets its egonet to the empty set.
\end{itemize}

\begin{figure}
    \subfigure[empirical distribution of $p_{u}$\label{fig:24a}]{
        \includegraphics[width=\columnwidth]{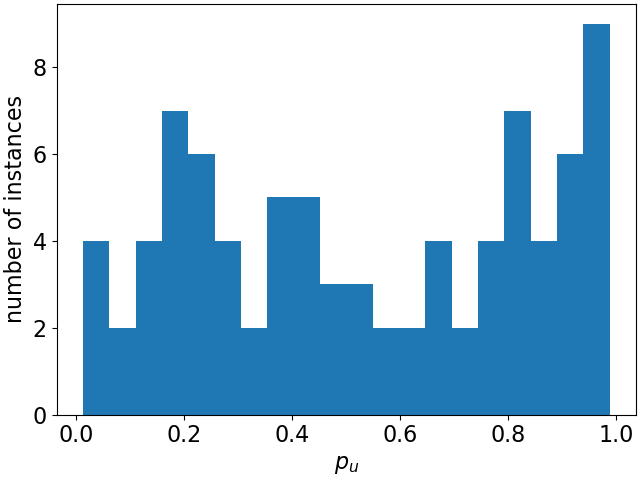}}
    \subfigure[empirical distribution of $\alpha$\label{fig:24b}]{
        \includegraphics[width=\columnwidth]{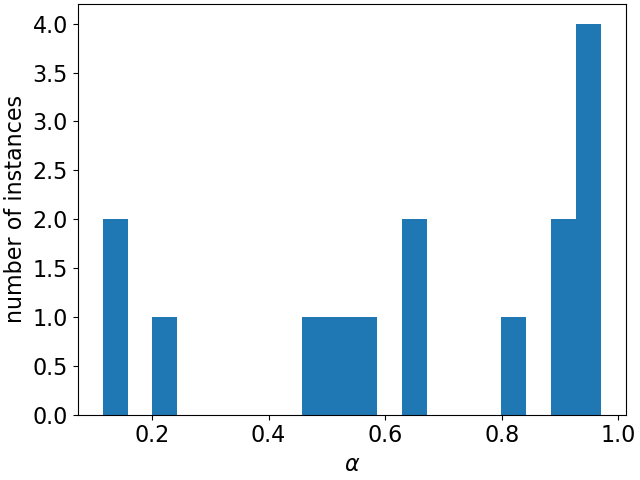}}
    \caption{\label{fig:024}\textbf{Distributions of $p_{u}$ and $\alpha$.} 
    }
\end{figure}

For each parameter we perform a Kolmogorov-Smirnov test to check whether or not its distribution has been altered by the genetic tuning.
As expected, we find that most of the float parameters are strongly affected by the genetic tuning, since their distribution over instances is far from the uniform case after tuning.
Only two parameters are viewed by the KS test as possibly having a uniform distribution: $p_{u}$ and $\alpha$.
To check whether there are false positives, we display their distribution after tuning on Supplemental Figure \ref{fig:024}.
The distribution of $p_{u}$ (panel \ref{fig:24a}) over all instances seems  to be compatible with the uniform case. However it is less clear for the distribution for $\alpha$ (panel \ref{fig:24b}), as many values are missing and a value close to 1 seems preferred. Further investigation would be needed to decide whether or not $p_{u}$ and $\alpha$ are relevant parameters in the  class of models we have considered.

\clearpage
\section{The node weight observable}
\label{sec:4}
In the main text we have considered 13 different observables, in particular the weight of edges in the aggregated network.
We did not consider the node weight (the number of timestamps a node is present with degree non-zero), although this observable has a strong link with an important parameter of the activity-driven class: the distribution of the nodes intrinsic activities.
Recall that the node intrinsic activity of a node $i$ is the probability $a_{i}$ that this node emits intentional interactions at each time step.
The form of the distribution for $a_{i}$ has proven to be relevant, since in this paper we showed that a power-law distribution for node intrinsic activities yielded more realistic statistical properties than a uniform activity $a_{i}=a,\forall i$ (cf V12).
Although it would thus be of fundamental interest to be able to measure the empirical distribution of $a_{i}$, we do not have direct access 
to it but only to the node weight. We can however study 
how the node weight and the node intrinsic activity distributions are related in the models, and on the other hand we can compare the distributions 
of the node weight in the models and in the empirical data sets.
\begin{figure*}
    \subfigure[conf16\label{fig:26a}]{
    \includegraphics[width=\columnwidth]{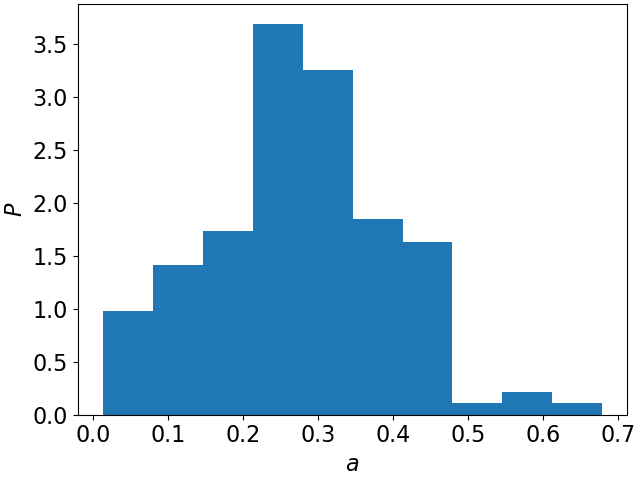}}
    \subfigure[utah\label{fig:26b}]{
    \includegraphics[width=\columnwidth]{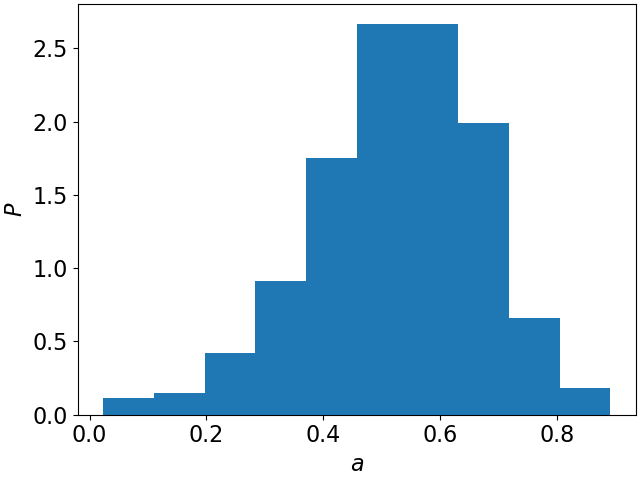}}
    \subfigure[V1, conf16\label{fig:26c}]{
    \includegraphics[width=\columnwidth]{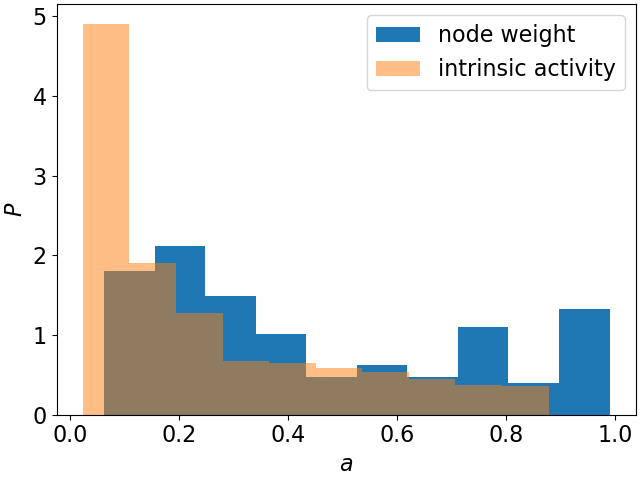}}
    \subfigure[V12, utah\label{fig:26d}]{
    \includegraphics[width=\columnwidth]{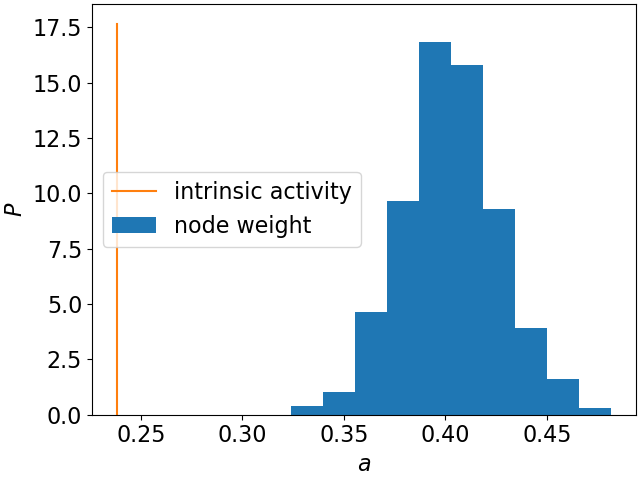}}
    \caption{\label{fig:026}\textbf{Node weight distribution for some data sets and model instances.}
    }
\end{figure*}

We show in Supplemental Figures \ref{fig:026} and \ref{fig:028} the node weight distributions for some empirical and artificial data sets (rescaled to have a support between 0 and 1). 
These distributions differ from one empirical reference to another and between models. Empirical references tend to have a more peaked distribution than the models (except V12). 
%For the models, except for
%version 12, the distributions tend to be broader.

Moreover, Supplemental Figures \ref{fig:026} and \ref{fig:028} show that the distributions of intrinsic activity and node weight differ. A shift towards higher values is apparent, as expected 
since an active node can  interact with an inactive node. The shape of the distribution is also different, and the 
 node weight distribution is closer to the intrinsic activity one when the reference contains less interactions.

\begin{figure*}
    \subfigure[V1, conf16\label{fig:28a}]{
    \includegraphics[width=0.18\textwidth]{distr_both_V1conf16.png}}
    \subfigure[V1, conf17\label{fig:28b}]{
    \includegraphics[width=0.18\textwidth]{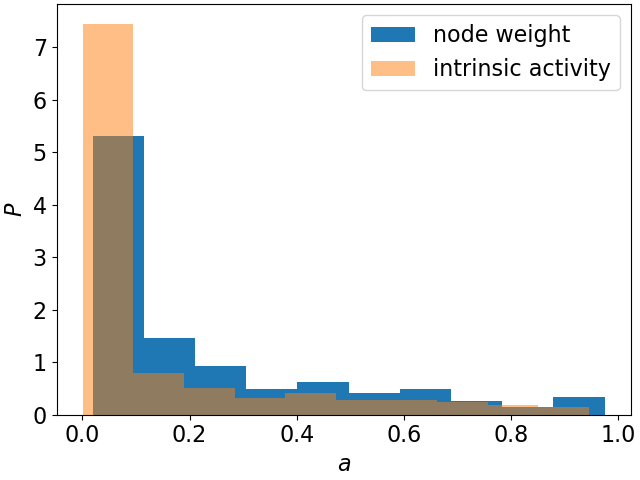}}
    \subfigure[V1, highschool3\label{fig:28c}]{
    \includegraphics[width=0.18\textwidth]{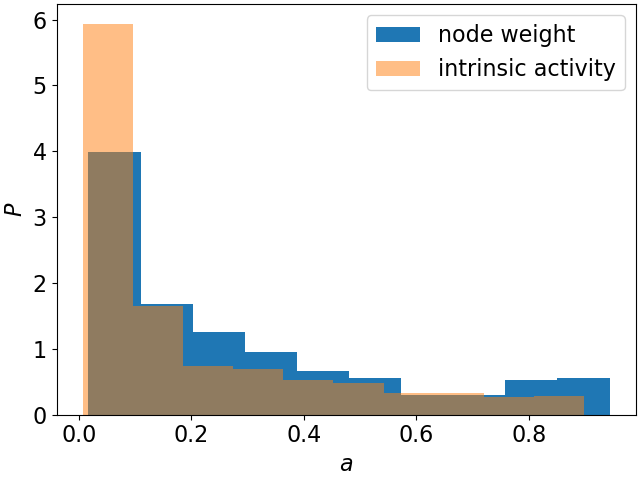}}
    \subfigure[V1, utah\label{fig:28d}]{
    \includegraphics[width=0.18\textwidth]{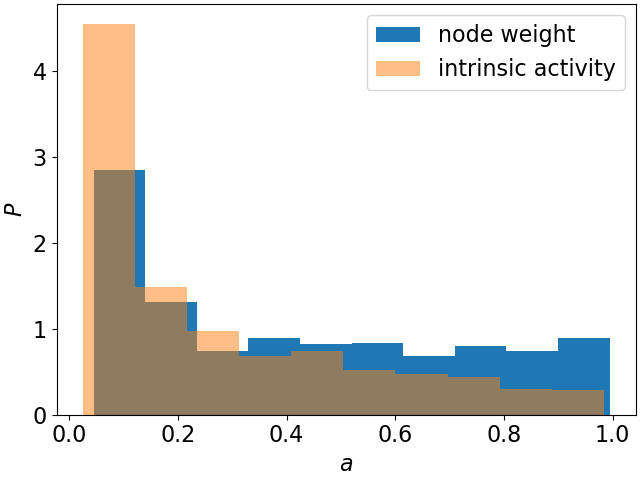}}
    \subfigure[V1, work2\label{fig:28e}]{
    \includegraphics[width=0.18\textwidth]{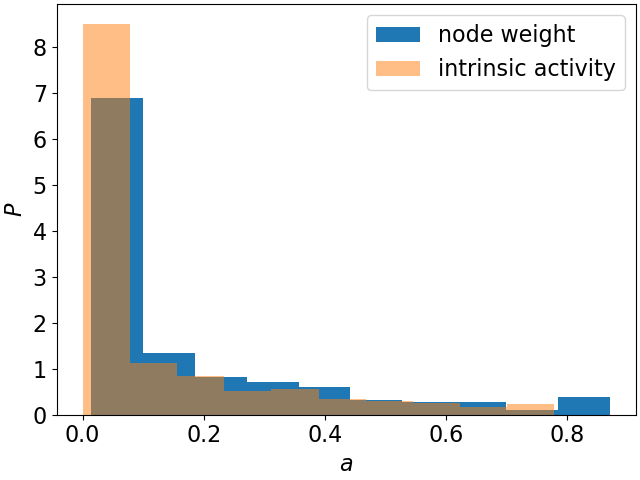}}
    \vskip\baselineskip
    \subfigure[V14, conf16\label{fig:28f}]{
    \includegraphics[width=0.18\textwidth]{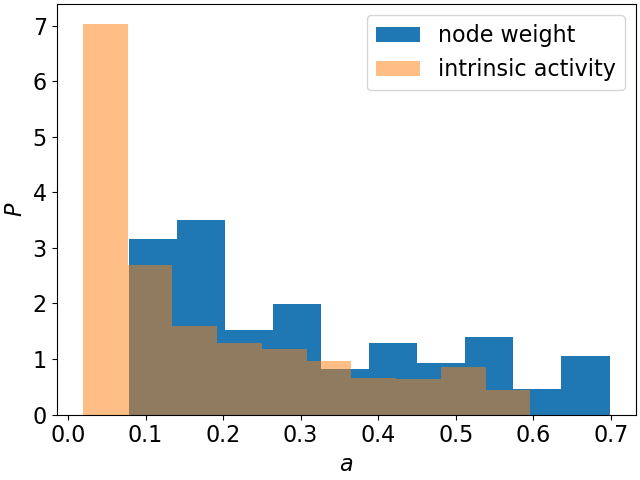}}
    \subfigure[V14, conf17\label{fig:28g}]{
    \includegraphics[width=0.18\textwidth]{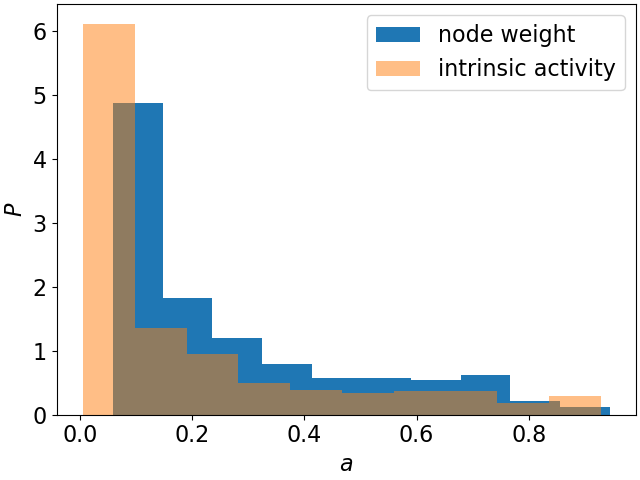}}
    \subfigure[V14, highschool3\label{fig:28h}]{
    \includegraphics[width=0.18\textwidth]{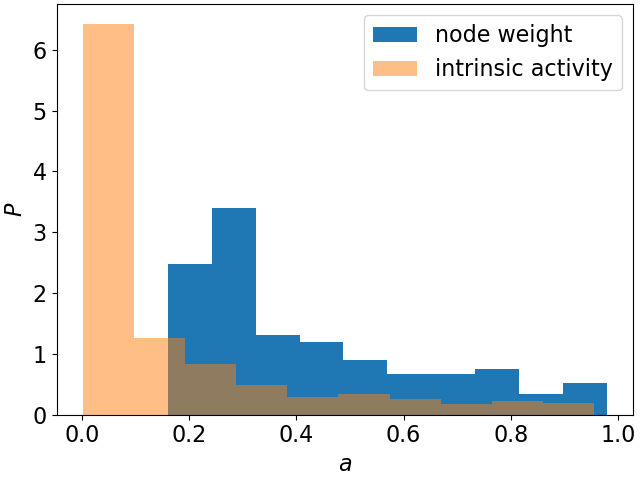}}
    \subfigure[V14, utah\label{fig:28i}]{
    \includegraphics[width=0.18\textwidth]{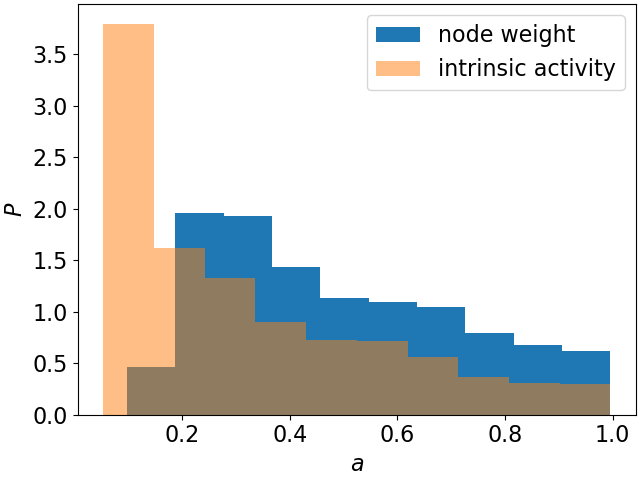}}
    \subfigure[V14, work2\label{fig:28j}]{
    \includegraphics[width=0.18\textwidth]{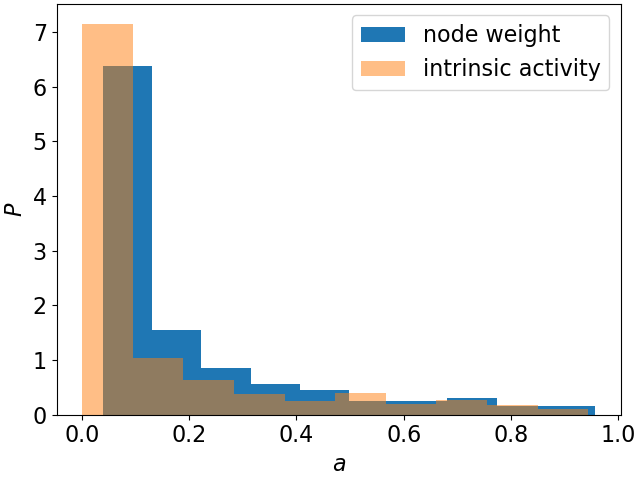}}
    \caption{\label{fig:028}\textbf{Comparison of node weight and intrinsic activity distributions for versions 1 and 14 tuned to the various empirical data sets.}
    }
\end{figure*}

To summarize:
\begin{itemize}
    \item the node weight distribution varies across empirical data sets;
    \item it is different between references and model instances;
    \item for a chosen version, the denser a reference data set, the more different are the node weight and intrinsic activity distributions for the associated model instances.
\end{itemize}

We confirm these observations using Kolmogorov-Smirnov (KS) tests between observations. We
 define similarity matrices between data sets as follows:
A data set is declared similar to another if the p value returned by the KS test of the two-sided hypothesis of identity between their node weight CDF is greater than 0.05. 
In this case, their similarity is set to 1. Otherwise they are declared dissimilar, and their similarity is set to 0.
Results are displayed in Supplemental Figure \ref{fig:029}.
\begin{figure*}
    \subfigure[comparison of node weight raw distributions in empirical data sets\label{fig:29a}]{
    \includegraphics[width=0.32\textwidth]{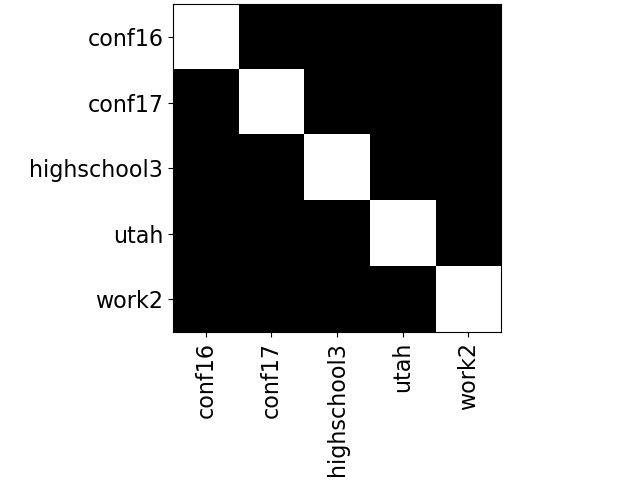}
    }
    \subfigure[comparison of node weight linearly rescaled distributions in empirical data sets\label{fig:29b}]{
    \includegraphics[width=0.32\textwidth]{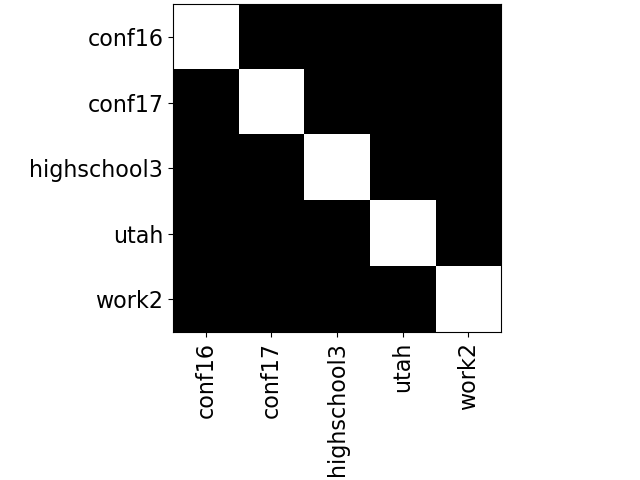}
    }
    \subfigure[similarity matrix between model instances associated to the reference 'conf16'\label{fig:29c}]{
    \includegraphics[width=0.32\textwidth]{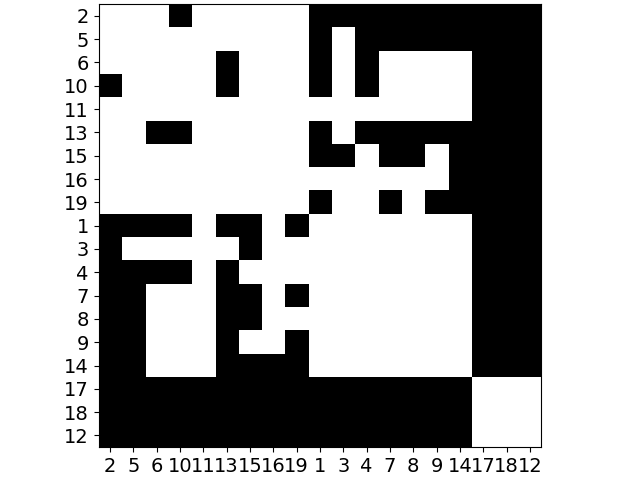}
    }
    \subfigure[similarity matrix between model instances associated to the reference 'highschool3'\label{fig:29d}]{
    \includegraphics[width=0.3\textwidth]{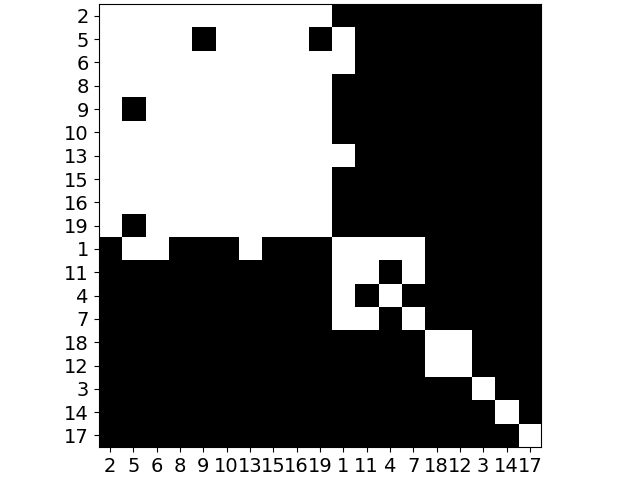}
    }
    \subfigure[similarity matrix between model instances associated to the reference 'work2'\label{fig:29e}]{
    \includegraphics[width=0.3\textwidth]{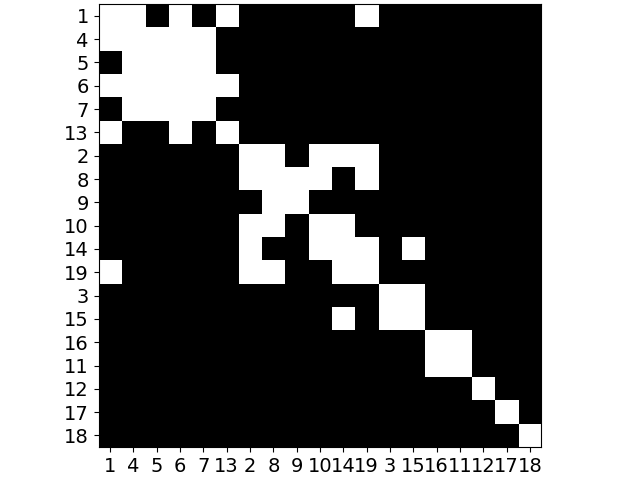}
    }
    \caption{\label{fig:029}\textbf{KS test of the two-sided hypothesis of identity between the node weight CDF in models and empirical data sets.}
    For each similarity matrix, a white box refers to a similarity of 1 and a black box refers to a similarity of 0.
    Panels \ref{fig:29a} and \ref{fig:29b}: According to the KS test, empirical data sets are all different from each other, in accordance with the observed fact that the node weight is not a universal observable.
    In panel \ref{fig:29b}, the node weight has been linearly rescaled between 0 and 1 before performing the KS test.
    Panels \ref{fig:29c} to \ref{fig:29e}: Before performing the test, the node weight has been linearly rescaled between 0 and 1. Contrary to the case of empirical references, we observe large communities, meaning the node weight observable is shared between many models despite their mechanistic differences.
    The models for which the intrinsic activity is a Dirac are the versions 12, 17 and 18.
    }
\end{figure*}
We also consider rescaled node weight (between 0 and 1 or not) to compare the shapes of the distributions.
Even after rescaling, empirical data sets are found to be all dissimilar to each other (cf fig \ref{fig:29b}).
For the models instead, groups with similar rescaled node weight distributions are obtained (cf fig \ref{fig:29c}, \ref{fig:29d} and \ref{fig:29e}).
It seems that models separate into two main families: models with a Dirac law for the intrinsic activity (models 12, 17 and 1) and models with a power-law.
Thus similar distributions for the intrinsic activity are associated to similar distributions for the node weight observable.

\clearpage
\newpage

\section{Composite models}

To do better than versions 1 and 9, we need to explore the ADM class further than one hypothesis away from the baseline. However testing all possible combinations of hypotheses would require large computational resources. Instead, we want to check whether the signatures combine when we combine hypotheses, or more precisely, if $\Delta s(\text{hyp})$ denotes the score variation of hypothesis ``hyp'' with respect to the baseline for a given group of observables, whether:
$$
\begin{cases}
\forall~\text{hyp1},\text{hyp2}, \Delta s(\text{hyp1})>0,\Delta s(\text{hyp2})>0,\\
\Delta s(\text{hyp1},\text{hyp2})\underset{??}{>}\max(\Delta s(\text{hyp1}),\Delta s(\text{hyp2}))
\end{cases}
$$

As we cannot check this assumption for all possible combinations of observables, we will test it on some combinations that we expect to perform better and on some combinations that we expect to perform worse than the baseline. We can determine such expectations from what we learned in the previous sub-subsection:
\begin{itemize}
    \item contextual interactions are relevant as pure noise: $R=W=I$ (as V7 has a better score than the baseline V1);
    \item social context as implemented here is relevant: $c_{ij}$ should \textit{not} be set to one (V1 has a better score than V5);
    \item the social tie rate should be shared by all nodes: $\alpha_{ij}=\beta_{ij}=\alpha$ (as V9 has a better score than V1);
    \item all active nodes should emit the same number of intentional interactions: $m_{i}=m$ (as V13 has a better score than V1);
    \item the node activity should be drawn from the power-law of the baseline, and \textit{not} be shared by all nodes (as V1 has a better score than V12);
    \item edges have to be pruned, rather than nodes (as V1 has a better score than V3);
    \item the egonet growth rate should be constant (as V1 has a better score than V4).
\end{itemize}

Thus, by adding to the baseline model only the changes in mechanisms that yield separately an improvement of the performance, we deduce that the composite model
7+9+13 is expected to perform the best. Moreover, we consider the following combinations:
\begin{itemize}
    \item V15: 2+5+8+13 (same as the original ADM but with edge pruning and constant egonet growth rate);
    \item V16: 5+8+11+13 (same as V15 but with a heterogeneous exponential Hebbian process instead of linear);
    \item V17: 3+5+8+9+12+13 (simplest version with an exponential Hebbian process);
    \item V18: 2+3+5+8+12+13 (simplest version with a linear Hebbian process).
\end{itemize}
We denote as V19 the best expected case 7+9+13 and we include V14 in the set of composite versions.

\begin{figure}
\subfigure[group I]{
        \includegraphics[width=0.8\columnwidth]{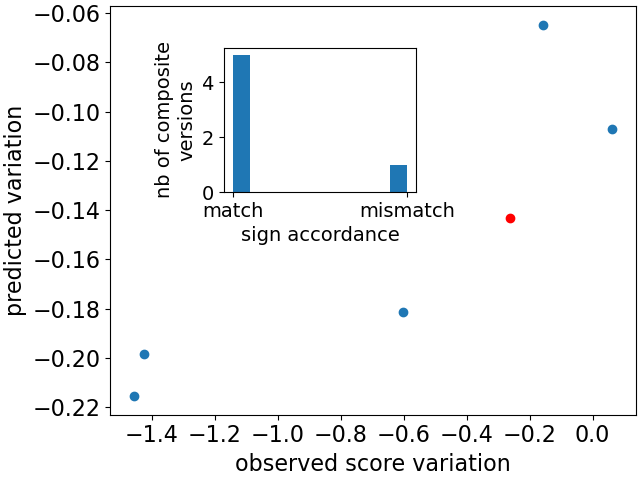}}
 %       \label{fig:18a}
 %   \end{subfigure}
\subfigure[group II]{
        \includegraphics[width=0.8\columnwidth]{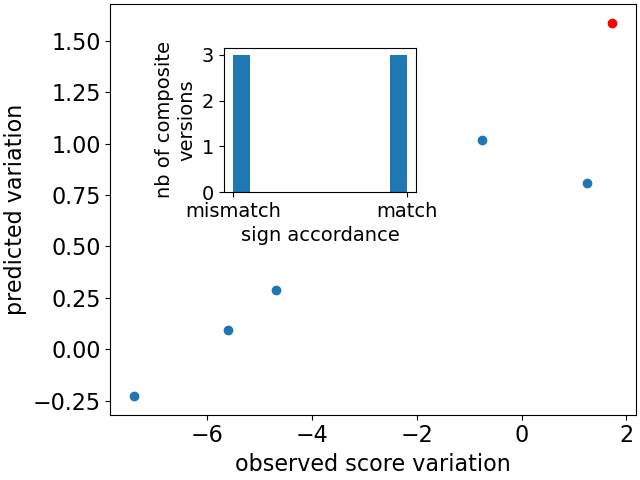}}
  %      \label{fig:18b}
\subfigure[group III]{
        \includegraphics[width=0.8\columnwidth]{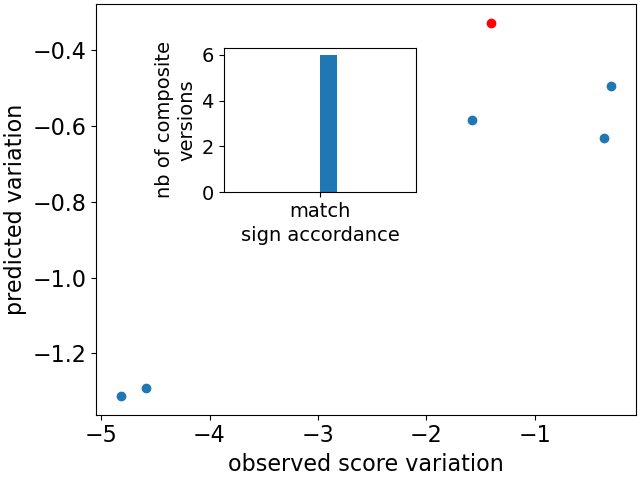}}
  %      \label{fig:18c}
  %  \end{subfigure}
    \caption{\label{fig:018}\textbf{Properties of composite models V14 to V19.} The version 19 is highlighted by a red dot. In this figure, the term ``score'' refers to the variation in score with respect to the baseline model.
    (a): group I observables. (b): group II. (c): group III.
In each panel, the predicted variation ($y$-axis) is obtained by summing the group scores of each component of the composite model. The $x$-axis is the actual score variation. The inset of each panel gives the number of cases in which the predicted and actual variation have the same sign (match) or opposite signs (mismatch).
    An important fraction of mismatches occurs only for group II. However this mismatch in sign is compensated by a linear relation between the predicted and actual scores. Hence it is enough to compute the parameters of this relation to deduce the sign of the score of the composite version from its adjacent components.
    }
\end{figure}

Supplemental Figure \ref{fig:018} leads to the conclusions that (i) the sign of the summed scores of the adjacent versions is a good indicator of the sign of the score of the resulting composite version, and (ii) a linear relation with positive slope approximates well the relation between the actual score of a composite version and the summed scores of its adjacent components.
Hence the relation between the statistical properties of a composite model and its adjacent components seems to be rather simple.

However this relation is not trivial (Supplemental Fig.~\ref{fig:018}): for group II in particular, there is a shift between the score of a composite version and the combination of the scores of its components. 
%To predict the best model, we should have taken this shift into account. 
As a result, V19 is not the best version, as seen in Supplemental
Figure \ref{fig:019} which shows the global rankings of adjacent and composite models (see also next section).

\begin{figure}
    \subfigure[averaging the ranks]{
        \includegraphics[width=\columnwidth]{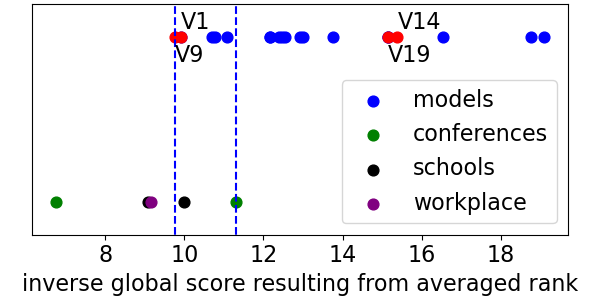}}
    
    \subfigure[averaging the scores]{
        \includegraphics[width=\columnwidth]{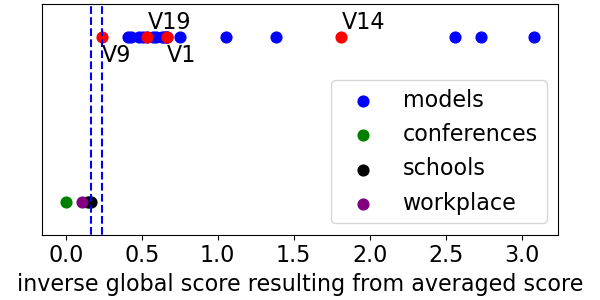}}

    \caption{\label{fig:019}\textbf{Ranking of adjacent and composite models.} The best version is now V9 according to both ranking strategies.
    We highlight the baseline V1, the original ADM V14 and the composite version V19, that was expected to be the best under the hypothesis that the score of a composite version is the sum of the group scores of its adjacent components. As this hypothesis does not fully hold (Supp. Fig.~\ref{fig:018}), V19 has a good score but is not the best.
    (a): The $x$-coordinate is given by the model rank averaged over all observables.
    (b): The $x$-coordinate is given by the opposite of the averaged score, shifted by the maximum averaged score to take positive values.
    }
\end{figure}

\clearpage
\newpage

\section{Rankings of models}
\label{sec:5}
In the main text we have defined a global ranking for models. However, as a different score is computed w.r.t. each observable, we also have one ranking for models per observable.
In  the main text, we have shown that the Kendall similarities between those rankings are small.
However, as the Kendall similarity puts on an equal footing the head and the tail of the rankings, we also give the complete rankings
 w.r.t. each observable in Supplementary Tables \ref{tab:rank14} and \ref{tab:ranks}.

\begin{table}[thb] \begin{tabular}{m{1cm}|m{1cm}|m{1cm}|m{1.2cm}|m{1cm}|m{1cm}|m{1.2cm}|m{1.3cm}|m{1.4cm}|m{1.4cm}|m{1cm}|m{1cm}|m{1cm}}
Clust. & Assort. & ETN3 & cc size & edge act. & nb of events & edge newborn act. & node act. duration & edge inter-activity duration & node inter-activity duration & edge weight & ETN2 W & ETN3 W  \\ \hline
12 & 3  &10 & 6 & { \textcolor{red}{9}}&  { \textcolor{blue}{14}} & 3   &  { \textcolor{red}{9}}& 4  & 7  & 12 & { \textcolor{red}{9}}&  { \textcolor{blue}{14}}\\
 \textcolor{red}{1} &{ \textcolor{blue}{14}} & \textcolor{red}{1}& 10& 8  &   \textcolor{red}{1}  & { \textcolor{red}{9}} &   4 & 13 & 10 & 4  & \textcolor{red}{1}& 8 \\
8 & 12 & 2  &  \textcolor{red}{1} & 2  &  6    & 7   &   6 & 2  &  5 & 11 & 13 & 13\\
11 & 7 & 7  &  \textcolor{red}{9} & 7  & 7     & 13  &  11 &  \textcolor{red}{1}& 13 & 10 & 5  & 7\\
13 & 5 & 11 & 13&  \textcolor{red}{1}  & 2     &  5  &   3 & { \textcolor{red}{9}}& 6  & 5  &  7 &  \textcolor{red}{1}\\ \hline
7 & 13 & { \textcolor{red}{9}}& 2 & 11 & 10    & 11  &   \textcolor{red}{1}&  10& 2  & 8  & 11 &  { \textcolor{red}{9}}\\
3 & 2  & 6  & 4 & 6  & 4     & 12  & 8   &   7& 8  &  \textcolor{red}{1}  & 8 & 6\\ 
10 & 6 & 8  & 5 & 5  & 13    & 8   & 10  & 11 & 3  &  \textcolor{red}{9}  & 3 & 5\\
4 &  \textcolor{red}{1} & 13 & 3 & 10 & 8     & 10  &  2  &  5 & 11 &  6 & 4 & 2\\
5 &  \textcolor{red}{9}  & 5  &11 & 3  &  \textcolor{red}{9}     &  \textcolor{blue}{14}  &  5  &  6 &  4 & 13 &  \textcolor{blue}{14}& 3\\ \hline
 \textcolor{red}{9} & 11 & 3  &12 & 13 & 3     &  2  & 13  &   \textcolor{blue}{14}&  \textcolor{red}{9}  & 3  & 10 & 4\\
2 & 8  &  \textcolor{blue}{14} & \textcolor{blue}{14} & 4  & 11    &  6  & 7   & 8  &   \textcolor{red}{1} &  \textcolor{blue}{14} & 2 &  11 \\
6 & 10 & 12 & 8 &  \textcolor{blue}{14} & 5     & 4   &  \textcolor{blue}{14}  &  3 &  \textcolor{blue}{14} & 2  & 6 & 10 \\
 \textcolor{blue}{14} &  4 &  4 & 7& 12 & 12    &  \textcolor{red}{1}  & 12  & 12 & 12& 7   & 12 & 12\\
        \end{tabular}
 \caption{\textbf{Rankings of the 14 adjacent models tested w.r.t. each observable.}\label{tab:rank14}}
\end{table}

\begin{table}[thb] \begin{tabular}{m{1cm}|m{1cm}|m{1cm}|m{1.2cm}|m{1cm}|m{1cm}|m{1.2cm}|m{1.3cm}|m{1.4cm}|m{1.4cm}|m{1cm}|m{1cm}|m{1cm}}
Clust. & Assort. & ETN3 & cc size & edge act. & nb of events & edge newborn act. & node act. duration & edge inter-activity duration & node inter-activity duration & edge weight & ETN2 W & ETN3 W  \\ \hline
18 & 3 & 10 & 15 & \textcolor{red}{9}& \textcolor{blue}{14} & 3 & \textcolor{red}{9}& 4 & 16 & 12 & \textcolor{red}{9}& \textcolor{blue}{14}\\
12 & \textcolor{blue}{14} & \textcolor{red}{1}& 6 & 19 & \textcolor{red}{1}& \textcolor{red}{9}&4 & 13 & 7 & 4 & \textcolor{red}{1}& 8 \\
\textcolor{red}{1} & 18 & 2 & 10 & 8 & 6 & 7 & 6 & 2 & 15 & 11 & 16 & 13\\
17 & 17 & 7 & \textcolor{red}{1}& 2 & 7 & 13 & 11 & \textcolor{red}{1}& 10 & 10 & 13 & 7\\
8 & 12 & 11 & \textcolor{red}{9}& 7 & 15 & 19 & 3 & \textcolor{red}{9}& 5 & 5 & 5 & \textcolor{red}{1}\\ \hline
11 & 7 & \textcolor{red}{9}& 13 & 15 & 2 & 5 & \textcolor{red}{1}& 16 & 13 & 15 & 7 & \textcolor{red}{9}\\
13 & 5 & 15 & 2 & \textcolor{red}{1}& 10 & 16 & 8 & 15 & 6 & 8 & 11 & 6\\ 
7 & 15 & 6 & 4 & 11 & 4 & 11 & 10 & 10 & 2 & \textcolor{red}{1}& 15 & 5\\
3 & 13 & 8 & 5 & 6 & 13 & 15 & 19 & 7 & 8 & \textcolor{red}{9}& 8 & 2\\
10 & 2 & 13 & 3 & 5 & 8 & 12 & 15 & 11 & 3 & 6 & 3 & 3\\ \hline
4 & 6 & 5 & 16 & 10 & \textcolor{red}{9}& 18 & 2 & 5 & 11 & 19 & 4 & 4\\
5 & 19 & 3 & 11 & 3 & 3 & 17 & 5 & 6 & 19 & 16 & \textcolor{blue}{14} & 19  \\
15 & \textcolor{red}{1}& 16 & 12 & 13 & 11 & 8 & 13 & \textcolor{blue}{14} & 4 & 13 & 19 & 17\\
9 & \textcolor{red}{9}& 19 & \textcolor{blue}{14} & 16 & 5 & 10 & 7 & 8 & \textcolor{red}{9}& 3 & 18 & 11\\
2 & 16 & \textcolor{blue}{14} & 19 & 4 & 19 & \textcolor{blue}{14} & 16 & 19 & \textcolor{red}{1}& \textcolor{blue}{14} & 10 & 15\\
19 & 11 & 12 & 8 & \textcolor{blue}{14} & 16 & 2 & \textcolor{blue}{14} & 3 & \textcolor{blue}{14} & 2 & 2 & 10\\
6 & 8 & 18 & 7 & 17 & 12 & 6 & 12 & 12 & 17 & 7 & 6 & 18\\
16 & 10 & 17 & 17 & 12 & 18 & 4 & 18 & 17 & 18 & 17 & 12 & 12\\
\textcolor{blue}{14} & 4 & 4 & 18 & 18 & 17 & \textcolor{red}{1}& 17 & 18 & 12 & 18 & 17 & 16  \\
        \end{tabular}
 \caption{\textbf{Rankings of the 19 models tested w.r.t. each observable.}\label{tab:ranks}}
\end{table}

\clearpage
\newpage

\providecommand{\noopsort}[1]{}\providecommand{\singleletter}[1]{#1}%